\documentstyle[aas2pp4,lscape,astrobib]{article}

\lefthead{Alcock et al.}
\righthead{Microlensing Optical Depth}

\begin{document}

\title{The MACHO Project: Microlensing Optical Depth towards the Galactic Bulge from Difference Image Analysis}

\author{C. Alcock\altaffilmark{1,2}, R.A. Allsman\altaffilmark{3}, D.R. Alves\altaffilmark{4},
T.S. Axelrod\altaffilmark{5}, A.C. Becker\altaffilmark{6}, D.P. Bennett\altaffilmark{2,7},\\
K.H. Cook\altaffilmark{1,2}, A.J. Drake\altaffilmark{1,5}, K.C. Freeman\altaffilmark{5}, 
M. Geha\altaffilmark{1}, K. Griest\altaffilmark{2,8}, 
M.J. Lehner\altaffilmark{9},\\ S.L. Marshall\altaffilmark{1,2}, D. Minniti\altaffilmark{1,10}, 
C.A. Nelson\altaffilmark{1,2}, B.A. Peterson\altaffilmark{5}, P. Popowski\altaffilmark{1},\\
M.R. Pratt\altaffilmark{11},
P.J. Quinn\altaffilmark{12}, C.W. Stubbs\altaffilmark{2,5,6}, W. Sutherland\altaffilmark{13},
A.B. Tomaney\altaffilmark{6},\\
T. Vandehei\altaffilmark{8}, {\sc and} D.L. Welch\altaffilmark{14}}
\author{\bf(The MACHO Collaboration)}
\vspace*{0.2cm}

\altaffiltext{1}{Lawrence Livermore National Laboratory, Livermore, CA 94550} 
\altaffiltext{2}{Center for Particle Astrophysics, University of California, Berkeley, CA 94720} 
\altaffiltext{3}{Supercomputing Facility, Australian National University, Canberra, ACT 0200, Australia}
\altaffiltext{4}{Space Telescope Science Institute, 3700 San Martin Dr, Baltimore, MD 21218}
\altaffiltext{5}{Research School of Astronomy and Astrophysics, Weston Creek, Canberra, ACT 2611, Australia}
\altaffiltext{6}{Department of Astronomy and Physics, University of Washington, Seattle, WA 98195}
\altaffiltext{7}{Department of Physics, University of Notre Dame, Notre Dame, IN 46556}
\altaffiltext{8}{Department of Physics, University of California, San Diego, CA 92093}
\altaffiltext{9}{Department of Physics, University of Sheffield, Sheffield S3 7RH, UK}
\altaffiltext{10}{Departmento de Astronomia, P. Universidad Cat\'olica, Casilla 104, Santiago 22, Chile}
\altaffiltext{11}{Center for Space Research, MIT, Cambridge, MA 02139}
\altaffiltext{12}{European Southern Observatory, Karl Schwarzchild Str.\ 2, D-85748 G\"{a}rching bei M\"{u}nchen, Germany}
\altaffiltext{13}{Department of Physics, University of Oxford, Oxford OX1 3RH, UK}
\altaffiltext{14}{Department of Physics and Astronomy, McMaster University, Hamilton, ON L8S 4M1, Canada}

\newpage
\begin{abstract}

  We present the microlensing optical depth towards the Galactic bulge based
  on the detection of 99 events found in our Difference Image Analysis (DIA)
  survey. This analysis encompasses three years of data, covering $\sim 17$
  million stars in $\sim 4$ deg$^2$, to a source star baseline magnitude
  limit of $V = 23$. The DIA technique improves the quality of photometry
  in crowded fields, and allows us to detect more microlensing events with
  faint source stars. We find this method increases the number of detection
  events by $85\%$ compared with the standard analysis technique.
  DIA light curves of the events are presented and the microlensing fit
  parameters are given. The total microlensing optical depth is estimated to
  be $\tau_{total}= 2.43^{+0.39}_{-0.38}\times 10^{-6}$
  averaged over 8 fields centered at $l=2\fdg68$ and $b=-3\fdg35$. 
  For the bulge component we find $\tau_{bulge}=3.23^{+0.52}_{-0.50}\times 10^{-6}$ 
  assuming a $25\%$ stellar contribution from disk sources.  These optical
  depths are in good agreement with the past determinations of the MACHO
  \shortcite{ALC97a} and OGLE \shortcite{USKK94d} groups, and are higher
  than predicted by contemporary Galactic models.  We show that our observed
  event timescale distribution is consistent with the distribution expected
  from normal mass stars, if we adopt the stellar mass function of Scalo
  (1986) as our lens mass function. However, we note that as there is still
  disagreement about the exact form of the stellar mass function, there is
  uncertainty in this conclusion.  Based on our event timescale distribution
  we find no evidence for the existence of a large population of brown
  dwarfs in the direction of the Galactic bulge.

\end{abstract}


\section{\sc Introduction}

Over the past seven years the MACHO group has been making observations
of the Galactic bulge in order to determine some of the
fundamental properties of our Galaxy. 
The Milky Way is expected to be an
SAB(rs)bc or SAB(r)bc type spiral galaxy \shortcite{DEV64,Fux97} with four
spiral arms \shortcite{VA95}. However, very little is known about the mass
distributions of the various components of our Galaxy (bulge, spheroid,
disk, halo).  Galactic microlensing surveys provide some insight into the
structure and dynamics of the inner Galaxy, spiral arms and the halo.
Unlike most types of observation, the presence of lensing objects can be
detected independent of their luminosities.  Microlensing is sensitive to
the mass distribution rather than light, this makes microlensing a powerful
way of investigating the mass density within our Galaxy. Furthermore,
microlensing can be used to investigate the stellar mass function to the
hydrogen burning limit, both within our Galaxy and other nearby galaxies.

The amplification of a source star during gravitational microlensing
is related to the projected lens-source separation $u$ normalised 
by the angular Einstein Ring radius $R_{E}$. This is given by

\begin{equation}\label{chisw}
A(u) = \frac{u^{2}+2}{u \sqrt{u^{2}+4}}.
\end{equation}

\noindent
The timescale of a microlensing event, $\hat t$, is characterised by the time
it takes for the Einstein ring associated with a foreground compact lensing
object, to transit a background source star at velocity $v_{\perp}$.  The
size of the Einstein ring, for a lens with mass $M$ (in $\rm M_{\odot}$), an
observer-lens distance $D_{d}$, and a source-observer distance $D_{s}$, is
given by

\begin{equation}
R_{E} = 2.85 {\rm AU} \sqrt{\frac{M D_{d}(1-\frac{D_{d}}{D_{s}})}{1\,\rm kpc}}.
\label{chiswolsonx}
\end{equation}

\noindent
The lensing timescale is $\hat t \equiv 2R_{E}/v_{\perp}$.
Hence, if $R_{E}$ were known, this would enable us to constrain some of the
physical parameters of a microlensing event ($M$, $D_{s}$, $D_{d}$).
However, $R_{E}$ is not known, generally, so it is not possible to determine
the lens masses from individual microlensing events.  Nevertheless, under
special circumstances it is possible to impose additional constraints on these
microlensing event parameters when quantities such as, the physical size of
the source star \cite{AAAA99} or the projected transverse velocity of the
lens \cite{ALC95a} are measured.

Photometry of the stars monitored by the MACHO project has previously
been carried out using a fixed position PSF photometry package SoDoPhot (Son
of DoPhot, \shortciteNP{BEN93}).  In 1996 we introduced a second reduction
method, Difference Image Analysis (hereafter DIA). The DIA technique enables
us to detect microlensing events which go undetected with the SoDoPhot
photometry because the events are due to stars which are too faint to be detected
when unlensed. This technique follows on from the work of \citeN{Crotts92},
\citeN{PHL95} and \citeN{TC96}, and allows us to detect and perform accurate
photometry on these new microlensing events found in the reanalysis of bulge
images.

Recently, the MACHO and OGLE groups reported that the microlensing optical
depth towards the Galactic bulge was a factor of 2 larger than expected from
stellar number density.  That is, the optical depth found by OGLE is
$3.3\pm{1.2} \times 10^{-6}$ \shortcite{USKK94d} and by MACHO is
$3.9^{+1.8}_{-1.2} \times 10^{-6}$ for 13 clump giant source star events out
of a 41 event sample \shortcite{ALC97a}.  It was suggested that the size of
these measurements could be explained by the presence of a bar oriented
along our line-of-sight to the bulge (\shortciteNP{PSUS94b,ZSR95}). The
density profile of the proposed bar is given by

\begin{equation}\label{pt1}
\rho_{b} = \frac{M}{20.65 abc}\; {\rm exp}\left (-\frac{w^{2}}{2}\right),
\end{equation}

\noindent
where

\begin{equation}\label{pt2}
w^{4} = \left[\left(\frac{x}{a}\right)^{2} + \left(\frac{y}{b}\right)^{2}\right]^{2} 
+ \left(\frac{z^{\prime}}{c}\right)^{4}.
\end{equation}

\noindent
For the bulge galactocentric coordinates (x,y,z$^{\prime}$):
$x = \cos{\theta} - \eta \cos{b} \cos{(l-\theta)}$,
$y=  -\sin{\theta} - \eta \cos{b}\sin{(l-\theta)}$,
$z^{\prime} = \eta \sin{b}$.
The bar inclination angle $\theta$ is oriented in the direction
of increasing $l$, and $\eta=D_{s}/D_{8.5}$ is the ratio of the 
source distance relative to a galactocentric distance, taken to be
8.5 kpc. The terms $a$, $b$ and $c$ define the bar scale lengths.

The idea that our Galaxy harbours a bar at its centre is not a new one as it
was first suggested by \shortciteN{DEV64} because of the similarity of the
gas dynamics observed in our galaxy with other barred galaxies.
\citeN{BGSBU91} provided further evidence for a bar from star counts.
The DIRBE results of \citeN{DAH95} were also found to be consistent with
this prediction. The presence of such a bar is an important way of
explaining the interaction of the disk, halo and the spiral density waves in
the disk.

A number of authors have adopted a bar into their Galactic models and have 
adopted various values of the bar orientation \shortcite{PSUS94b,Peale98,ZM96} 
and bar mass \shortcite{Peale98,ZM96}. Other authors have also proposed that
large optical depth contributions could come from the disk component \shortcite{EGTB98}, 
or the Galactic stellar mass function (M\'era, Chabrier, \& Schaeffer 1998; 
\shortciteNP{HC98,ZM96}).\nocite{MER98a} 

In this paper we present a new value for the microlensing optical depth and
investigate what is known about the Galactic parameters with most influence
the optical depth. In the next section we will detail the
observational setup. In \S 3, we will outline the reduction procedure.  We
will next review the microlensing event selection process in \S 4.  The
results of our analysis are presented in \S 5, and we will discuss how the
microlensing detection efficiency was calculated in \S 6.  The microlensing
optical depth for the sample of fields presently analysed with DIA and for
each of the individual field will be presented in \S 7.  In \S 8, we will
review what is known about the most important factors affecting the observed
optical depth and discuss the implications of our results.  In the final
section we summarise the results of this work.

\section{\sc Observations}

The MACHO observation database contains over 90000 individual observations
of the Galactic bulge and Magellanic Clouds.  The Galactic bulge
dataset consists of $\sim 30000$ observations of 94, $43\arcmin$ $\times$
$43\arcmin$ fields.  The largest set of microlensing events reported to
date was given by \citeN{ALC97e}. This consisted of 41 events
detected from one year of observations in 28 of those Galactic bulge fields.
The observations in this work were taken between March 1995 and August 1997.
We consider $\sim3000$ Galactic bulge observations from eight fields.
The central location of the eight fields 
is
$l = 2\fdg68$, $b =-3\fdg35$.

All observations were taken with the Mount Stromlo and Siding Spring Observatories'
1.3m Great Melbourne Telescope with the dual-colour wide-field {\em Macho
  camera}.  The Macho camera consists of two sets of four CCDs, one for red
band ($R_{M}$) images, and the other for blue band ($B_{M}$) images. These
observations were taken simultaneously by employing a dichroic
beam splitter. Each CCD is 2k by 2k with an on-the-sky pixel size of
$0.63\arcsec$. All bulge observations used 150 second exposures.

The median seeing of the data subset is $\sim2.1\arcsec$ and the median 
sky levels are $B_{M} \sim 1300$ ADU and $R_{M} \sim 2200$ ADU. To improve 
the average data quality of our light curves we have chosen to reject 
$\sim 350$ observations where the seeing of the observation 
was $>4\arcsec$ or $B_{M}$ band sky level was $> 8000$ ADU.
The number of observations for each field varies. The smallest number of
observations in this subset is 204 for field 159 and the largest is 
334 for field 104. Low Galactic latitude fields such as field 104 have a 
higher observing priority than, for example field 159, because the 
microlensing optical depth is expected to be higher closer to the 
Galactic center.

\section{\sc Reduction Technique}

The DIA technique involves matching a good seeing {\em reference} image to
other images of the same field, so called {\em test} images.  The test
images are first spatially registered to the reference images and the PSFs
of the images are matched.  The test images are then photometrically
normalised (flux matched) to the reference image and the images are
differenced to reveal variable stars, asteroids, novae, etc. These variables
and microlensing events are searched for in each difference frame and
photometry is performed on the entire set of difference images (see
\shortciteNP{ALC99b}).  Initial results of our Galactic bulge DIA were
presented for a single field in Alcock {et~al.\ }(\citeyearNP{ALC99c,ALC99b}). In this analysis we combine that data with images from seven
additional fields to determine an accurate value for the microlensing
optical depth towards the Galactic bulge.

Our implementation of the DIA technique closely follows that given in
\citeN{ALC99b}. In short, this involves the initial selection of
$200-300$ bright ``PSF'' stars from the Macho star database.  The centroids
of these stars are coordinate-matched in the reference and test images. The
coordinate transform between the observations is calculated and used to
spatially register the test observation to the reference image.  The PSF
stars are ``cleaned'' of neighbouring stars and then combined to provide a
high signal-to-noise ratio (hereafter S/N) PSF for each observation. Next, the image
matching convolution kernel is calculated from the Fourier Transforms of the
two PSFs employing the IRAF task PSFmatch. The reference image is convolved
with the empirical kernel to seeing-match the images.  The fluxes in
the two observations are matched to account for variances in the atmospheric
extinction, and differences in the observed sky level of the two images. Lastly
a difference image is formed by subtracting one of these images from
the other.  An object detection algorithm is applied to the images with a
three $\sigma$ threshold, in both $R_{M}$-band and $B_{M}$-band
difference images.  The coordinates of objects found within each passband
are matched to sort the real sources from spurious ones, such as
cosmic rays and noise spikes.  The systematic noise is determined from
the residuals of the PSF stars in each difference image. The photon
noise is calculated from the test images and the PSF-matched reference
image.  For further detail the reader is referred to \shortciteN{ALC99b}
(see also \shortciteNP{AL98,ALA99a} for similar type of difference 
image technique).

\subsection{\it Photometry}

The MACHO project has focussed on the detection of gravitational
microlensing events by repeatedly imaging millions of stars in the Galaxy.
The search for microlensing events has provided a consistent set of
photometry for these stars spanning the duration of the experiment. This
data is thus also extremely valuable for studying the properties of the
stellar populations present.

In this analysis of difference images, aperture photometry is performed on
all the objects detected in the difference images. For each field this
aperture photometry database consisted of $\sim 40000$ objects.  As all
images of each field were registered and matched to a single reference
image, the centroid coordinates for each photometry measurement are the same
in each image. No extinction corrections are needed as image fluxes are also
matched in the DIA process. With DIA we are in the unique situation of being
able to perform aperture photometry in a field which is usually very
crowded with non-variable stars. All these constant sources were removed,
leaving behind only the residual noise and the variations from the
relatively small number of variable stars and other transient phenomena,
such as microlensing events (\shortciteNP{ALC99b},
\citeyearNP{ALC99c}).

\begin{figure}[ht]
\epsscale{1.0}
\plotone{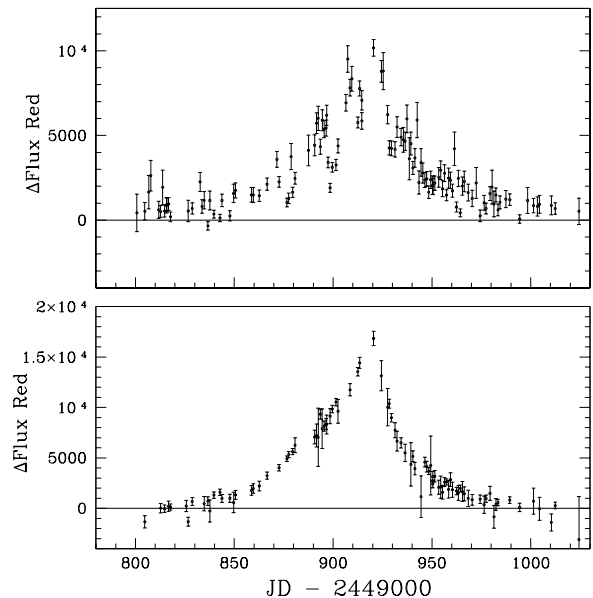}
\figcaption{Red bandpass difference flux light curves
for one of the exotic microlensing events detected in this 
analysis. Top: baseline subtracted SoDoPhot (PSF) photometry.
Bottom: DIA aperture photometry.\label{figBin}}
\end{figure}

At present the photometry data is put into a simple ASCII database since
there are relatively few variable objects in each field relative to the
total number of stars ($\sim 4e4/1e6$).
In Figure \ref{figBin} we present an example of the photometry for one of
the microlensing events\footnote{The asymmetry is most likely due to either
binary lensing or parallax. A similar microlensing event was presented in
\shortciteN{AAAA99}.}.  The oscillations in the SoDoPhot photometry
light curve are due to the offset between the nearest SoDoPhot object centroid
and the centroid of the microlensing source star.  Changes in the seeing can
cause varying amounts of flux to be sequestered into the nearest SoDoPhot PSF.
The SoDoPhot centroid is fixed in position, whereas DIA uses the actual event's 
centroid.

\section{\sc Event Selection}

There are a number of well known theoretical properties of microlensing
which can be used to select events. That is to say, the light curves of
stars affected by microlensing should exhibit: a single, symmetric,
achromatic excursion from a flat baseline. In reality these properties
serve only as a guide, since there are a number of exceptions to each of
them (see \citeNP{DrakeTh}).  For instance, the amplification may not be
achromatic when various types of blending are considered \cite{ALC99b}.  
Multiple peaks can, and do occur for many types of exotic
microlensing events, such as binary lensing and binary source events.
Furthermore, all microlensing events are affected to some degree by the
parallax induced by the Earth's orbital motion around the Sun. In most cases
this is negligible, but in some circumstances the magnified peak shows a
significant asymmetry (see \citeNP{ALC95a}). Therefore, a more
rigorous set of selection criteria is necessary.

In our DIA microlensing event selection process we firstly required that the
events had a total S/N $> 10$, in three or more photometry measurements
bracketing the maximum amplification, in each bandpass. We selected only
those light curves exhibiting a positive excursion from the baseline
flux\footnote{The baseline itself can be negative.}. An initial estimate of
the baseline flux level was determined from the median flux, $F_{med}$, of
the difference flux light curve.  In addition to these criteria, we required
that each light curve passed a set of level-1.0 criteria presented in Table
\ref{tab1}.  These level-1.0 ``cuts'' use the flux values $F_{i}$, and
uncertainties $\sigma_{i}$, measured at time $t_{i}$, to discern whether a
light curve is following a microlensing-like profile, or a more
variable-like curve. These initial cuts are targeted at removing particular
types of variables from the event candidate list based on the general
characteristics of a variable type.

In the level-1.0 selection, one cut is aimed at removing variables by the nature
that they repeat. A cut on the existence of a second peak is an efficient
way of removing variables from a candidate list. However, this also can have
the negative effect of removing binary lensing events.  For this reason we
apply only a loose cut on the occurrence of a second deviation from
baseline.  This cut can only remove binary events with high S/N and a long
duration between caustic crossings ($> 110$ days). The standard of the light
curve photometry was also accessed in this level-1.0 selection. These t
level-1.0 cuts remove the high S/N variable stars from the candidate
lensing event list, but a number of the lower S/N variables remained, so
further cuts were necessary.

The DIA source light curves passing level-1.0 cuts were then fed through a
stricter set of level-1.5 cuts. These cuts remove lower S/N variables and
are based on the microlensing goodness-of-fit statistic $\chi^{2}_{m}$, and
a constant baseline goodness-of-fit statistic $\chi^{2}_{c}$ (performed in
the region $t_{max} \pm 2 \hat{t}$). Here, $\chi^{2}_{m}$ and $\chi^{2}_{c}$
denote reduced chi-squared statistics.  To give weight to higher S/N events
we enforced what we call an $\Omega\chi^{2}$ cut, this was defined as
$\Omega\chi^{2}$ =$1000/pf \times (\chi^{2}_{c} + \chi^{2}_{m})$.  The
symbol $pf$ refers to the flux at peak amplification. For all light
curves we required $\Omega\chi^{2} < 3.6$.

\begin{figure}[ht]
\epsscale{1.0}
\plotone{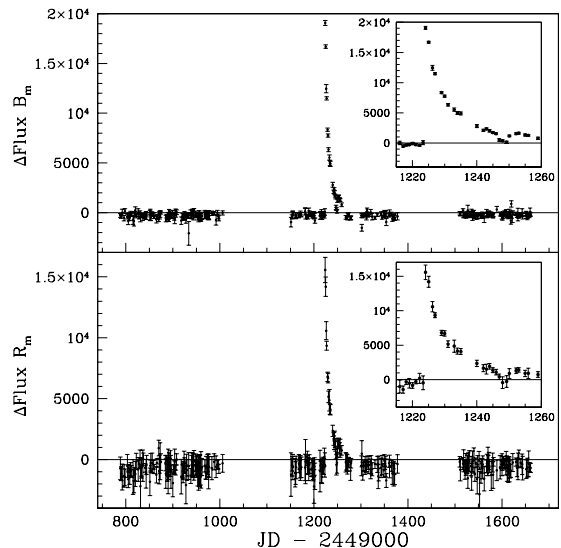}
\figcaption{An example of a DIA light curve for one of the dwarf novae
  removed by our colour and fit microlensing cuts. For this event 
 $V-R \sim 0$.  Inset is a blow up of the outburst event.
\label{figDN}}
\end{figure}

During our analysis we found that the DIA technique was very sensitive to
the detection of dwarf novae. Most of these variables can be rejected during
event selection based on their poor microlensing fits. An example of these
dwarf novae is presented in Figure \ref{figDN}.  However, some faint dwarf
novae light curves have large uncertainties and can thus pass a microlensing
fit $\chi^{2}$ cut.  Because of this, we found it necessary to impose a
colour cut $(V-R)_{108} > 0.55$ and a $V$-magnitude cut for $V > 17$.
$(V-R)_{108}$ denotes the colour an event would have if seen at the average
reddening of field 108 ($E(V-R) = 0.51$). To determine this we have made use
of the reddening values given in \shortciteN{ALC98a}.
Transformations from our $B_{M}$ and $R_{M}$ colours to {\it Cousins} $V$
and $R$ are given in \shortciteN{Alves99}. We note that, although the
colours of the dwarf novae companion stars can vary, the difference flux
colour is the hue of the outburst flux component. These outburst fluxes
appear to generally be bluer than $V-R = 0.55$.  This colour cut removes
very few stars potential sources of microlensing events since the stars
observed on the blue side of $V-R = 0.55$ are generally brighter than $V = 17$.
In this analysis we do not apply any colour cut on the red stars in the CMD
where a large number of variables stars are known to lie.

Within the analysed observations there are a number of spurious detections
due to satellite trails and other transient objects.  These trails generally
cause multiple object detections within a given observation.  We remove
these spurious objects by requiring that the time of peak flux $t_{max}$ for
a microlensing event $i$, in image section $j$, does not occur at the same time
as another event $k$, in the same image section ($t_{max_{ij}} \neq
t_{max_{kj}}$). The probability that two microlensing events will occur with
the same value of $t_{max}$ in close proximity, is very small. 
This cut also removes any spurious detections due to systematic effects
which can occur when the seeing in an image is poor, or the telescope 
slips during an exposure.

Long timescale events are subject to significant parallax effects. To
determine the efficiency of detecting such events we would have to a priori
assume a distribution of sources and lenses towards the Galactic bulge.
Therefore, in this analysis we impose an upper-limit on $\hat{t}$ of one
year.  Similar restrictions are true for binary lensing events since the
distribution of parameters for binaries is not well known. However, we
think our selection should not be significantly biased against these
events as our final selection contains all the events we believed were
binaries in the loose level-1.0 cuts.

To summarise our level-1.5 cuts, all events must meet the following
criteria: $V-R > 0.55$, $A > 1.34$, $\Omega\chi^{2} < 3.6$, $\hat{t} < 365$
days, $t_{max_{ij}} \neq t_{max_{kj}}$, $\chi^{2}_{c} < 30$, plus one
level-1.5 cut which is based on the event's microlensing fit peak-flux in
the $R_{M}$-band\footnote{The light curves have higher S/N in the $R_{M}$
band since $B_{M}-R_{M} > 0$ in general.}, $F_{pr}$, from Table \ref{tab2}.

The last set of selection cuts are the level-2.0 cuts. These cuts are
designed to remove the low S/N variables which are well fitted by
microlensing light curves, but are nevertheless obviously variable by eye.
For final selection the candidate microlensing event must pass all the
level-2.0 cuts where its peak flux ($F_{pr}$) is in the range of the cut.
The level-2.0 cuts are also specified in Table \ref{tab2}.  To quantify the
effect of our cuts let us mention that there were $\sim 300000$ variable
objects detected in the eight fields, only 776 of these passed our level-1.0
selection. 219 then passed the level-1.5 cuts and 99 passed the final
event selection cuts (level-2.0).

We note that it is possible to express these level-1.5 and level-2.0 cuts in
a more direct form. However, the procedure presented here reflects the real
selection process which is progressive and most easily accomplished in
stages. Furthermore, the selection of final microlensing event candidates
this way is necessary in order to quantify the detection efficiency of the
analysis system. The experimental determination of the microlensing optical
depth requires that this detection efficiency is known.  Some microlensing
groups still select final candidates by eye, for instance, see \citeN{AUB99}.


\section{\sc Results}

\begin{figure}
  \vskip 19cm
  \includegraphics{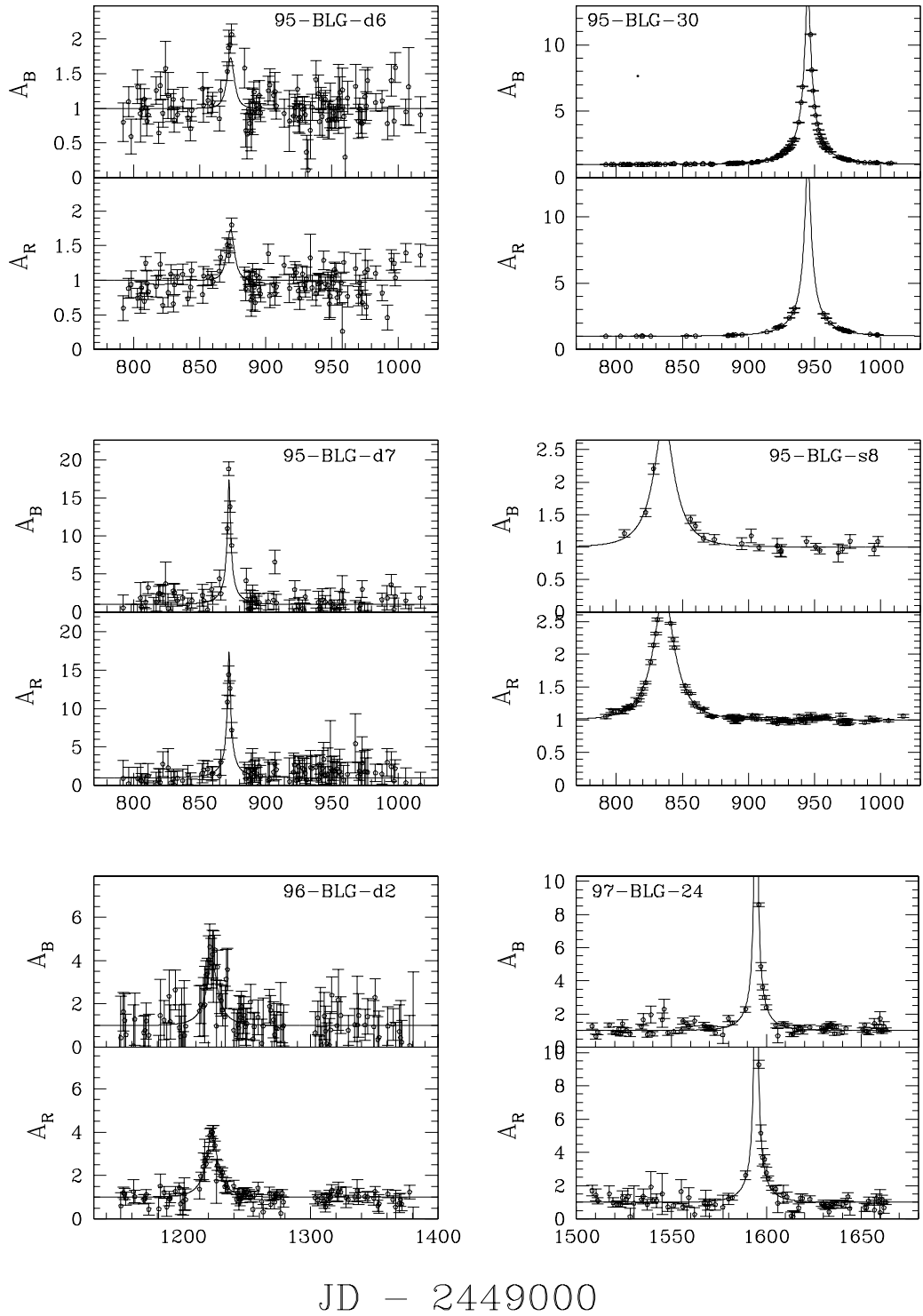} 
  \figcaption{Events from Field 101\label{lc1}.  New DIA events have Ids
    containing ``d''. New events selected with this analysis and a 
    reanalysis of the SoDoPhot data have Ids containing ``s''.}
\end{figure}

We have discovered 99 microlensing event candidates.  The light curves and
microlensing fits for 83 events are presented in Figure \ref{lc1}. The
actual amplification of each event was obtained by fitting the light curves
with the baseline source flux as an additional parameter. The other 16 events
in field 108 were shown in \citeN{ALC99c}.  In each of these figures only a
single season of data for each light curves is displayed.  However, all the
DIA light curves span three observing seasons.


The numbers of events detected in each field are given in Table \ref{tab3},
along with results from the SoDoPhot analysis of the same images. The
coordinates and labels used for identifying the events in the datasets are
given in Table \ref{tab4}.  We will refer to events primarily using their
microlensing event ``alert'' ID\footnote{The details of microlensing alerts
  can be found at\\ {\em http://darkstar.astro.washington.edu/}}.  Each
alert ID includes the year, observation target (BLG, LMC or SMC), and the
order of the detection. For example, the first alert event detected in the
bulge in 1995 is labeled 95-BLG-1. Microlensing events which were detected
by DIA, but not in the alerts or SoDoPhot re-analysis, have a ``d'' before
the event number (e.g.  95-BLG-d1).  Events which were not found as alerts,
but were detected in the subsequent analysis of SoDoPhot photometry, are
labeled with an ``s'' (e.g.  95-BLG-s1).

\newpage
\begin{figure}
\vskip 19cm
\figurenum{3}
\includegraphics{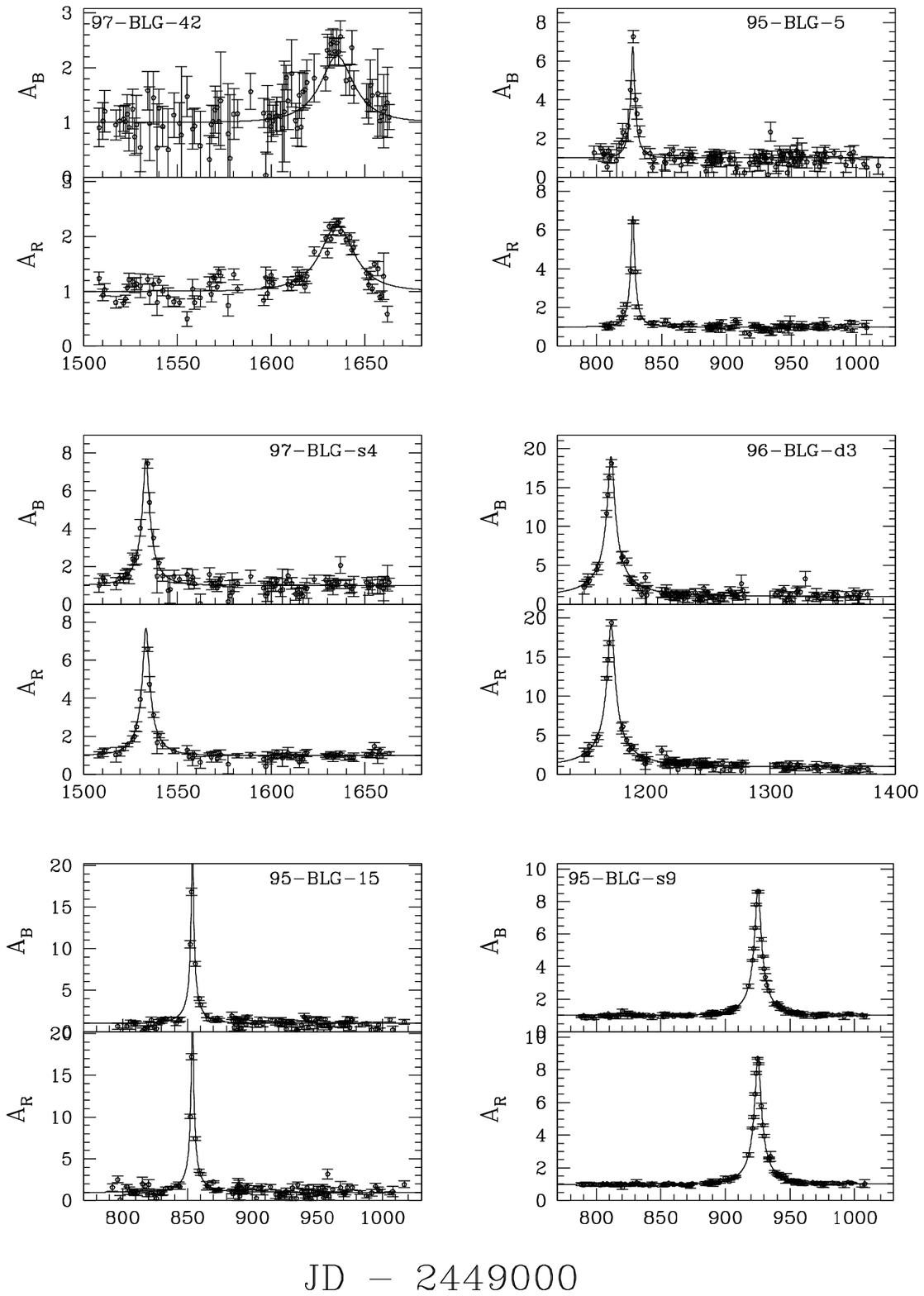} 
\figcaption{Events from
  Fields 101 \& 104.  New DIA events have Ids containing ``d''. New events
  selected with this analysis and a reanalysis of the SoDoPhot data have Ids
  containing ``s''.}
\end{figure}

.
\newpage
\begin{figure}
\vskip 19cm
\figurenum{3}
\includegraphics{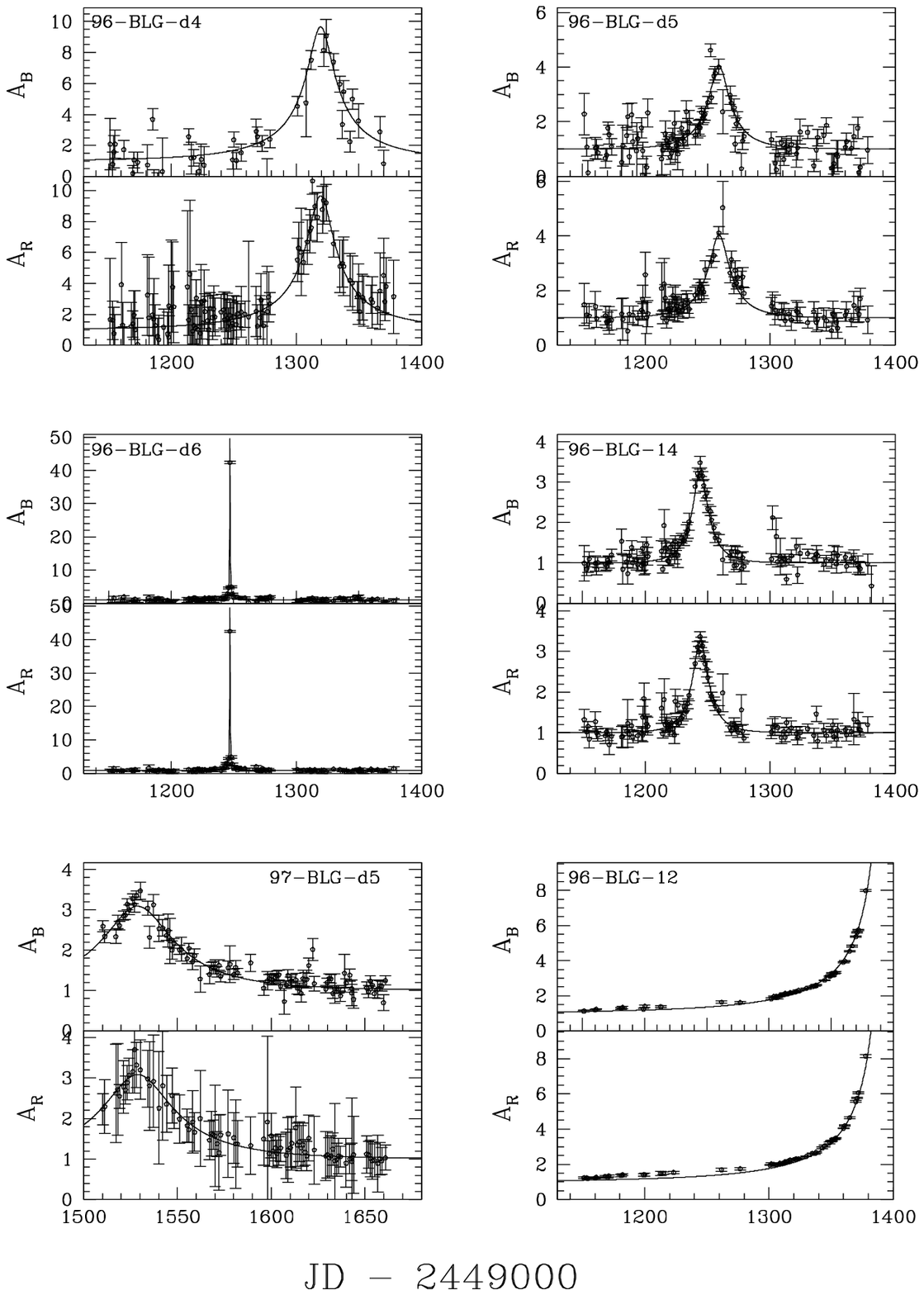} 
\figcaption{Events from
  Field 104. New DIA events have Ids containing ``d''. New events selected
  with this analysis and a reanalysis of the SoDoPhot data have Ids
  containing ``s''.}
\end{figure}

.
\newpage
\begin{figure}
\vskip 19cm
\figurenum{3}
\includegraphics{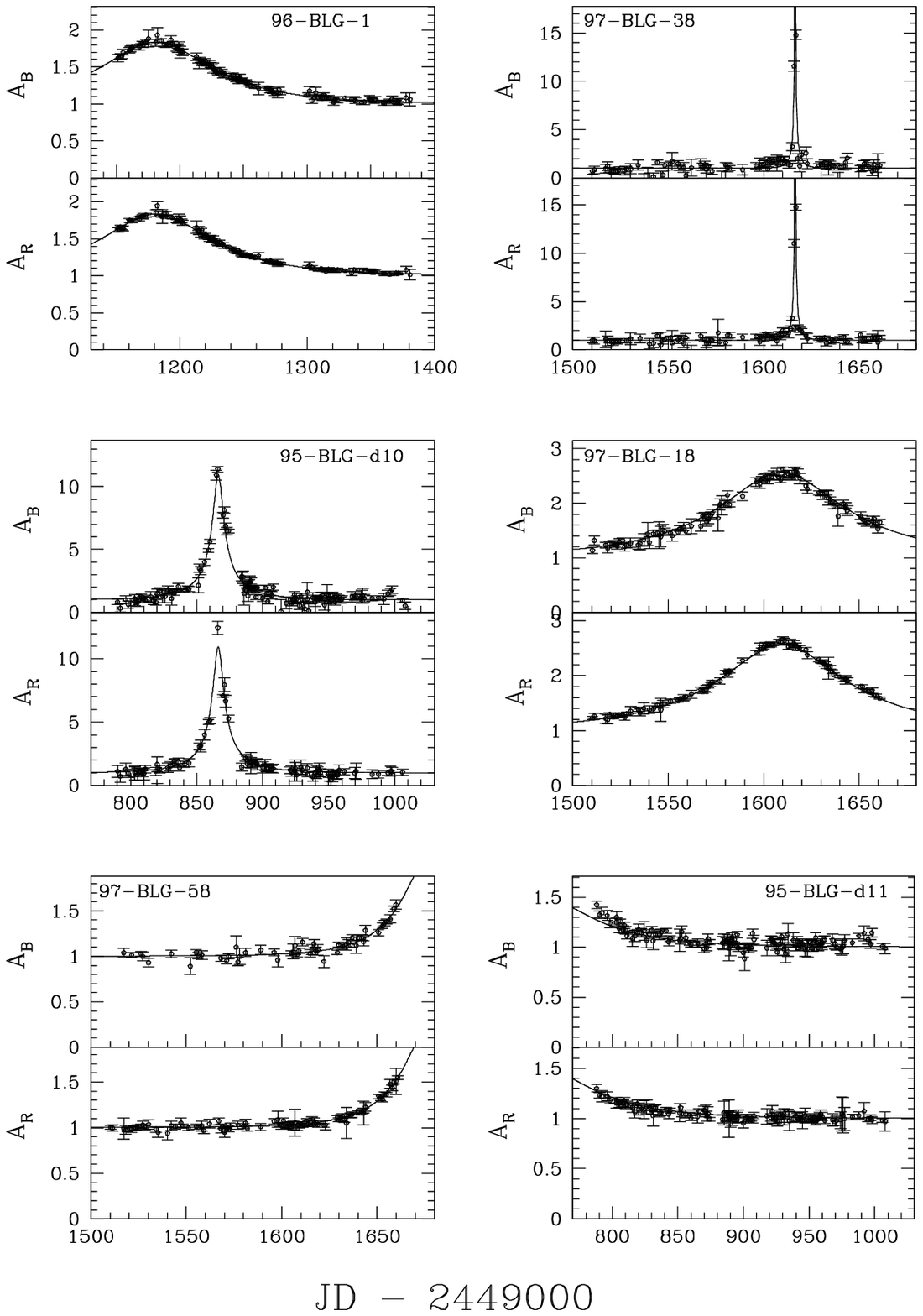}
\figcaption{Events from Field 104 cont. New DIA events are have Ids
    containing ``d''. New events selected
  with this analysis and a reanalysis of the SoDoPhot data have Ids
  containing ``s''.}
\end{figure}

.
\newpage
\begin{figure}
\vskip 19cm
\figurenum{3}
\includegraphics{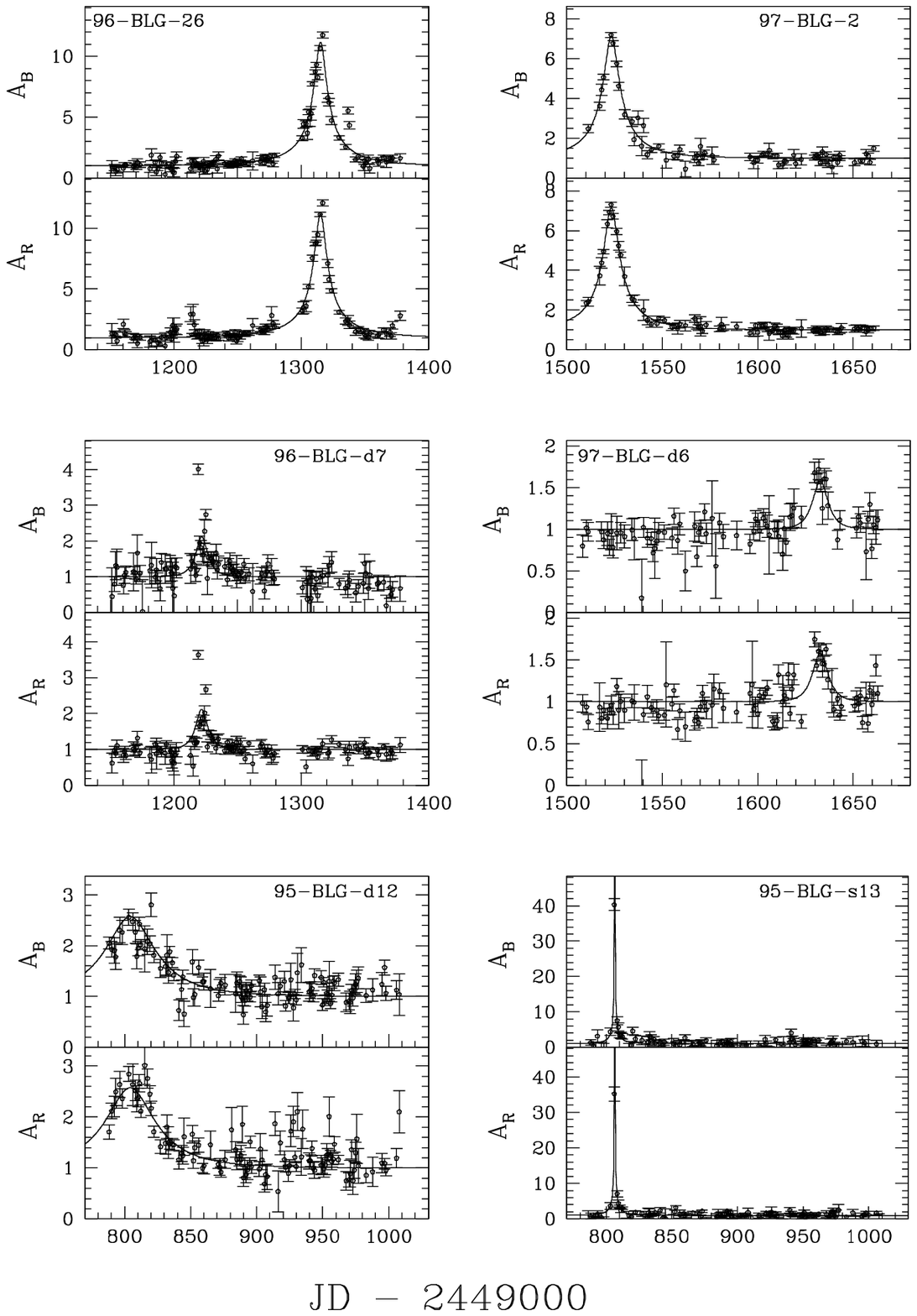}
\figcaption{Events from Fields 104 \& 113. New DIA events have Ids
    containing ``d''. New events selected
  with this analysis and a reanalysis of the SoDoPhot data have Ids
  containing ``s''.}
\end{figure}

.
\newpage
\begin{figure}
\vskip 19cm
\figurenum{3}
\includegraphics{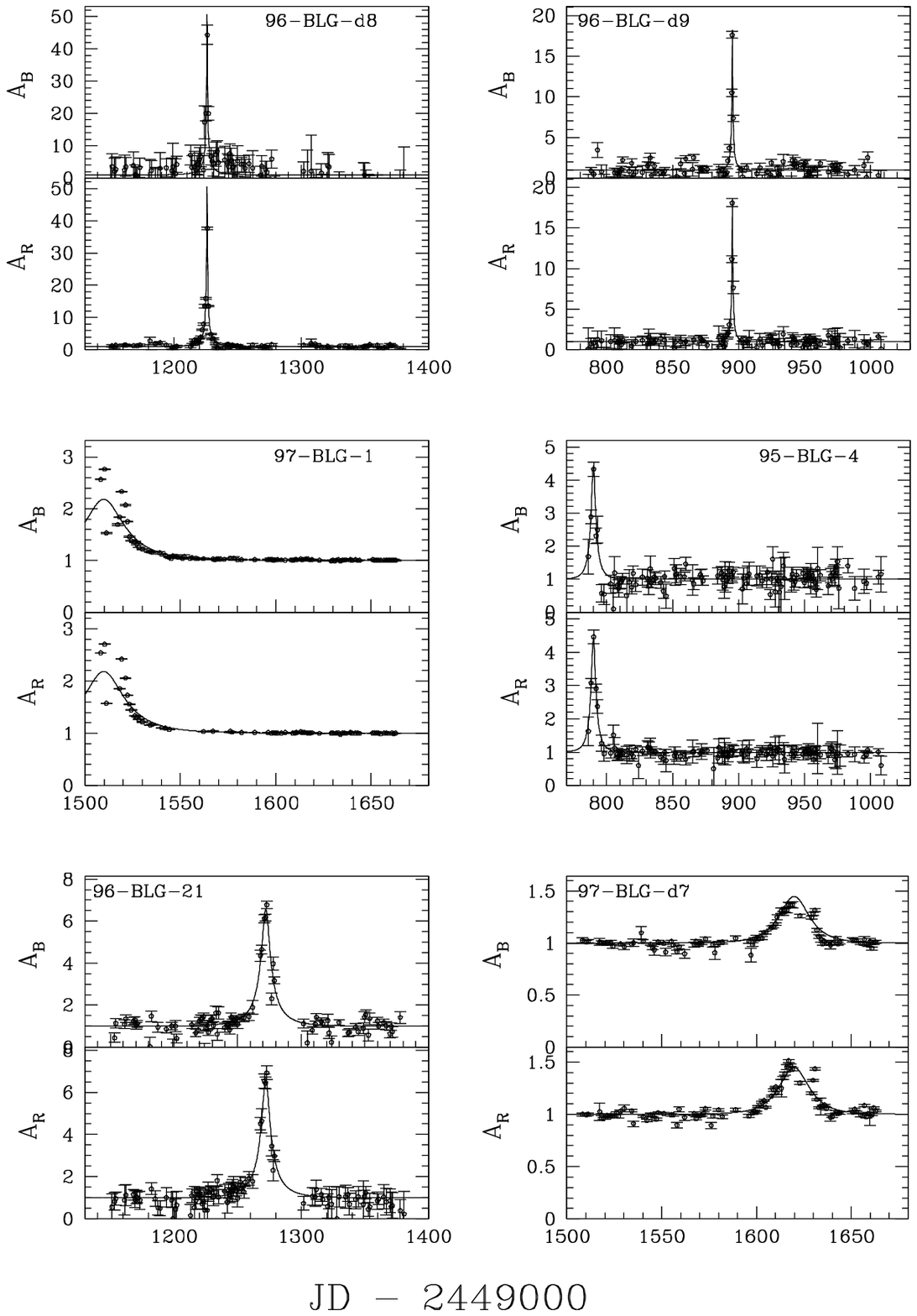}
\figcaption{Events from Field 113. New DIA events have Ids
    containing ``d''. New events selected
  with this analysis and a reanalysis of the SoDoPhot data have Ids
  containing ``s''.}
\end{figure}

.
\newpage
\begin{figure}
\vskip 19cm
\figurenum{3}
\includegraphics{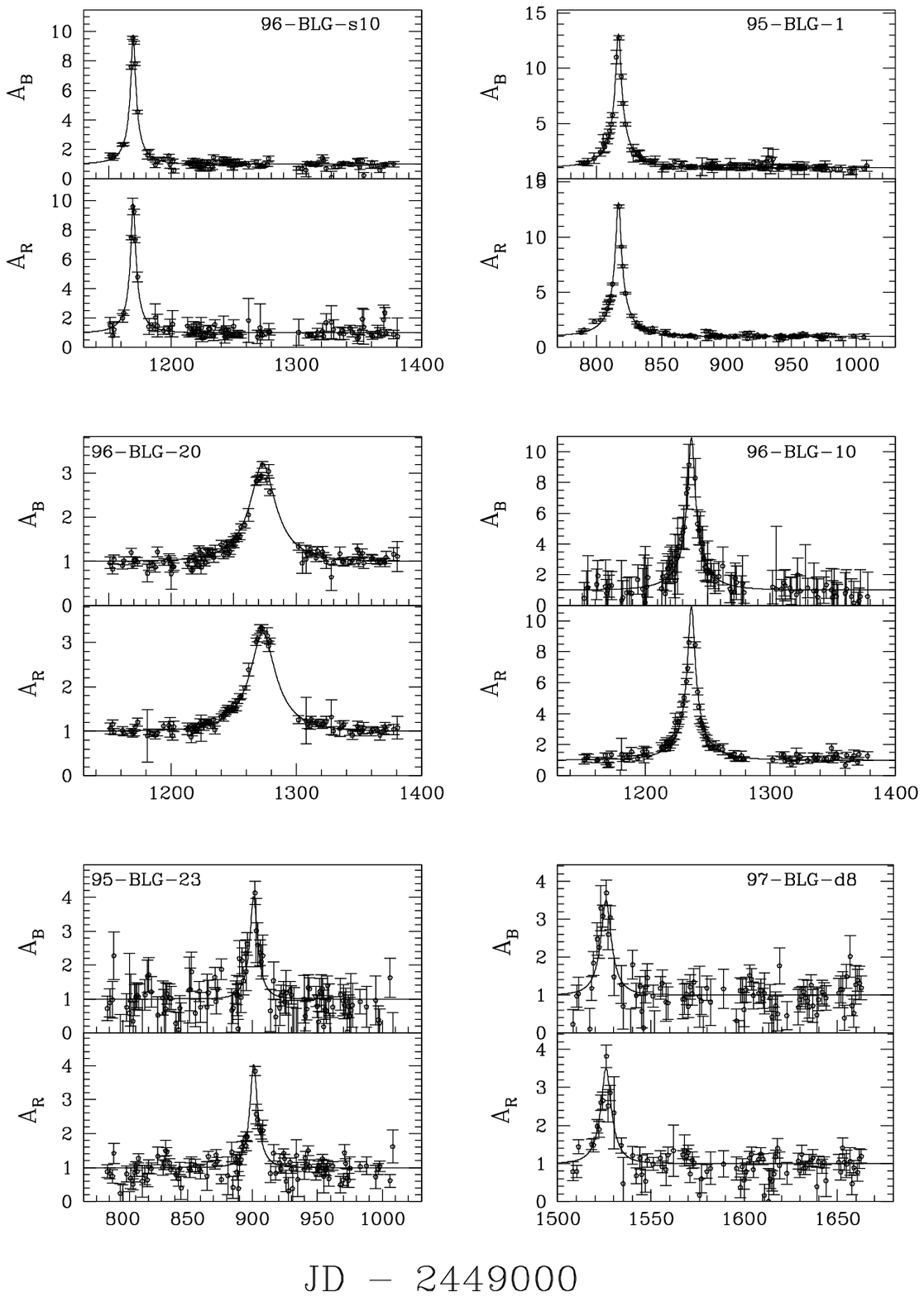}
\figcaption{Events from Field 113 cont. New DIA events have Ids
    containing ``d''. New events selected
  with this analysis and a reanalysis of the SoDoPhot data have Ids
  containing ``s''.}
\end{figure}

.
\newpage
\begin{figure}
\vskip 19cm
\figurenum{3}
\includegraphics{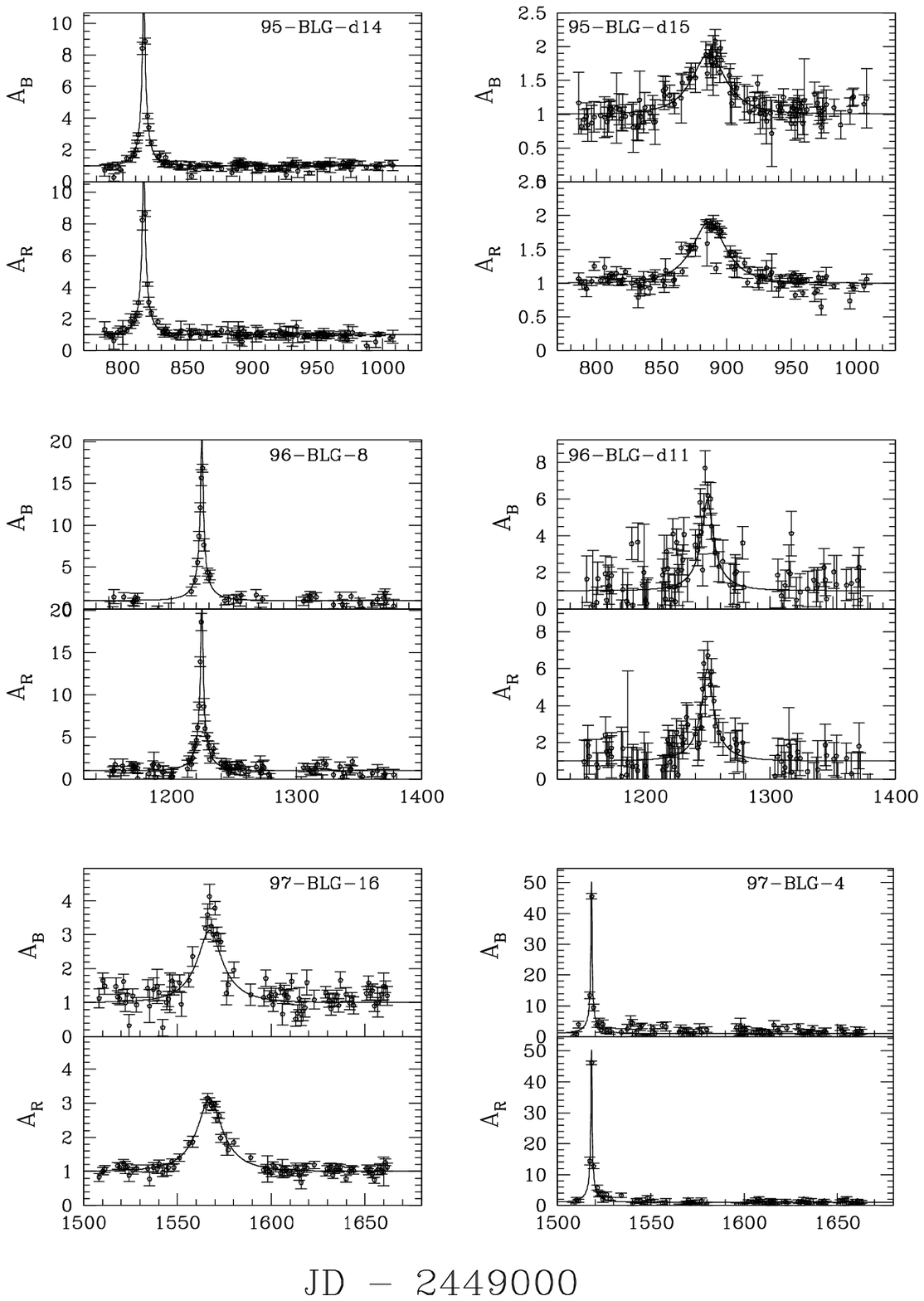}
\figcaption{Events from Fields 113 \& 118. New DIA events have Ids
    containing ``d''. New events selected
  with this analysis and a reanalysis of the SoDoPhot data have Ids
  containing ``s''.}
\end{figure}

.
\newpage
\begin{figure}
\vskip 19cm
\figurenum{3}
\includegraphics{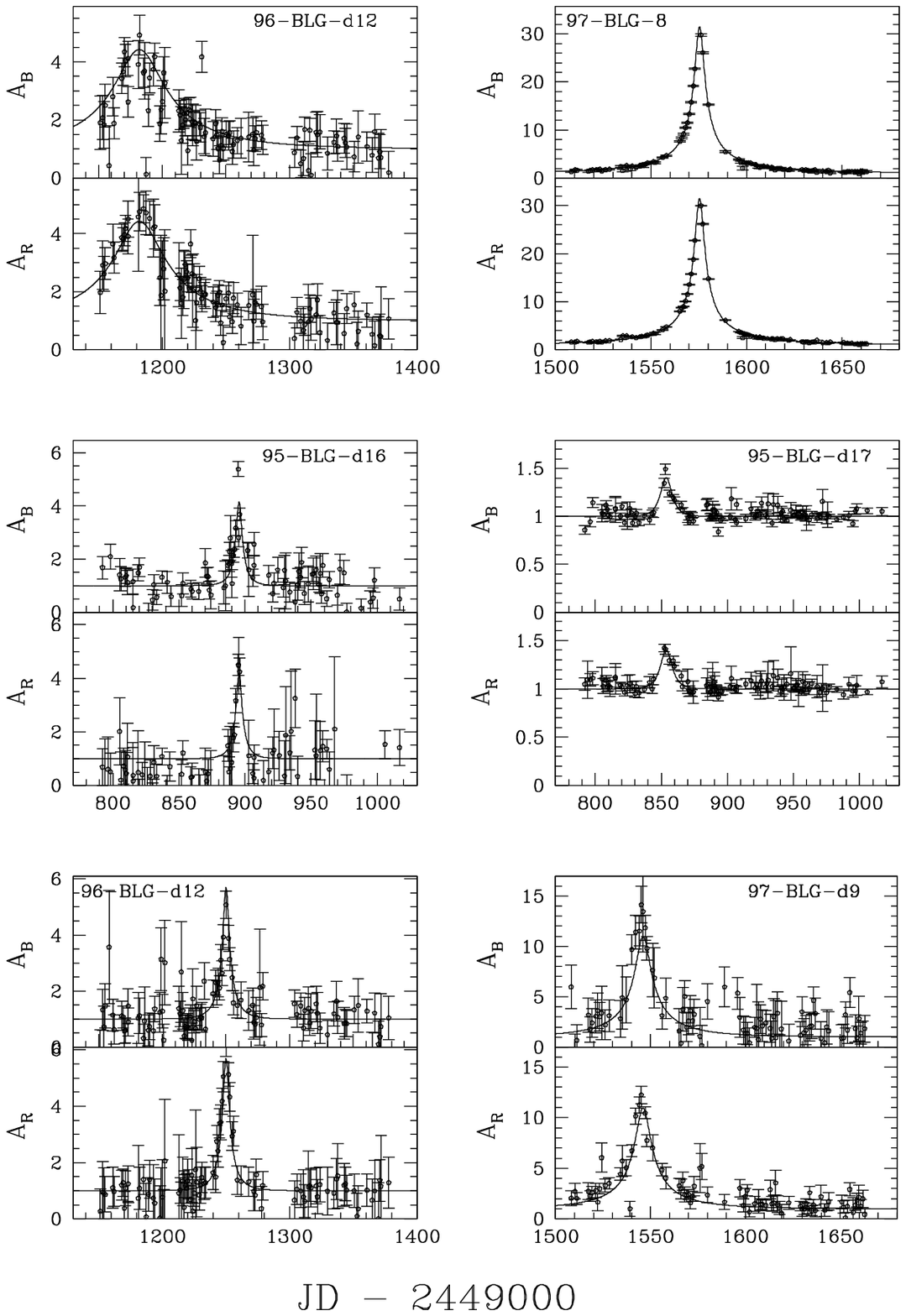}
\figcaption{Events from Field 118. New DIA events have Ids
    containing ``d''.   New events selected
  with this analysis and a reanalysis of the SoDoPhot data have Ids
  containing ``s''.}
\end{figure}

.
\newpage
\begin{figure}
\vskip 19cm
\figurenum{3}
\includegraphics{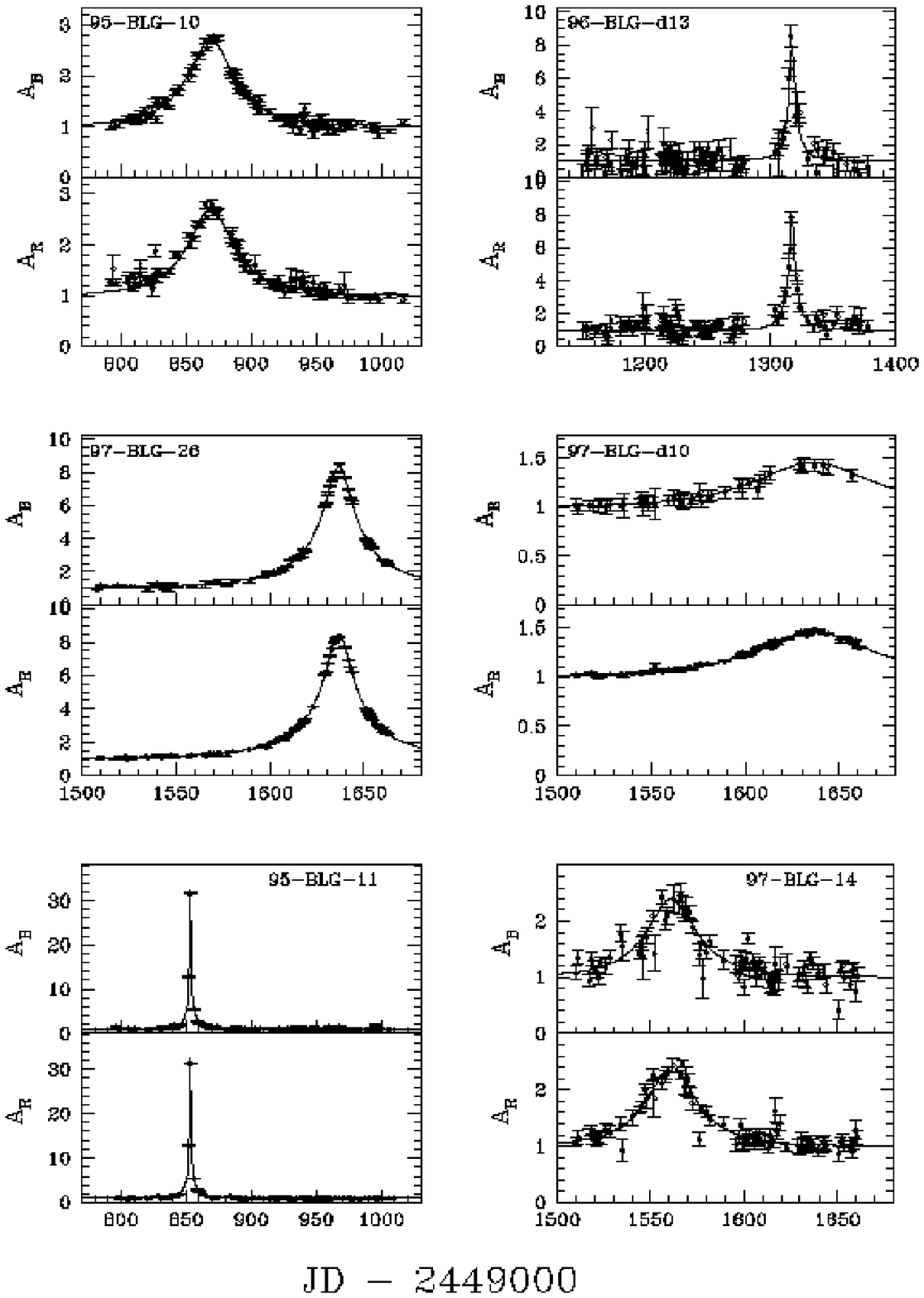}
\figcaption{Events from Fields 118 \& 119. New DIA events have Ids
    containing ``d''. New events selected
  with this analysis and a reanalysis of the SoDoPhot data have Ids
  containing ``s''.}
\end{figure}

.
\newpage
\begin{figure}
\vskip 19cm
\figurenum{3}
\includegraphics{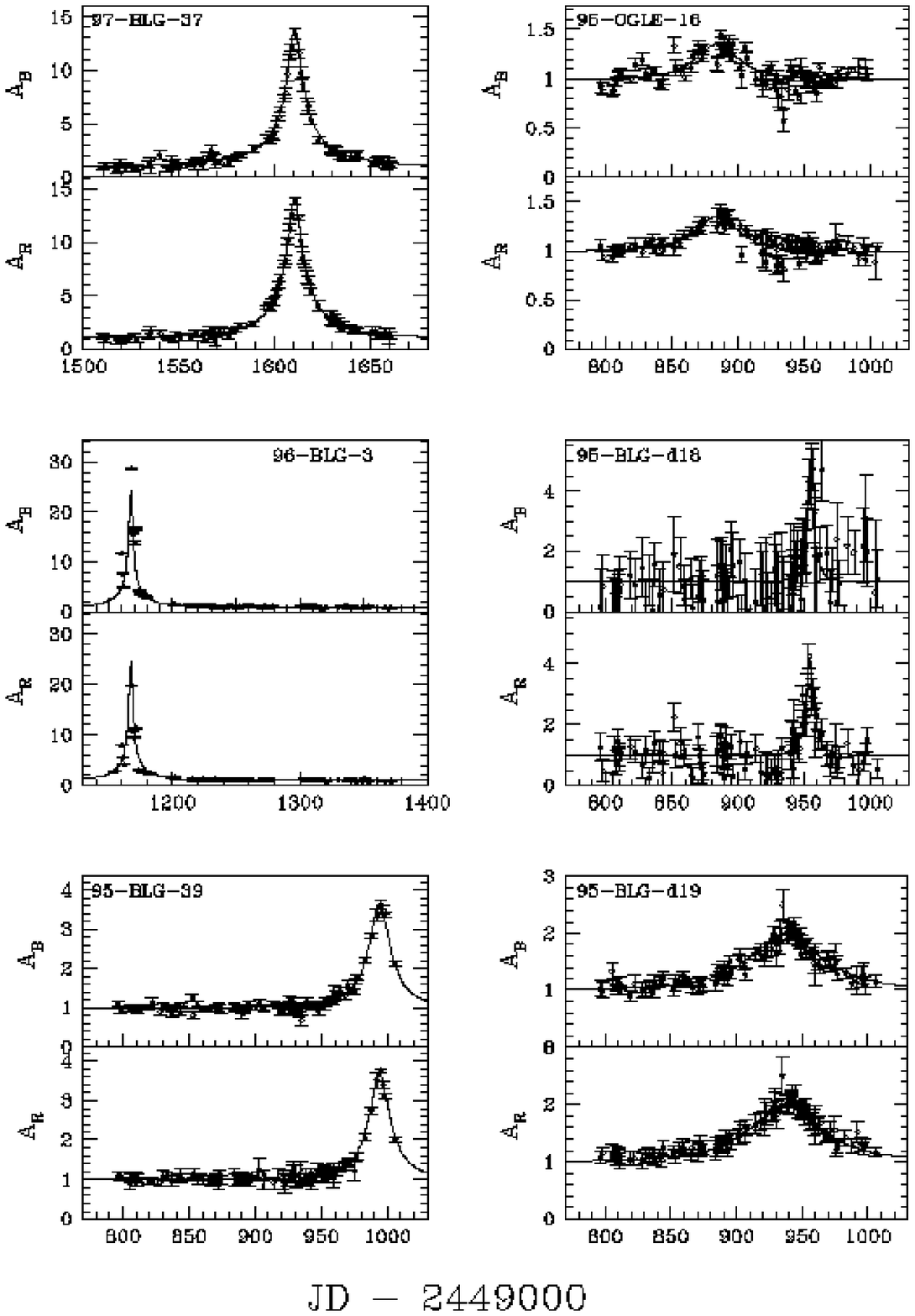}
\figcaption{Field 119 events. New DIA events have Ids
    containing ``d''.  New events selected
  with this analysis and a reanalysis of the SoDoPhot data have Ids
  containing ``s''.}
\end{figure}

.
\newpage
\begin{figure}
\vskip 19cm
\figurenum{3}
\includegraphics{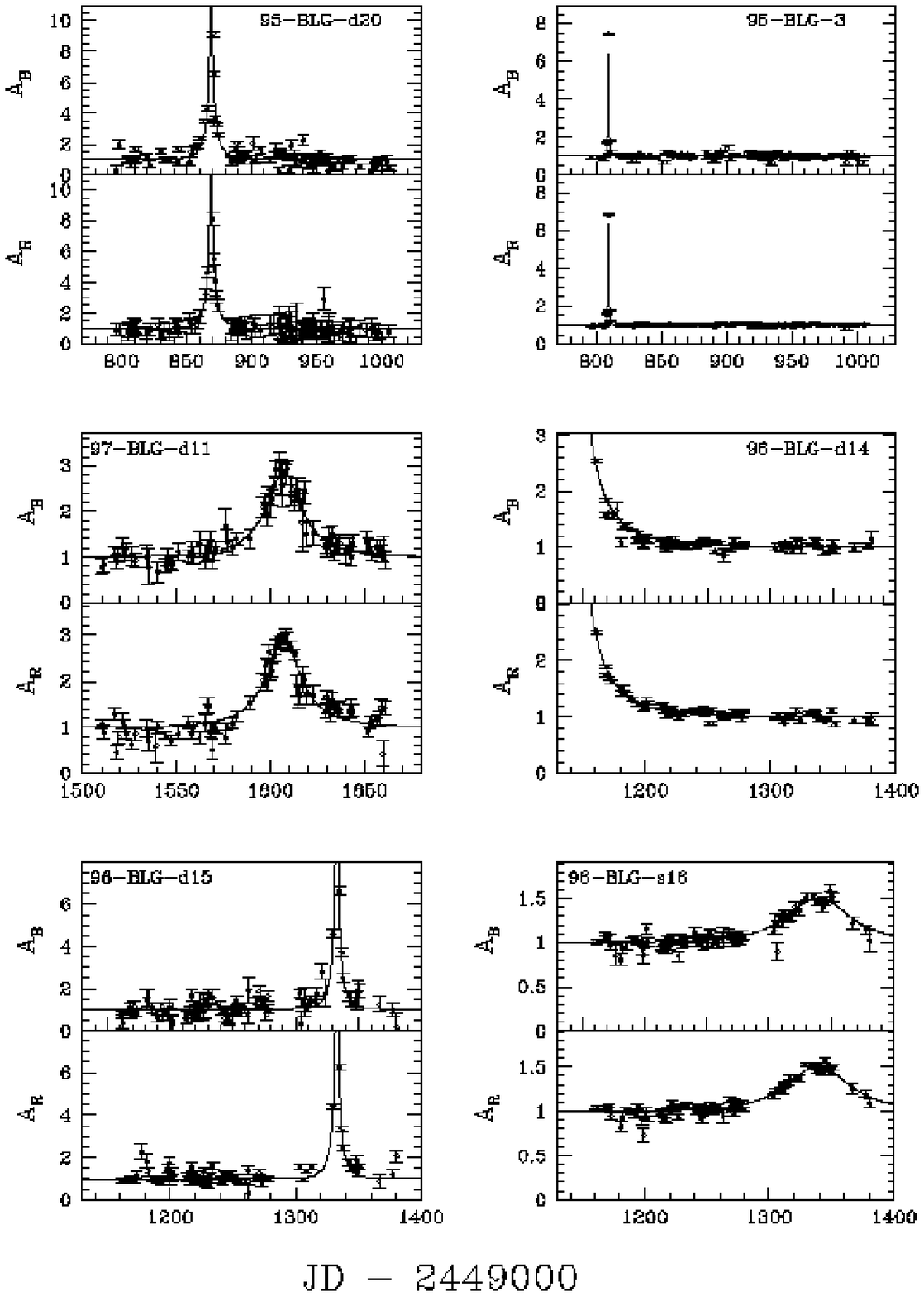}
\figcaption{Events from Fields 119 \& 128. New DIA events have Ids
    containing ``d''. New events selected
  with this analysis and a reanalysis of the SoDoPhot data have Ids
  containing ``s''.}
\end{figure}

.
\newpage
\begin{figure}
\vskip 19cm
\figurenum{3}
\includegraphics{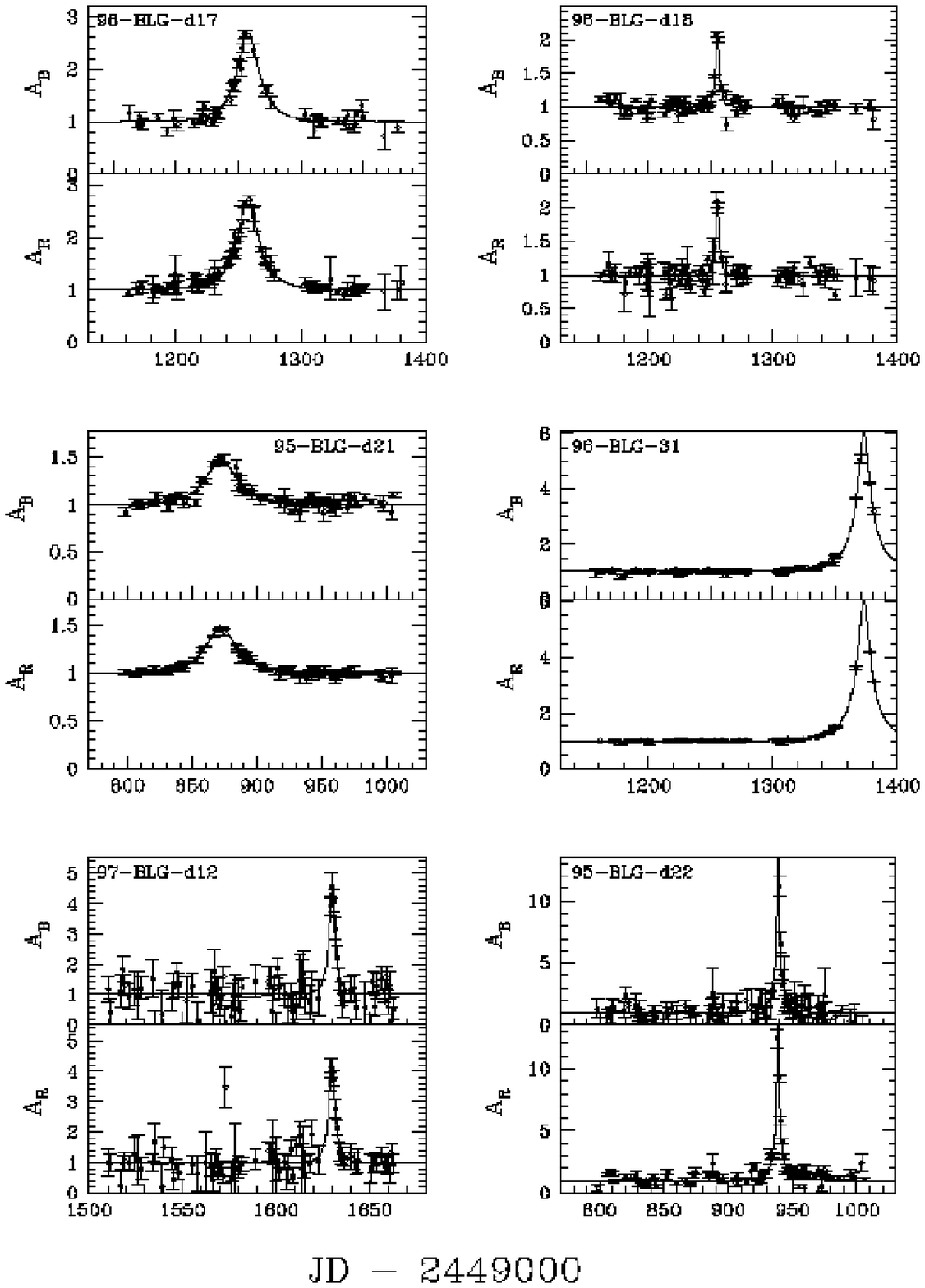}
\figcaption{Events from Field 128. New DIA events have Ids
    containing ``d''.  New events selected
  with this analysis and a reanalysis of the SoDoPhot data have Ids
  containing ``s''.}
\end{figure}

.
\newpage
\begin{figure}
\vskip 19cm
\figurenum{3}
\includegraphics{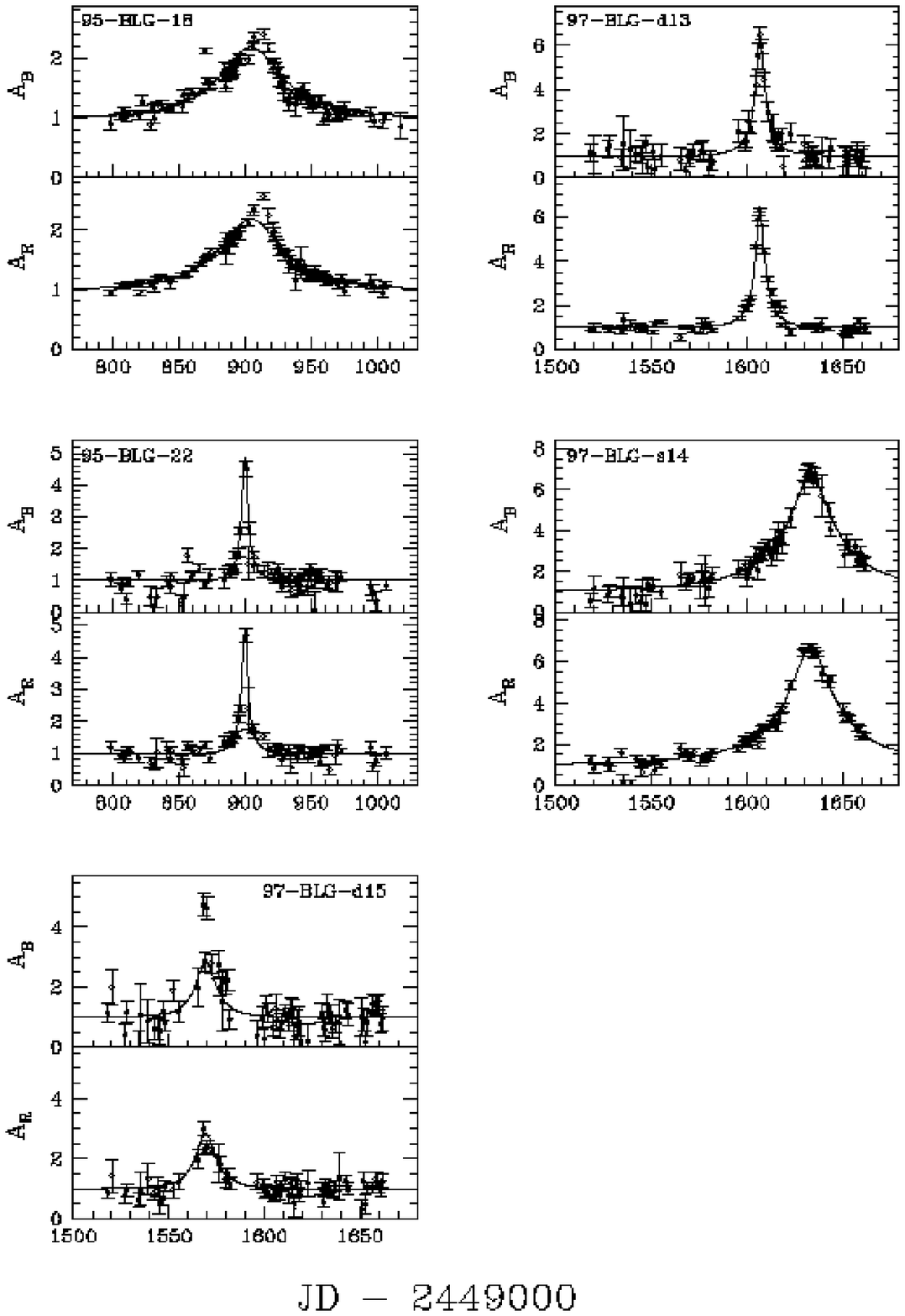}
\figcaption{Events from Fields 128 \& 159\label{lclast}. New DIA events
  have Ids containing ``d''.  New events selected
  with this analysis and a reanalysis of the SoDoPhot data have Ids
  containing ``s''.}
\end{figure}
.
\newpage
.
\newpage

\begin{figure}[ht]
\epsscale{1.0}
\plotone{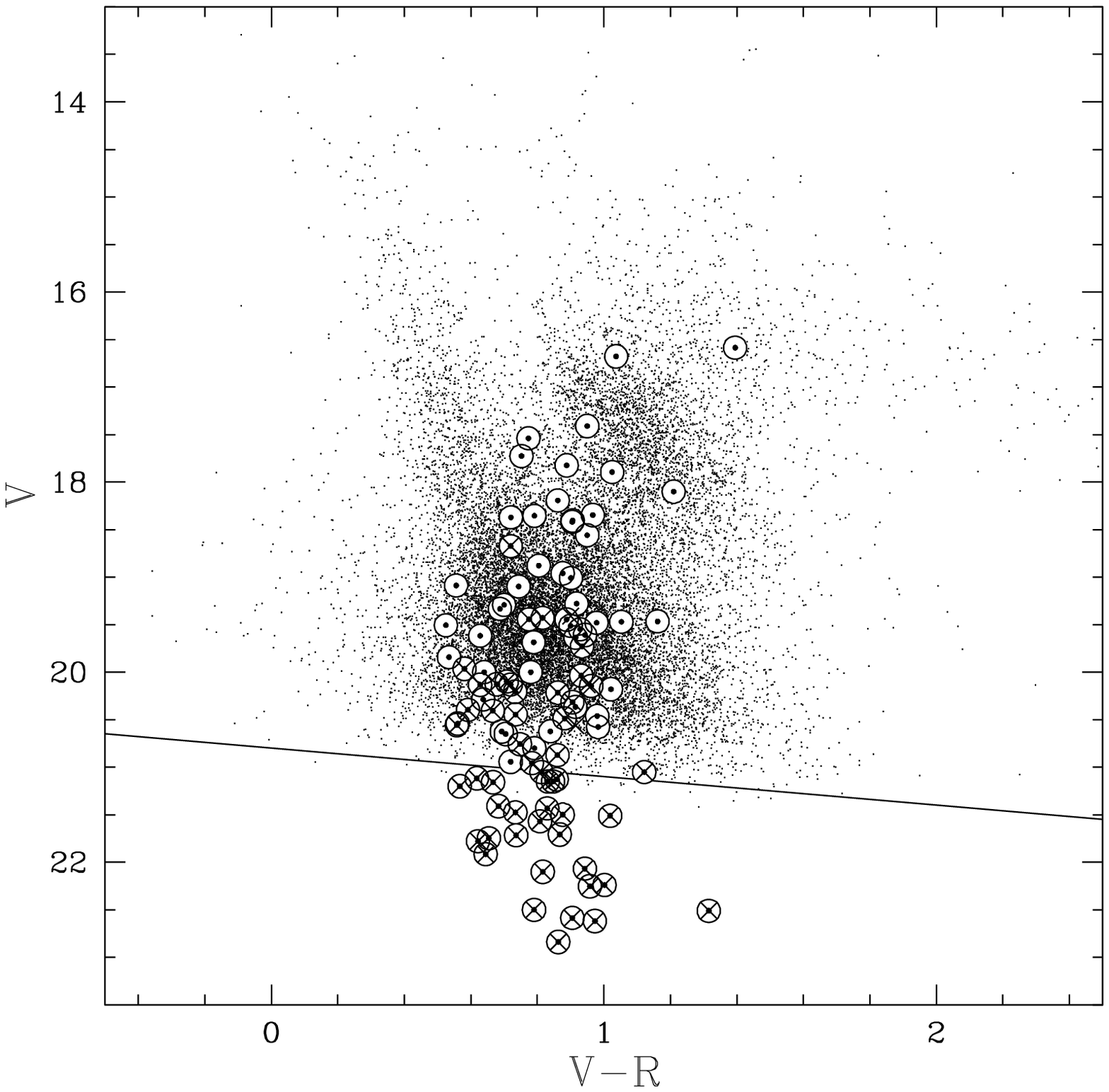}
\figcaption{The fitted $V$ magnitudes and colours of the microlensing events 
overlaid on a CMD of neighbouring resolved stars. The solid line corresponds
to a pixel lensing cut. Circles with dots represent classical events
and circles with crosses are pixel lensing events.
Some pixel lensing event sources appear above the cut because the 
are not associated with monitored SoDoPhot sources.\label{figCMD}}
\end{figure}

The microlensing event's source star V-magnitudes and $V-R$ colours are
displayed in Figure \ref{figCMD}. This figure is an average CMD, which
has been constructed by combining the SoDoPhot photometry for the
$\sim 250$ stars nearest each microlensing event source star.  Notice that
there are very few events with bright main sequence sources. This is to be
expected, since not only are there few of these bright stars, but most of
them are foreground disk stars which ``see'' a much lower microlensing
optical depth.
The SoDoPhot analysis is sensitive to microlensing events occurring with
relatively bright source stars ( $V < 21$) or fainter stars which are blended
with these monitored bright stars. The DIA technique on the other hand is
sensitive to events independent of the locations of the brighter stars.
In this way the DIA technique is expected to mainly detect events where
the source star is faint and not blended with a bright star monitored by
SoDoPhot. Events with unresolved sources are termed pixel lensing events.

\placetable{tab1}
\placetable{taball}

\subsection{\it Pixel Lensing Events}

Our set of microlensing events contains both classical microlensing events
and pixel lensing events. In recent years, the division between these two
groups of events has been unclear because of the difference between pixel
lensing and the pixel method. We will attempt to clarify this division in
regard to our events.  Our definition of pixel lensing derives from
the theory of \shortciteN{Gould96a} who defined Pixel Lensing as the
``gravitational microlensing of unresolved stars''. Whereas, the ``Pixel
Method'' is a method of binning images, which is used by the AGAPE group to
detect microlensing events towards M31 \shortcite{AABB97,GON97}.

Throughout this work when we refer to pixel lensing, we shall mean:
``gravitational microlensing events where the source stars are unresolved in
initial template/reference images of a field''.  This seems the most logical
definition as it is equally applicable for events seen towards extra-galactic 
targets, such as M31, as it is for events in the line-of-sight to the 
Magellanic clouds and Galactic bulge. In contrast to this, a classical 
microlensing event is taken to be one where the source is resolved both 
before and after the lensing has taken place.  
As a division between classical and pixel microlensing still remains somewhat
unclear, we will further refine our concept of pixel lensing.

For all the microlensing surveys presently being carried out,
there is a significant amount of crowding and blending.
Specifically, there are many faint unresolved stars within the seeing disk
of any star bright enough to be detected. A microlensing event detected by
monitoring such a {\em group} of stars, could be due to either a bright
detectable star (classical lensing), or one of the many faint unresolvable
neighbouring stars blended with it (pixel lensing).

When events are due to the unresolved faint stars, there can be a measurable
shift in the flux centroid position of the event as the event brightens
\shortcite{ALC97d}.  However, the offset between the faint source's centroid
and the group centroid is a random quantity.  Such an offset will take an
arbitrarily small value in some events.  Hence, this approach can not quantify
the degree of blending for all events.  Another property of these events is
that they can exhibit a significant chromatic signature when the event's
source star colour varies from the overall colour of the group.  Sources on
the main sequence near our detection limit ($V\sim21$) make up the bulk of
the monitored stars and have a narrow range of colours. For this reason, in
many cases, there may be little difference between an individual star's
colour and that of the blended group of stars it resides amongst.  The final
piece of information about blended events comes from the shape of the light
curve. That is, a given amplification microlensing event has a specific
shape. Once again there is only a slight difference between the shapes of
events with different amplifications if various timescales are considered
(see \shortciteNP{GH92}, \shortciteNP{WP97} \& \shortciteNP{ALC99c}).

From these considerations, a sensible method of differentiating between
classical and pixel microlensing events is to put limits on the
allowable centroid offset, the true amount of flux (amplification), and the
event's colour.  Beyond such hypothetical limits an event could be defined
to be pixel lensing with significant confidence.  For this reason we will
set further limits on our definition of pixel lensing events.  To quantify
this decision we define that, for classical lensing the centroid offset
between the event centroid and the nearest photometered object centroid
should be less than three times the centroid uncertainty.  Otherwise this is
a pixel lensing event. The average uncertainty in our event centroid
positions is $\sim 0.2 \arcsec$ for the results presented here. This
uncertainty includes the error in the transformation between DIA and
SoDoPhot templates.  We thus adopt an offset of 1 pixel ($0.63\arcsec$) as
our classical event limit.  In addition, if the microlensing event fit gives
a source $V$ magnitude which is significantly below the detection threshold
(for isolated stars), the event is deemed a pixel lensing event.  Because
none of the stars in the observed fields are truly isolated, the actual
detection threshold will be higher.  Thus, this serves as a robust lower
limit on classical lensing event sources. However, the uncertainty in the
fitted baseline source flux increases as the brightness of the source
decreases.  For the analysed fields we have set this threshold limit to be
$V_{pix} < 20.8 + 0.3(V-R)$ (see Figure \ref{figCMD}).

Measured variations between the microlensing event's colour, and the
source's baseline colour, do not provide sufficient evidence to distinguish
whether an event is due to pixel lensing or classical lensing. For some
events the chromatic signature of blending can be measured in the light
curve of the event, although the source is bright enough to be detected if
it were isolated. Such colour changes can tell us the magnitude of the
difference in the colours, and the colour of the group. But colour
information does not guarantee an event is due to an unresolvable star.
However, the larger the colour differences, the more likely an event is
pixel lensing.

Applying these definitions to our events we separate the classical events
and pixel lensing events.  Slightly different information is available for
classical and pixel lensing events. With classical events there is a
measured baseline flux of the source star. This baseline flux is a blend of
all the stars within the seeing disk. For these events one can perform blend
fits on the photometry in order to determine the true baseline source flux
and hence amplification and event timescale.  The fit parameters for the
individual events are given in Table \ref{tab5}. A blended baseline flux is
not applicable for pixel lensing events. However, the baseline source flux,
the amplification, and the timescale can still be determined from the
microlensing light curve fit, these are given in Table \ref{tab6}.

Two types of fits were performed for classical events: one which assumed the
nearest SoDoPhot source star flux was the true source flux, and the other
where the source star flux was a free parameter. The results of the first
approach was expected to yield biased results because of the blended flux.
However, this set is useful to examine timescale and other biases
caused by blending.  We fitted with the baseline source flux as a free
parameter for all our pixel lensing events. This enabled us to determine the
amplification of each lensing event relative to a fitted baseline flux
value.


\subsection{\it Comparison with the Standard Analysis}

During our normal data reduction procedure we carry out a detailed
reanalysis of all the SoDoPhot photometry. This reanalysis is designed to
find new microlensing events and better characterise the known {\em alert}
events.  The reanalysis typically finds a few more events than the alert
system as the alert scheme was designed for real time notification,
and as such only includes the data reduced at the time the alert 
is issued.

\begin{figure}[ht]
\epsscale{1.0}
\plotone{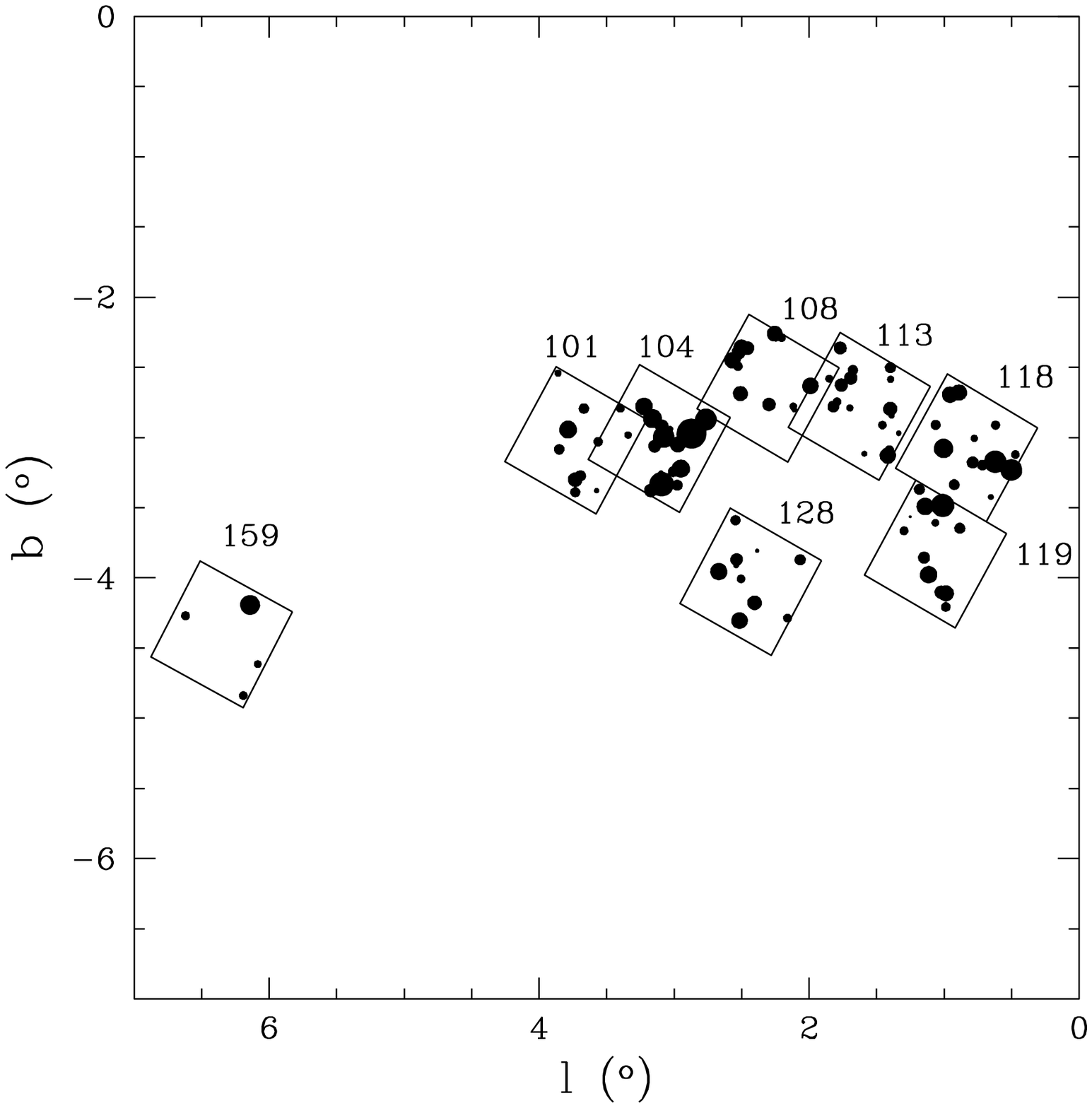}
\figcaption{The locations of the 8 Galactic bulge fields presented in this
  paper are shown in Galactic coordinates (with their corresponding MACHO Id
  numbers).  The 99 microlensing events are represented as dark spots at the
  event locations.  The area of each spot is proportional to the $\hat{t}$
  value of the event.  Baade's Window is in field 119.\label{fig1}}
\end{figure}

Implementing the DIA technique we have found 41 more events than the
SoDoPhot analysis using the same image data. This method thus gives us $\sim
71\%$ more events than our approach based solely on PSF photometry.  We
believe this is not due to the failure of our PSF analysis, but instead is a
tribute to the advantage of the new technique.  To emphasise this, we point
out that there are in fact 57 new events in the DIA which were not found in
the standard SoDoPhot analysis (see Table \ref{tab3}). However, 16 events
detected with the SoDoPhot analysis were not found with the DIA reduction.
These missed events fall into a number of categories. Seven events
were missed because of differences between the microlensing cuts used (1 low
S/N event, 3 poor photometry events, 1 event with $\hat{t} > 300$ days, 2
events fail the $\chi^{2}_{m}$ cut).  A further 4 of these events were never
detected because of the slight differences between the pointing of the
SoDoPhot and DIA reference images.  Lastly, there is a dead amplifier on one
of our CCD mosaics. Five SoDoPhot microlensing candidates fall into this
location in our fields, thus photometry for these events is only available for
one colour ($B_{M}$). However, in the difference image analysis we have only
analysed regions where two colour photometry is possible.

For a fair comparison of the two techniques it is necessary to compare
equivalent areas of the target fields. Small differences in the pointing
should not affect the overall number of events and any variation in the cuts
applied is part of the reduction technique.  Therefore, to fairly compare
the two sets we set aside the five single colour events in the region not
analysed with DIA.  We thus conclude that the SoDoPhot analysis yields 53
events compared to 99 events from the DIA technique. This method thus
provides {$\sim \bf 87\%$} more events than the standard SoDoPhot analysis.

The entire MACHO dataset consists of around 350 candidate microlensing
events.  A complete DIA-based reanalysis of the Galactic bulge database
could therefore provide $> 600$ events.  The number of events detected
towards the LMC, by SoDoPhot analysis, is still relatively small
\cite{ALC00p}.  A factor of $2$ increase in the number of events detected
towards the LMC would be an important way of reducing the statistical
uncertainty in the microlensing optical depth and hence the baryonic
fraction of the Galactic halo.  We believe this factor of 2 could be
achieved in a future reanalysis of LMC data with DIA.

\section{\sc Microlensing Detection Efficiency}

To determine the microlensing detection efficiency of this analysis we
produced a Monte-Carlo simulation for each of the eight fields reduced.
In this simulation we attempted to include all the known aspects of the
analysis, such as the seeing, sky-level, transmission, systematic noise, 
etc., of each observation.

\begin{figure}[ht]
\epsscale{0.8} 
\plotone{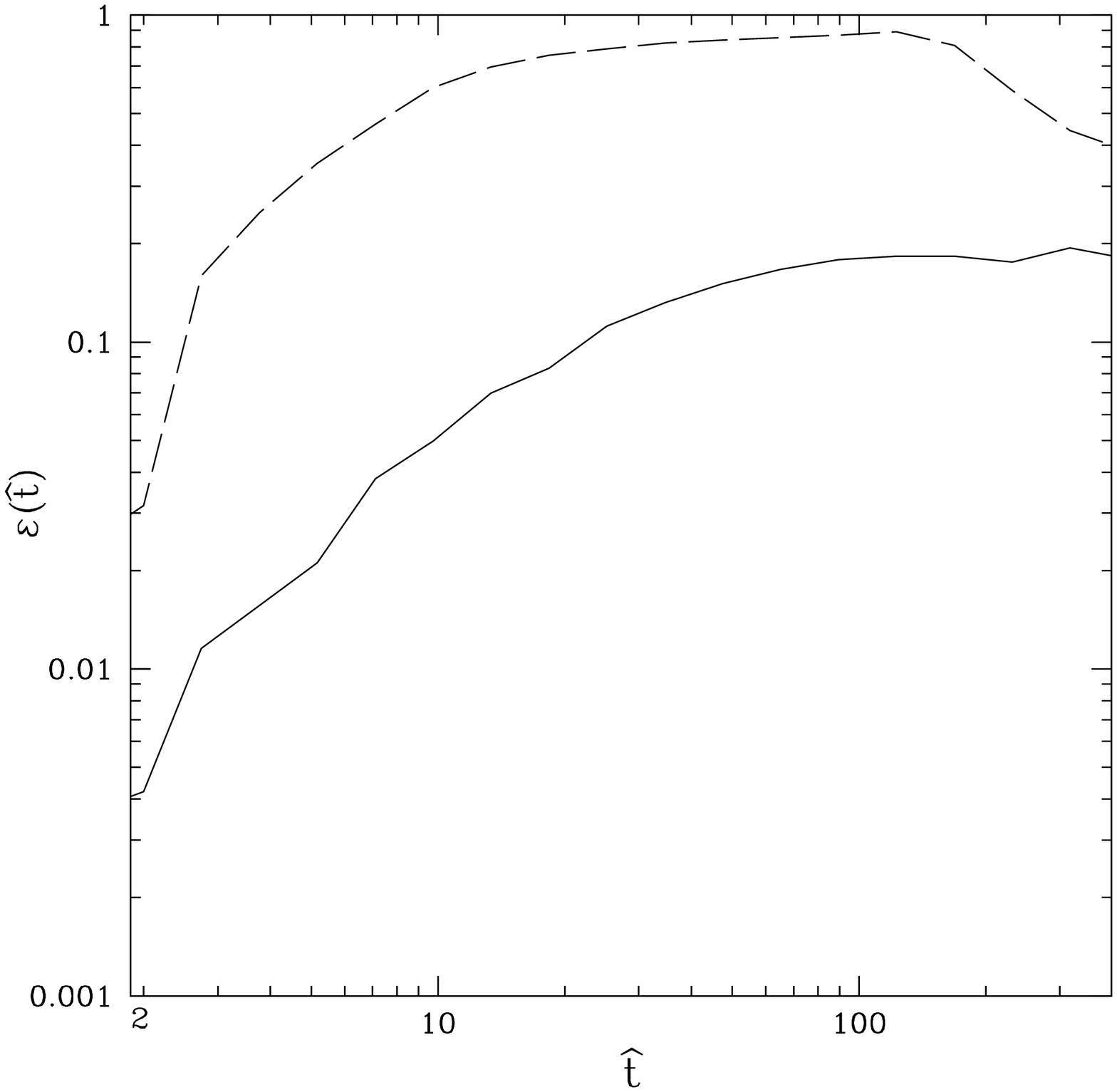} 
\figcaption{Combined
    microlensing detection efficiency for source stars to a limiting
    magnitude of $V \sim 23$ (solid line), and for clump sources (dotted
    line). Efficiencies differ from these lines for individual fields
    because of variations in sampling, reddening seeing, etc.\label{figEff}}
\end{figure}

\noindent
The combined average detection efficiency, as a function of event timescale,
is given in Figure \ref{figEff}. The calculation of the optical depth
contribution from each microlensing event uses the efficiency for the field
where the event was found.

\subsection{\it The Combined Luminosity Function}

The observed detection efficiency is dependent on the number and colour
distribution of target source stars. In the case of pixel lensing events
these stars may lie below the detection threshold of our experiment.  To
determine the properties of stars below our detection threshold we have used
the Hubble Space Telescope (HST) Luminosity Function (LF) and
colour-magnitude diagrams (CMDs) for Baade's Window of \citeN{HWBG98}.  These observations were taken with the HST Wide Fields Planetary
Camera 2 (WFPC2) which has a field-of-view of $\sim 5\Box\arcmin$. 
There are very few bright stars ($ V < 16$) in these HST data, so the
bright end of the luminosity function is not well defined from these data
alone.  However, the MACHO camera's field-of-view is hundreds of times larger
than that of WFPC2 ($43\arcmin \times 43\arcmin$), so many bright stars are
in the MACHO observations of Baade's Window.  We thus combine the luminosity
function and CMD of \citeN{HWBG98} with those of our field $119$ 
(our Baade's Window field).  

To combine these data we transformed the \citeN{HWBG98}'s HST photometry
from $(F555-F814)$ to Landolt's $(V-I)$ with the calibration given in \citeN{HBCTWW95}. 
The HST magnitudes were converted to $Kron-Cousins$ $V-R$
with the transformations of \citeN{Bessell95}.  For MACHO data we used the
conversion from $R_{M}$ and $B_{M}$ to $Kron-Cousins$ $V$ and $R$
given in \citeN{ALC99c}. We note this calibration varies slightly from the
most recent determination given in \shortciteN{Alves99}.  The two data sets
were combined at $V$ $\sim 18.5$ where the MACHO detection efficiency starts
to decrease from unity. The uncertainty in the number of stars from the
combined LF is $9.5\%$.  This uncertainty carries through to our optical
depth estimate with the same magnitude. The number of stars in each of our 8
fields is calculated using this combined luminosity function and the
measured stellar density of clump stars relative to field 119.  By taking
this approach we have implicitly assumed that the stellar density does not
vary greatly within Baade's Window.

A major cause of variations in the apparent stellar density towards the
Galactic center is the patchy extinction caused by dark dust clouds.
However, Baade's Window is well known as a region of low extinction.
Furthermore, the extinction in Baade's Window appears to be relatively
homogeneous \cite{ALC97b}.  Therefore, we believe that the
combined LF should only be slightly affected by extinction inhomogeneities.
We have also made the assumption that the morphology of the CMD does 
not vary between the HST field and the MACHO one. We expect that the 
presence of the Galactic bar \cite{PSUS94b} could cause 
the number of clump giants to vary slightly across our 
Baade's Window field.

\subsection{\it Artificial Events}

To simulate microlensing events, firstly we weighted the $V$-magnitude axis
of the combined (HST+MACHO) CMD to reflect the luminosity function
distribution.  We then binned this CMD and determine the relative
probability of selecting a source star in any given bin.  Next, we randomly
selected a source star for the event based on the bin probability. A random
impact parameter $u_{min}$, timescale $\hat{t}$, and time at maximum
$t_{max}$ were then assigned to the artificial event.  An artificial image
was produced for each observation and the artificial source star was added.
Each image is only $\sim 25\arcsec \times 25\arcsec$ in size.  These images
were further populated with neighbouring so called {\em blend-stars}.  These
stars were selected from the CMD in the same way as the source star and
allow us to simulate the effect of blending on light curves.  The number of
blend-stars we added to the images was based on the observed stellar density
of the field we were simulating.  Each of the artificial stars was placed at
a random location within the artificial image. An artificial image was made
for each observation of the event. These images were produced with seeing
conditions that matched those observed during the data reduction.

For each simulated lensing event we performed photometry on the set of artificial
images making up the light curve. This photometry included the
uncertainties produced by the photon and systematic noise of all the stars
contributing to the flux aperture. The photometry of each of the simulated
events was piped through the detection and selection processes used in the
real analysis to determine the overall efficiency.

The simulation was performed separately on each of the eight Galactic bulge
fields using the seeing conditions, sky background level, etc., of the field
selected.  For each field we produced $100,000$ artificial microlensing
events each with $\sim 300$ observations. The seven fields adjacent to
Baade's Window were simulated using the combined Baade's Window luminosity
function. In each case the LF and the combined CMD were reddened to the
average reddening of each field.  Since the fields are at Galactic latitudes
spanning $\sim 2\arcdeg$ we expect the contributions from the number of disk
stars and bulge stars to vary between fields.  The stellar densities in
simulation of each field were adjusted to reflect the observed densities.
These field-dependent stellar densities were employed because the amount of
blending is dependent on the observed density.

\begin{figure}[ht]
\epsscale{1.0} 
\plotone{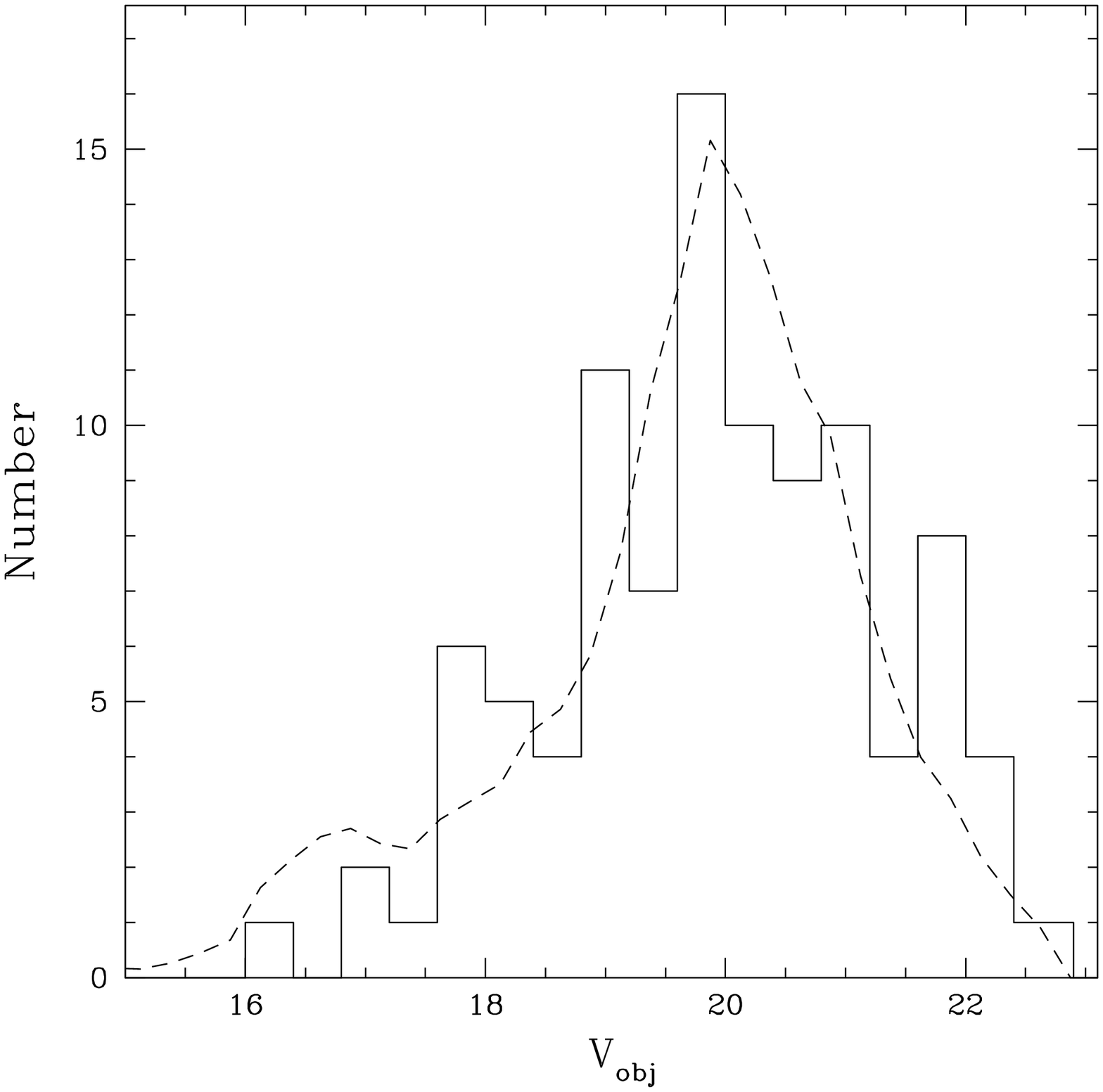}
\figcaption{The distribution of source
    magnitudes.  The solid histogram shows the distribution of the events in
    $V_{obj}$ corrected to the average reddening of field 119 (Baade's
    Window). The dashed line is the luminosity function modified by the
    efficiency and normalised to the area of the histogram.\label{vdist}}
\end{figure}

To compare the results from our simulation with the observed distribution of
events, we have plotted the actual (histogram) and expected (dashed line)
$V$-magnitude distributions in Figure \ref{vdist}. In this plot we have
multiplied the Baade's Window LF by the detection efficiency, thus giving
the expected source star $V$-magnitude distribution. The microlensing event
source star magnitudes ($V_{obj}$) have been extinction corrected to the
average reddening observed in our Baade's Window field.  The good agreement
between these two distributions suggests that our simulated efficiency
analysis reproduces the actual detection sensitivity quite well.  For
further details of the detection efficiency simulation, see \citeN{DrakeTh}.

\section{\sc The Optical Depth}

The microlensing optical depth is defined as the probability that any given
star is microlensed with impact parameter $u_{min} < 1$ (i.e. $A_{max} >
1.34$) at any given time. This optical depth is independent of the mass of
the lensing objects, so no assumption is required about velocity
distributions and mass functions. The optical depth can be estimated by,

\begin{equation}
\tau_{est} = \frac{\pi}{4N_{*}T_{*}} \sum_{i} \frac{\hat t_{i}}{\varepsilon(\hat t_{i})}.
\end{equation}

\noindent
$N_{*}$ is the number of stars and $T_{*}$ is the exposure time of
the experiment in years, $\hat{t_{i}}$ is the Einstein ring crossing time for
the $i$th event, and $\epsilon(\hat{t_{i}})$ is the detection efficiency for
a given event timescale.  In this analysis we use the source fit $\hat{t}$
values, given in Tables \ref{tab5} and \ref{tab6}, to determine the
optical depth.  For the {\em exotic}: finite source, binary lensing and
parallax affected events, we use the $\hat{t}$ values given in Alcock 
{et~al.\ }(\citeyearNP{ALC97a}, \citeyearNP{Becker00} \&
\citeyearNP{Beck00b}).  For microlensing events with timescales which our
experiment is sensitive to ($2 - 300$ days), we obtain
$\tau^{300}_{2}=2.43^{+0.31}_{-0.29}\times 10^{-6}$.  When we include the
uncertainty in the effective number of stars monitored, we obtain
$\tau^{300}_{2}=2.43^{+0.39}_{-0.38}\times 10^{-6}$.

\begin{figure}[ht]
\epsscale{0.9}
\plotone{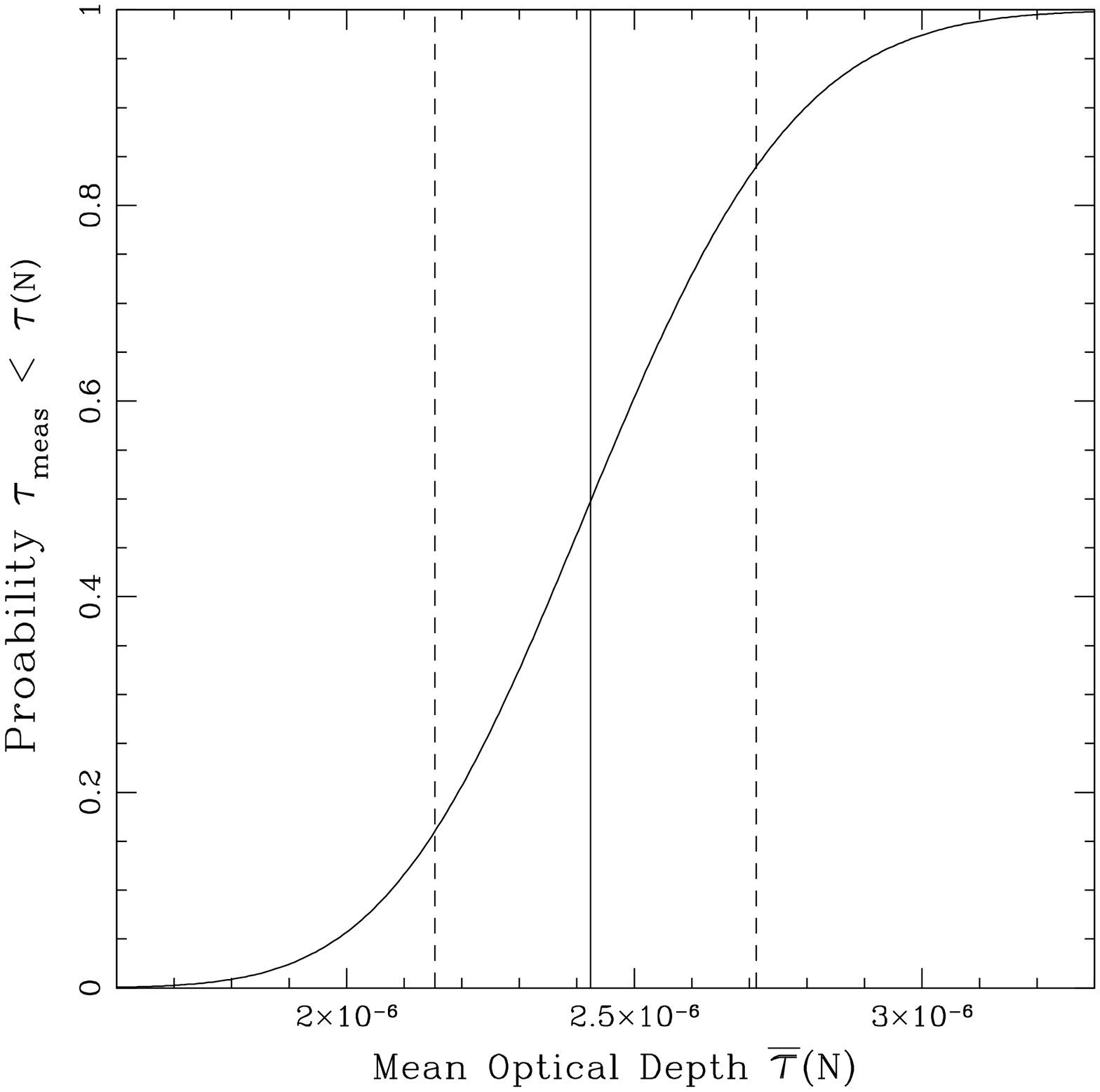}
\figcaption{The distribution of statistical uncertainty in the microlensing 
optical depth from Monte-Carlo simulation for our 99 event sample.
The dashed lines mark one $\sigma$ confidence limits. The solid line gives
the observed optical depth.\label{figOptUn}}
\end{figure}

The statistical uncertainties in the optical depth have been obtained by
using the fact that the number of events obey Poisson statistics. We have
simulated ``experiments'', where the number of observed events $N$, follows
a Poisson distribution. For each event in the simulated timescale
distribution we randomly assign one of our measured event timescales (see
also \shortciteNP{USKK94d,ALC97e}).  The optical depth
for each of these distributions is then evaluated.  Counting the fraction of
these experiments which yield a larger optical depth than our measured value
$\tau$, allows us to determine the statistical distribution of optical
depth.  This distribution is shown in Figure \ref{figOptUn}, from which we
determine confidence limits on our measured optical depth $\tau_{meas}$.

\begin{figure}[ht]
\epsscale{0.9} 
\plotone{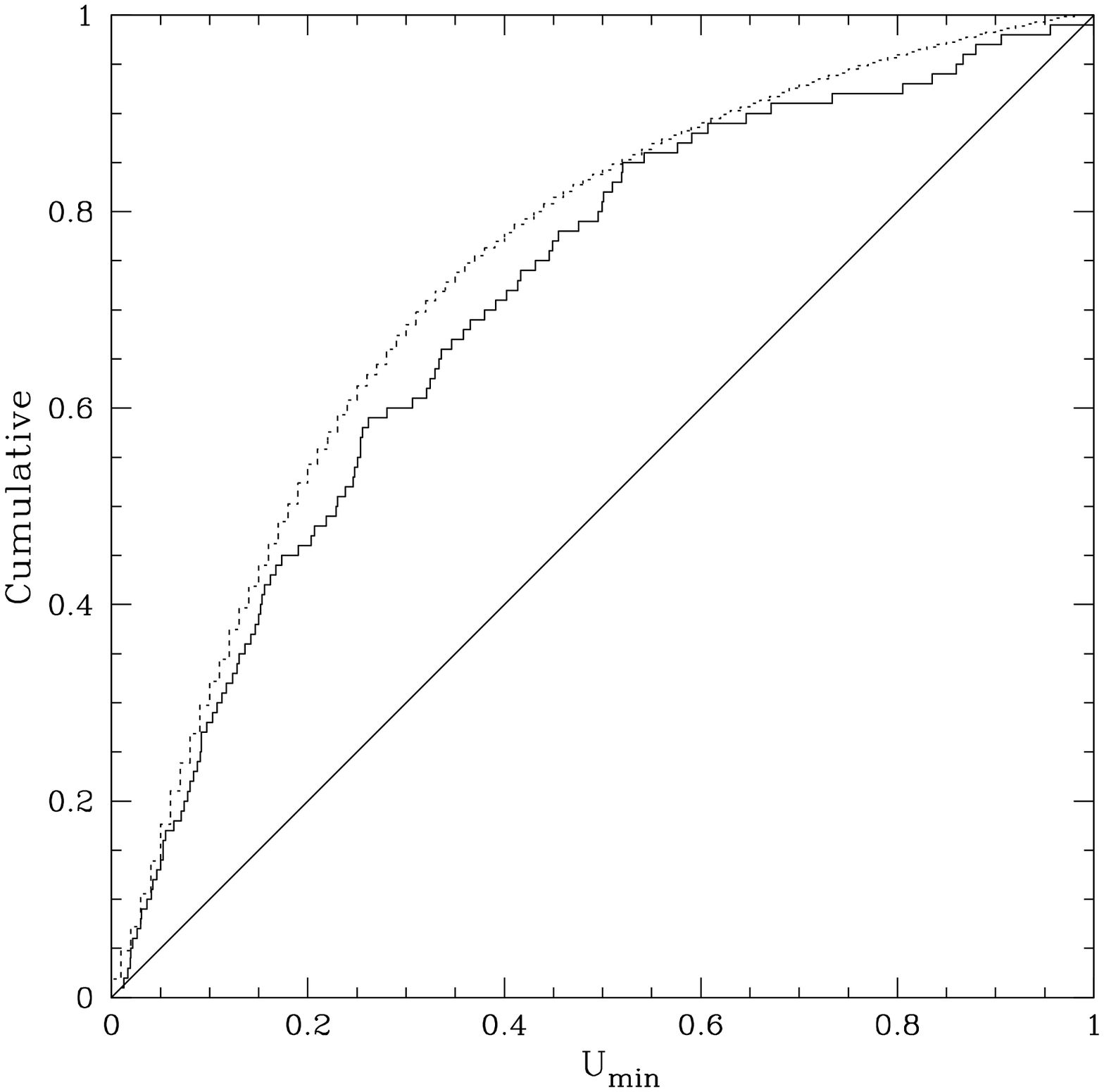} 
\figcaption{The cumulative distribution
    of impact parameter for the 99 microlensing event sample. Here the
    straight line shows the expected $u_{min}$ distribution for microlensing
    without the effect of efficiency.  The dotted histogram show the
    distribution for the efficiency simulation and the solid histogram the
    observed distribution.\label{figKS}}
\end{figure}

The simple geometry of microlensing events results in the theoretical
prediction of a linear distribution in impact parameters ($u_{min}$).
However, experimentally events with faint source stars require larger
magnifications to be detected. This tendency skews the impact parameter
distribution towards smaller $u_{min}$ values.  The effect of this is
clearly seen in the cumulative distribution shown in Figure \ref{figKS}.

A Kolmogorov-Smirnov (K-S) test is a useful method for comparing whether
samples of events are drawn from the same distribution.  A comparison
between the expected distribution from our efficiency analysis and that
observed, gives a K-S statistic of $D = 0.086$.  This corresponds to a
probability of $P(KS)=0.39$. In a comparison between the observed
distribution and a uniform distribution, we get a K-S statistic of $D=
0.338$ corresponding to $P(KS)= 5.78e^{-10}$.  The observed $u_{min}$
distribution is only compatible with that found using our efficiency 
simulation, which demonstrates the bias towards high amplification 
events.

\subsection{\it The Optical Depth in Individual Fields}

The microlensing optical depth is a measure of the mass in lensing objects
along a line-of-sight. This quantity is independent of the individual masses
of lensing objects, as long as they have characteristic timescales which lie
within the region to which the experiment is sensitive.  To calculate this
statistic from the measurements we do not need to know the velocity or mass
distributions of the object causing the lensing. This value is therefore
useful to compare the measurements with the predictions from Galactic
models, where the distribution are assumed.  However, any Galactic models
which match the observed total optical depth must also match the observed
event timescale distribution.  Event timescales provide a constraint on a
model's velocity and mass function. By splitting the dataset of observed
events into sub-regions, one can also compare the optical depth distribution
as a function of location.

Since we expect the relative number of disk stars and bulge stars to vary
between fields we decided to model this variation.  To account for this we
produced a disk-LF by combining the LFs of \citeN{WJK83p163} and
\citeN{GBF96} at $M_{V} = 9$. The result is a luminosity function for the
disk ranging from $M_{V} = -1$ to $M_{V} \sim 14$.  We assume that the disk
is well modeled by the standard double exponential disk density profile
given by,

\begin{equation}\label{disk}
\rho_{d} = \rho_{d0}\;{\rm exp}\left (\frac{-|z^{\prime}|}{h} 
+ \frac{(R_{0}-s)}{s_{d}} \right).
\end{equation}

\nocite{ZSR95}
\nocite{VA95}\noindent
For, the disk cylindrical galactocentric coordinates ($s,z^{\prime}$),
we have $s^{2}  =  1 + z^{2}\cos^{2}{\theta} - 2z \cos{b}\cos{l}$,
and $z^{\prime}  =  z\sin{b}$,
where $z  =  D_{d}/D_{8.5}$ and $s_{d}$ is the disk scale length 
($2.5-3.5$ kpc). The mass density in the solar neighbourhood 
is given by $\rho_{d0}$ (0.05 $\rm M_{\odot}pc^{-3}$), and $h$ is 
the scale height of the disk.


We use this double exponential disk model to determine the form of the
disk-LF that we expect in each of our fields, with proper consideration of
the reddening and the number of stars within the observed volume.  Here we
assumed a disk scale height $h$ of 325 pc and a scale length $s_{d}$ of 3
kpc.  We normalised the disk-model luminosity function for each field to the
number of disk stars, observed in the magnitude range $14 < V < 18$. This
allowed us to approximate how many disk stars there are in each field down
to the source star magnitude limit ($V=23$). The percentage of disk stars at
this limit, $p$, was given in Table \ref{tab3}.  By subtracting the
disk-model LF for the Baade's Window field from the HST+MACHO combined LF,
we were able to determine the bulge luminosity function. We then assumed
that the number of bulge stars in our fields were traced by the number of
clump giant stars.  The detection completeness for these bright clump stars
is nearly $100\%$, so these stars serve as a good tracer of the number of
fainter bulge main sequence stars, where our detection completeness is low
and uncertain. We scaled our disk-subtracted bulge luminosity function by
the ratio of the number of clump stars in each field relative to our Baade's
Window field, $r$ (Table \ref{tab3}). We finally determined the total number
of stars in each field by combining the number of disk stars from our disk
model, and bulge stars from the scaled Baade's Window bulge luminosity
function.

The optical depth for each individual field (given in Table \ref{tab3}) was
determined using the total number of stars in each field. From the tabulated
data it seems there is a trend in optical depth with Galactic latitude and a
weaker trend with longitude, as is expected. However, the observed optical
depth for field 104 appears to break from the general trend, although the
uncertainties for this field are quite large. The density of stars in field
104 is not significantly higher than the other fields ($N_{104} \approx
N_{108}$), and this evidence naturally leads us to believe that the number
of faint lensing stars should be similar. A group or cluster of low-mass,
faint stars in the foreground of the field could go undetected in our
photometry and would act as efficient lenses. There is a globular cluster in
field 104 and another nearby, but the locations of these do not appear to
coincide with the observed microlensing events.

\subsection{\it Uncertainties in the Optical Depth}

The major cause of uncertainty with the determined optical depth could be
erroneous values from the microlensing fits. For low S/N events it is
difficult to determine whether the correct amplifications and $\hat{t}$
values have been found.  For such events there are similarities between the
light curves of events with different timescales and amplifications
\cite{WP97}.  However, \citeN{Han99} has recently shown
that fits to high amplification pixel lensing events do give accurate
results.

In this analysis we have used a single CMD to determine the microlensing
detection efficiency of the analysis. This CMD is artificially reddened
to the average value determined for each field, which assumes that the
morphology of the CMD does not vary much between these fields.  From our
disk model it appears that the number of disk stars relative to bulge stars
varies little from field to field.  The gradient in the observed optical
depth is in part due to the dependence of the numbers of disk stars on the
Galactic latitude.
If our assumed model of the disk is in error, the number of stars we
determine for the individual field will be incorrect. This will also affect
our disk-subtracted bulge luminosity function. However, the calculated
number of disk stars is only about $10\%$ of the total number of stars in
any field, in good agreement with \citeN{ZCFGOR99}, so this effect
should be small.  Our total optical depth result is less dependent on the
assumed disk model than the individual fields, since the ratio of the total
number of disk stars ($14 < V < 18$) to clump stars in all 8 fields
combined, is very close to the ratio of disk to clump stars in Baade's
Window.

In addition, the extinction in each observed field is taken into account
using values determined from the RR Lyrae stars in each field
\shortcite{ALC98a}.  There are small uncertainties in the efficiencies due
to variations in reddening within a field. This should not affect the
overall optical depth but may be important for estimates in individual
fields.

\section{\sc The Structure of the Galaxy}

We will now review what has been reported about each of the Galactic
features that have the largest effect on the observed microlensing optical
depth. For each of these we will discuss whether our results are consistent
with models and previous determinations.

\subsection{\it Bar Orientation}

The optical depth is highly dependent on the position angle of the bar
\shortcite{Peale98} and bulge mass \cite{Gyuk99}.  The
observational results for the bulge inclination based on a number of
different types of observations give conflicting values ranging from
$16\arcdeg$ to $44\arcdeg$, see Table \ref{tab7}.  A bar inclined at the
large angle reported by \citeN{SSV99} is not an efficient source of
lensing events. The size of our observed optical depth favours the smallest
possible bar inclination angle.  However, the bar is insufficient to produce
an optical depth greater than $\sim 2.5 \times 10^{-6}$ even for models with
a small bar inclination angle, and a large bar mass (see \citeNP{Peale98}).

\subsection{\it Bar Mass}

Ideas to explain the observed microlensing optical depth with values of
bulge mass are also not clear-cut.  Based on COBE map data \citeN{ZM96}
found $M_{bulge} =(2.2\pm0.2) \times 10^{10}\,\rm M_{\odot}$, yet
\citeN{DAH95} found a mass of only $1.3 \times 10^{10}\,\rm M_{\odot}$.
\citeN{BEBG97} found that, based on DIRBE results, within 2.4 kpc of
the Galactic center the combined bulge plus disk mass is $1.9 \times 10^{10}\,
\rm M_{\odot}$.  However, only $0.72-0.86 \times 10^{10}\,\rm M_{\odot}$ of this is
attributed to the bulge mass. This is consistent with \citeN{HWBG98}'s
results ($0.74 - 1.5 \times 10^{10}\, \rm M_{\odot}$).

Predictions of the Galactic bar mass have also been discrepant.  Based on the virial
theorem, \citeN{HC95b} predicted that $M_{bar}= 1.6 \times
10^{10}\,\rm M_{\odot}$, as did \citeN{Kent92} based on a simple oblate
rotator model. But \shortciteN{Blum95} predicted $M_{bar}= 2.8 \times
10^{10}\,\rm M_{\odot}$ when pattern rotation of the Galactic bar is included. 
\citeN{ZM96} determined that a bar mass of at least 
$> 2.0 \times 10^{10}\,\rm M_{\odot}$ is required for the COBE G1 model 
(with $\theta=11$ degrees) to account for observed amount of lensing. 
The \shortciteNP{ZM96} model is consistent with \citeN{ALC97a}'s
microlensing data at the 2 $\sigma$ level, if a bar mass of $2.8 \times 10^{10}\,\rm M_{\odot}$ 
is used. \citeN{Gyuk99} advocated $M_{bulge}= 2.5 \times 10^{10}\,\rm M_{\odot}$
based on a maximum likelihood estimate of the COBE G2 model, where a small
inclination angle of $\theta= 12$ degrees was assumed.  However, if $\theta$
was instead taken to be 20 degrees (consistent with most of the
values in Table \ref{tab7}), the most likely bulge mass rises to 
$\sim 3.6 \times 10^{10}\,\rm M_{\odot}$. A heavy bar is favoured for our 
observed optical depth, but as yet there is no evidence that the bar is 
massive enough to produce the observed optical depth. It is clear that more 
accurate measurements are necessary to better constrain the bar mass 
used in models. In Table \ref{tab8} we present the optical depths for a 
number of Galactic models for comparison with our result.


\subsection{\it The Disk}

The estimates of the optical depth due to the disk also exhibit a range of
values. Spiral arms may contribute $0.5 \times 10^{-6}$ \shortcite{Fux97}.  
A truncated disk would contribute $0.37-0.47 \times 10^{-6}$
\shortcite{PAC91}, whereas a full disk would contribute
$0.63-0.87 \times 10^{-6}$ \cite{ZSR95}.  However, from these values it
is clear that the disk is expected to be a less important contributor to
the optical depth than bar mass or bar orientation. The only measurement 
of the contribution of disk lensing comes from the EROS II analysis
($0.38^{+0.58}_{-0.15} \times 10^{-6}$, \shortciteNP{DAF99}). 
This is in good agreement with predictions, but is based on just three 
microlensing events.

We have estimated a disk lensing contribution to the optical depth of
$f_{disk} \sim 25\%$.  This gives 
$\tau_{_{bulge}}= 3.23^{+0.52}_{-0.50}\times 10^{-6}[0.75/(1-f_{disk})]$.  
Disk lenses are not expected to contribute much more than $25\%$ since there
is little evidence of any disk-disk lensing of the foreground main sequence
stars in Figure \ref{figCMD}.  However, the optical depths we observed in
the individual Galactic bulge fields are quite high and are thus consistent 
with a large disk optical depth.

\subsection{\it The Timescale Distribution}

The mass function of the compact objects in the Galaxy has a direct effect
on the timescales of microlensing events.  To date, few authors have
attempted to reproduce simultaneously both the optical depth and the
timescale event distribution. The agreement between models and observations
under these two constraints is necessary, if any confidence is to be put in
the models.  \citeN{Peale98} and \shortciteN{MER98a}
have found that the Galactic models poorly reproduce the
observed timescale distribution of \citeN{ALC97a}.

\subsubsection{Short Timescale Events}

The geometry of microlensing produces short timescale events where the
lenses are either near the observer or the source.  However, variation in
the lens location has only a small effect, unless either the observer-lens
distance or lens-source distance is very small. Short timescale events can
also be produced by low mass lenses, but the timescale goes as $\rm
M^{1/2}$, so the effect is small. Distributions with a significant number of
short timescale events and a number of long timescale events, point to a
large range of lens masses.

In an attempt to reproduce the observed short timescale events of
\citeN{ALC97e}, mass functions with large numbers of low mass
objects, such as brown dwarfs, have been produced \shortcite{MER98a,Peale98,HC98}. This is supported by recent discoveries of
free floating brown dwarfs \cite{RKLBG99,CSMDL99}.  But there is still
controversy surrounding whether this will \cite{HC98}, or will not
\cite{Peale99} reproduce the expected timescales correctly.
If the short timescale events are not brown dwarf lensing events, 
then this indicates a large population of low mass stars in the bulge.
It is possible that a large number of M-type stars might 
better explain the timescale distribution \shortcite{Peale98}, 
but this does not appear to be consistent with the shallow slope
of the mass function found in recent deep observations of the 
Galactic bulge \cite{HWBG98,ZCFGOR99}.

\subsubsection{Long Timescale Events}

The timescale distributions from contemporary Galactic models also do not
reproduce the observed number of long timescale events, $\hat{t} > 140$ days
(for example, see \shortciteNP{ALC97e,HC96b,MER98a,Peale98}).  With the data
of \shortciteN{ALC97e} it was unclear whether these long timescale events
were a real population or a statistical anomaly. For long timescale events
we expect either large lens masses, low transverse velocities, or equal
observer-lens and lens-source distances. Disk-disk lensing events are
expected to give long timescale events because of the low velocity
dispersion of the disk. Long timescale events have been observed for known
disk-disk lensing events \shortcite{DAF99}. However, disk-disk lensing
events are considered in the Galactic models and are constrained by star
counts.  The variation in lens-observer and observer-source distances has a
relatively small effect on an events timescale (factor of $\sim 2$).  If the
lenses are normal main sequence stars, their luminosities can easily be
related to their masses.  Owing to this relation, for nearby lenses we can
impose an upper-limit to the lens mass, given an upper-limit to the lens'
luminosity (from the microlensing fits). This constrains the proximity of
bright lenses to the observer \shortcite{ALC97a}.  However, reddening
towards the Galactic bulge is patchy and weakens this argument.  One
possibility is that there exists some unknown population of dynamically cold
or massive, dark objects in the Galactic disk.

\subsubsection{The Observed Timescale Distribution}

\begin{figure}[ht]
\epsscale{1.0}
\plottwo{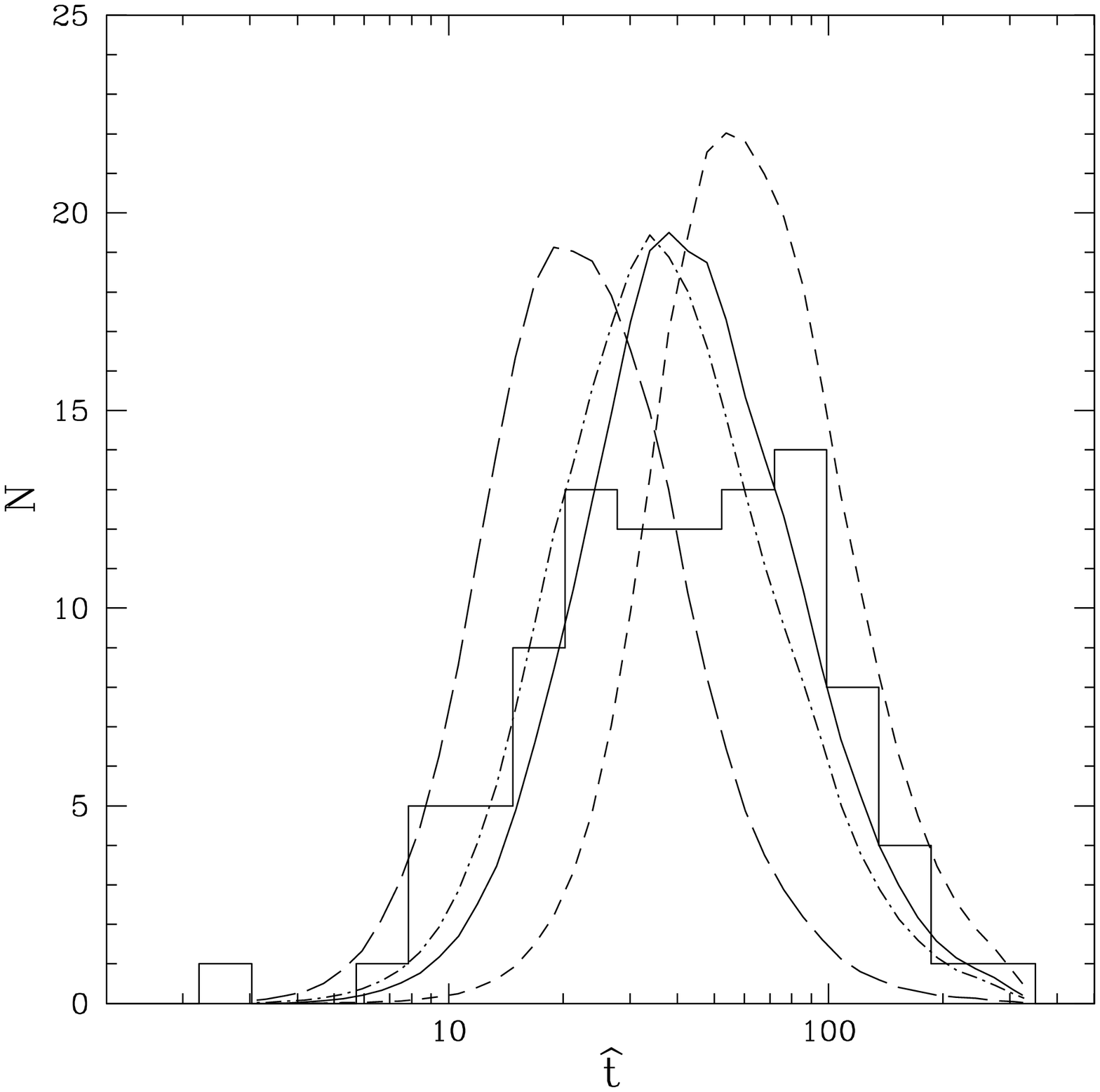}{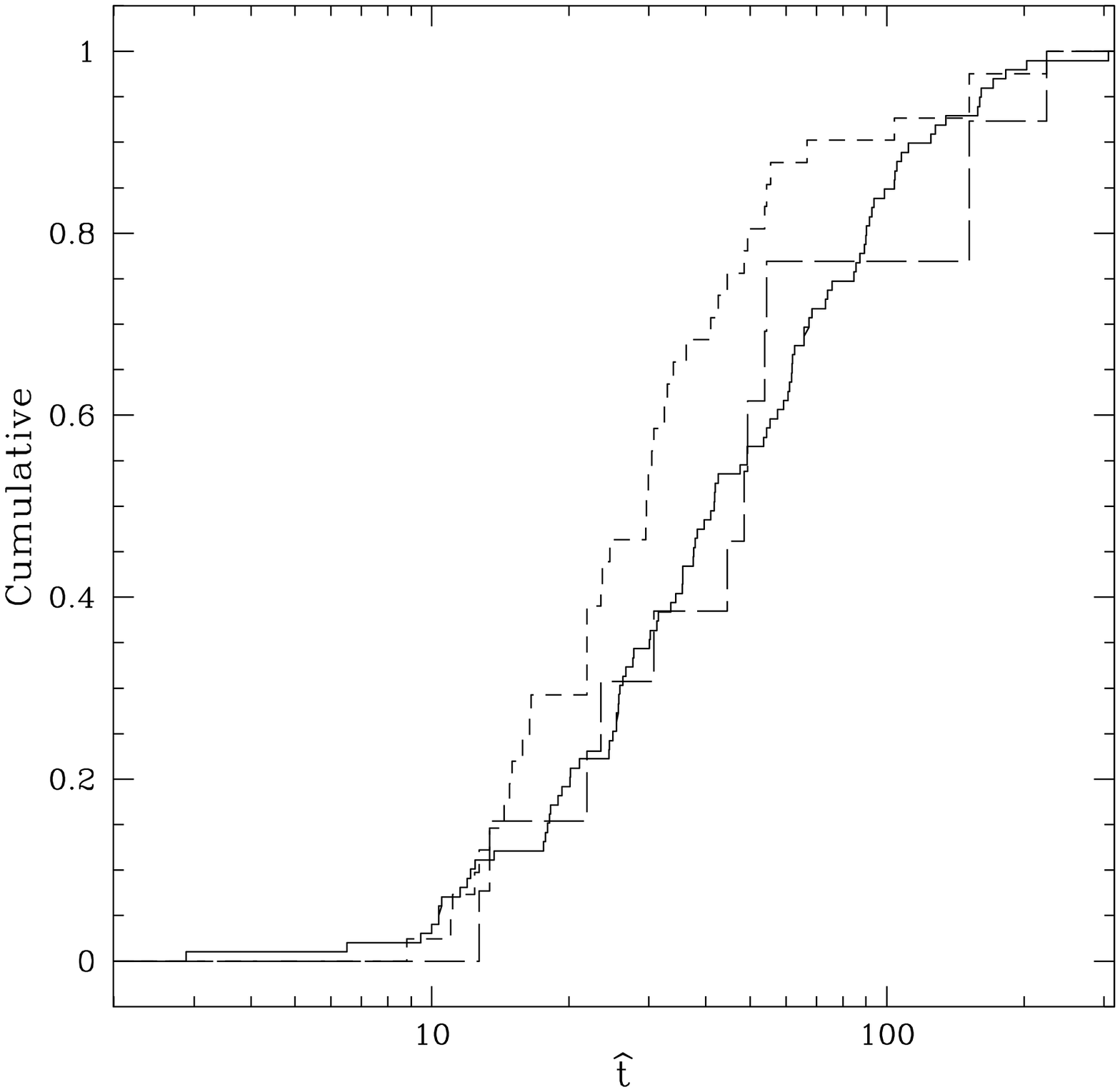}
\figcaption{Left: the timescale ($\hat{t}$) of the 99 candidate microlensing
  events compared to predictions from four mass models, normalised to the
  observed number of events.  The mass functions are: a $\delta$ function at
  0.1M$_{\odot}$ (long-dashed line); a $\delta$ function at 1M$_{\odot}$
  (the short-dashed line); a Scalo (1986) PDMF (solid line); the Han \&
  Gould (1996) power-law model with $\alpha = -2.3$ and $m_{lo} = 0.1$
  (dash-dotted line).  Right: the cumulative timescale distributions of the
  events from this work (solid line). The Alcock {et~al.\ }(1997e) full
  sample of 41 events (short-dashed line) and the Alcock {et~al.\ }(1997e)
  13 clump giants (long-dashed line).  The results are consistent with the
  13 clump sample but not with the 41 event sample which are affected by
  blending problems.\label{figdis}}
\end{figure}

The microlensing event timescale distribution is plotted in Figure
\ref{figdis}. Here we have over-plotted the timescale distribution expected
for four mass functions assuming the barred bulge, given in equations
(\ref{pt1}) and (\ref{pt2}), plus a double exponential disk density model
given in \shortcite{ALC97e}.  These four mass models are: a \citeN{Scalo86}
main sequence present day mass function (hereafter PDMF), two $\delta$
function distributions (0.1M$_{\odot}$ \& 1M$_{\odot}$), and the power-law
distribution of \citeN{HC96b} ($\alpha = -2.3$, $0.1 < m < 1.4M_{\odot}$).
The agreement between the \citeN{Scalo86} mass model and event timescale
distributions appears quite reasonable.  This might imply that the
microlensing events seen towards the Galactic bulge can be explained by the
observed distribution of stars, if they follow a \citeN{Scalo86} mass model.
However, this does not explain the large observed optical depth.  The
timescale distribution also appears to be somewhat broader and much less
peaked than expected. This difference could possibly be due to uncertainties
in the timescales of the fainter events.  There does not appear to be a
large population of short timescale events in the distribution. This implies
there is probably not a large population of brown dwarfs along the bulge
line-of-sight. However, there is some evidence for a population of long
timescale microlensing events, as there are seven events with timescales
$\hat{t} > 140$ days.

In the right panel of Figure \ref{figdis} we present the cumulative
distribution of event timescale for these results compared with our previous
results for the Galactic bulge.  Here we find relatively few short timescale
events compared to the previous results of \shortciteN{ALC97e}, where fits
did not include parameters for the blended flux component.  A K-S statistic
comparison between the cumulative distributions of the events from
\shortciteN{ALC97e} and this analysis gives a K-S statistic $D =0.296$,
corresponding to a probability $P(KS)= 0.0154$. This indicates that there is
a significant difference between the two timescale distributions.  The fact
that blending will bias the event timescale distribution to shorter values
has been known for many years \cite{Nemiroff94} and has been studied in
detail by a number of authors (\citeNP{ALC96d,WP97,Alard97,HAN97a};
\shortciteNP{HJK98}). Therefor, this result is not surprising.  We note that a
small fraction of the difference between these and the previous results is
also due to difference between the detection efficiencies of
\shortciteN{ALC97e} and those of this analysis.  In a comparison with the 13
clump giant sample of \shortciteN{ALC97e} we get a K-S statistic of $D =0.25$
corresponding to $P(KS)= 0.37$.  Our results are thus consistent with the
clump giant microlensing events. Clump giant microlensing events are less
affected by blending since they generally are much brighter than the stars
with which stars they are blended. This suggests that, as expected, our
microlensing timescale distribution is less biased by blending than the
distribution of 41 events given in \shortciteN{ALC97e}.

\begin{figure}[ht]
\epsscale{1.0} 
\plottwo{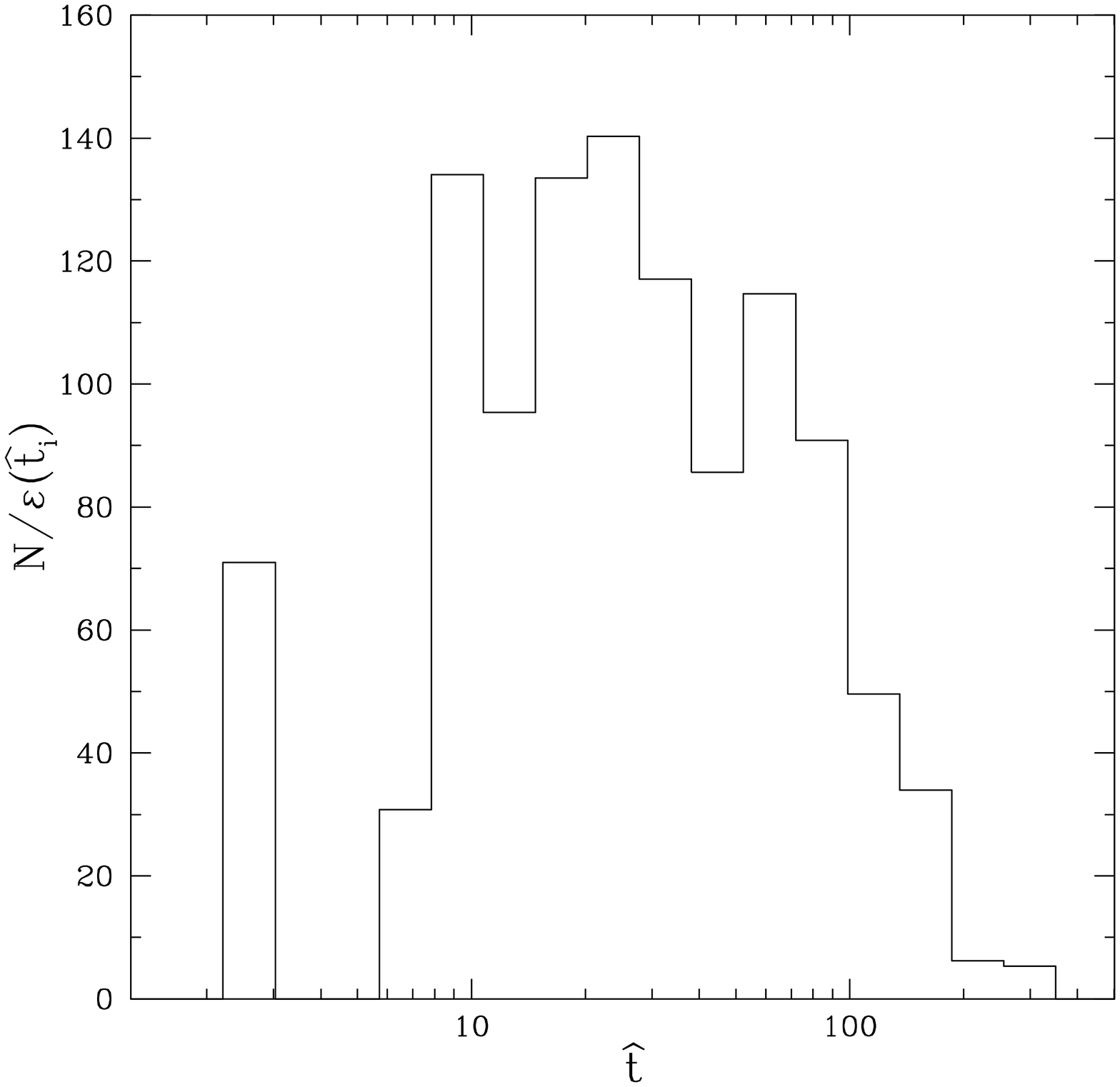}{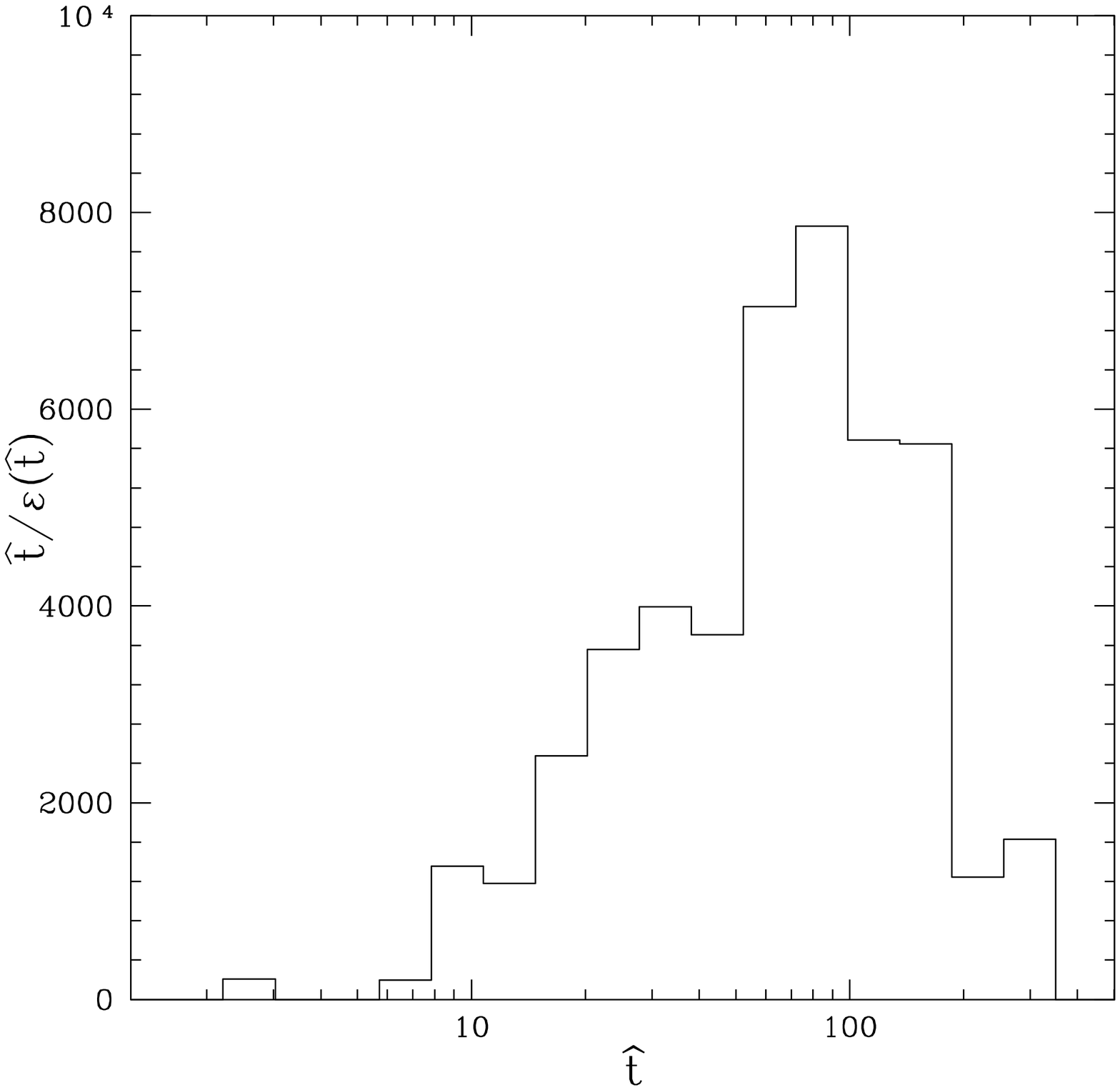}
\figcaption{Left: a histogram of the $\hat{t}$ distribution corrected to
    $100\%$ efficiency (The expected true event timescale
    distribution.)\label{figcor}.  The first bin in the distribution is a
    lower estimate as the efficiency has been truncated at 2 days where
    events are too short to be detected in this analysis. Right: the
    contribution to the overall microlensing optical depth ($\tau$) of the
    observed event timescale distribution.  }
\end{figure}

The efficiency corrected timescale distributions are presented in Figure \ref{figcor}.
From the right panel of Figure \ref{figcor} one can also see that the optical
depth for this sample of events is not dominated by the long timescale
events as it was in \citeN{ALC97e}.  The present optical depth
determination is less dependent on a small number of long timescale events.
However, the relative contribution of each individual long timescale event
is still large compared to short ones because of the detection efficiency.

\section{\sc  Summary}

We have presented the results from the DIA survey of MACHO Galactic bulge
data. In this analysis we detect 99 microlensing events in eight fields.
This survey covers three years of data for $\sim$ 17 million stars to a
limiting magnitude of $V \sim$23.  Our result is consistent with the
detection of $75-85\%$ more events than an analysis performed with PSF
photometry on the same data.

We have measured a microlensing optical depth of $\tau= 2.43^{+0.39}_{-0.38}
\times 10^{-6}$ for events with timescales between 2 and 300 days.  With
consideration of the disk-disk component we find a Galactic bulge
microlensing optical depth of $\tau_{_{bulge}}= 3.23^{+0.52}_{-0.50}\times
10^{-6}[0.75/(1-f_{disk})]$.
These optical depth determinations are consistent with the previous 45 event
analysis and 13 event clump giant sub-sample of \shortciteN{ALC97e}, and the
value determined by the OGLE group \shortcite{USKK94d}.

For the individual fields we find that there is a trend in optical
depth with longitude and latitude as expected, although there is some
evidence of fine structure within the optical depth spatial distribution.
However, it is difficult to set limits on this as the uncertainties for each
field are quite large.

We find that our timescale distribution is compatible with lenses having
masses distributed in the same way as the PDMF of Scalo (1986).  We
note that the timescale distribution of \shortciteN{ALC97e} was biased
towards shorter events by blending, making it appear that many low mass
objects were required to explain the observed distribution. With our new
unbiased sample, we do not require a large population of brown dwarfs towards
the Galactic centre to reproduce the measured timescale distribution. It is
still unclear whether or not there is an anomalous population of long
timescale events. 

The measured microlensing optical depth is a lower limit to the true value
as events shorter than few days or longer than a few hundred would not be
detected in this analysis.  Nevertheless, the values presented here
still appear to be larger than those predicted by most Galactic models with a bar
mass and inclination consistent with observations. Such models might be
better constrained by attempting to reproduce the optical depth, the
observed timescale distribution and the observed optical depth gradient
measured here.

We are grateful to S.~Chan, M.~MacDonald, S.~Sabine and the technical staff
at the Research School of Astronomy and Astrophysics\footnote{Formerly Mount
Stromlo Observatory.} for their skilled support of the project.  
This work was in part performed under the auspices of the U.S. Department 
of Energy by University of California Lawrence Livermore National 
Laboratory under contract No. W-7405-Eng-48.
Work at the Center for Particle Astrophysics at the
University of California, Berkeley is supported by NSF grants AST 88-09616
and AST 91-20005. Work at Mount Stromlo and Siding Spring Observatories is
supported by the Australian Department of Industry, Technology and Regional
Development.  K.~G. and T.~V. acknowledge support from DoE under grant
DEF0390-ER 40546.  W.~J.~S. is supported by a PPARC Advanced Fellowship.
C.~W.~S. is grateful for support from the Sloan, Packard and Seaver
Foundations. D.~M.  is supported by Fondecyt 1990440. This work was carried
out by A.~J.~D. in partial fulfillment of the requirements for the degree of
PhD at ANU.


\begin{deluxetable}{llll}
\footnotesize
\tablecaption{Parameters of Level-1.0 Cuts.\label{tab1}}
\tablewidth{0pt}
\tablehead{\colhead{Cut type} & \colhead{} & \colhead{Cut level} & \colhead{}}
\startdata
generic var. & $66.6\%$ pts, $\rm F_{i} < (F_{med} + 4 \sigma_{i})$\tablenotemark{a} &
$54.5\%$ pts, $\rm F_{i} < (F_{med} + 2.5 \sigma_{i})$\tablenotemark{a} &
$31\%$ pts, $\rm F_{i} < (F_{med} + 1.5 \sigma_{i})$\tablenotemark{a} \nl
transient var. & $90\%$ pts, $\rm F_{i} > (F_{med} - 4.5 \sigma_{i})$\tablenotemark{b} &
$93\%$ pts, $\rm F_{i} > (F_{med} - 8 \sigma_{i})$\tablenotemark{b} &
$<$11 pts, $\rm F_{i} < (F_{med} - 11 \sigma_{i})$\tablenotemark{b} \nl
pulsating var. & $<$5 pts\tablenotemark{d}$\;$, $\rm F_{i} < (F_{med} - 3.5 \sigma_{i})$\tablenotemark{b} &
$<$ 4 pts\tablenotemark{d}$\;$, $\rm F_{i} < (F_{med} - 5.5 \sigma_{i})$\tablenotemark{b} &
$<$5 pts\tablenotemark{d}$\;$, $\rm F_{i} < (F_{med} - 5.5 \sigma_{i})$\tablenotemark{c}\nl
2nd peak & $SN_{p3} < 13 \sigma_{i}$, $t_{i} > (t_{max} + 110)$\tablenotemark{b} &
$SN_{p4} < 16 \sigma_{i}$, $t_{i} > (t_{max} + 110)$\tablenotemark{c}\nl 
photometry std & $95\%$ pts, $\rm F_{i} < \rm F_{p}$\tablenotemark{a} &
$>$ 100 good photometry pts\tablenotemark{a}\nl
\enddata
\tablecomments{Microlensing event selection cuts.
We denote the uncertainty in an individual flux measurement 
$\rm F_{i}$, taken at time $t_{i}$, as $\sigma_{i}$. 
Event peaks occur with flux measurement $\rm F_{p}$ at $\approx t_{max}$.
The median flux value for a lightcurve is $\rm F_{med}$.
The total signal to ratio of three points in a second peak is denoted $SN_{p3}$.
For four points this is denoted $SN_{p4}$.
}
\tablenotetext{a}{ In $R_{m}$ or $B_{m}$ bands.}
\tablenotetext{b}{ In $R_{m}$ \& $B_{m}$ bands.}
\tablenotetext{c}{ In $R_{m}$ band.}
\tablenotetext{d}{ Consecutive good photometry points.}
\end{deluxetable}

\begin{deluxetable}{ccccccc}
\footnotesize
\tablecaption{Parameters of level-1.5 \& 2.0 Cuts.\label{tab2}}
\tablewidth{0pt}
\tablehead{\colhead{Level} & \colhead{$\rm F_{pr}$} & \colhead{$\rm A_{F(>)}$} &
\colhead{$\chi_{c(<)}$} & \colhead{$\chi_{m(<)}$} & \colhead{$\rm SN_{r(>)}$}}
\startdata
1.5 & \nodata & 1.34 & 9.3 & 3.0 & \nodata\nl
1.5 & $>$ 2.5e4 & 1.34 & 15 & \nodata & \nodata\nl
1.5 & $>$ 5e4 & 1.34 & 30 & \nodata & \nodata\nl 
1.5 & $>$ 3e2 & 10 &  9.5 & 4.0 & \nodata\nl
1.5 & $>$ 3e2 & 20 &  15 & 4.0  & \nodata \nl
1.5 & $>$ 3e3 & 20 &  15 & \nodata & \nodata \nl
1.5 & $>$ 3e4 & 10 &  30 & \nodata & 100\nl
2.0 & $<$ 8e3 & 1.34 & \nodata & $1.65 + 1.8e$-$4 \rm\, F_{pr}$  & \nodata\nl
2.0 & $<$ 1e4 & 1.34 & $3.75 + 6.0e$-$4 \rm\, F_{pr}$   & \nodata & \nodata\nl
2.0 & 1e4 $< \rm\, F_{pr} <$ 1e5 & 1.34 & \nodata & \nodata & $2 + 2e$-$4 \rm\, F_{pr}$ \nl
2.0 & $>$ 1e4 & 1.34 & \nodata & \nodata & \nodata \nl
\enddata
\tablecomments{Microlensing event selection cuts. 
All level 1.5 events must pass the cuts in the text
($V-R > 0.55$, $A > 1.34$, $\Omega\chi^{2} < 3.6$, $\hat{t} < 365$ days, 
$t_{max_{ij}} \neq t_{max_{kj}}$, $\chi^{2}_{c} < 30$) in addition to one line 
of level 1.5 cuts.
All events must also pass all 2.0 cuts for which their red fit peak flux $\rm F_{pr}$ 
is in the range of the cut. Unlisted values specify no cut is applied on that parameter.
Here $\rm A_{F>}$ specifies that the fitted microlensing event amplitudes $\rm A_{F}$ must be 
greater to pass this cut. Likewise for the other columns, signal to noise ratio 
in the red band ($SN_{r}$), microlensing fit reduced chi square
$\chi_{m}$ and constant baseline fit reduced chi square $\chi_{c}$.
}
\end{deluxetable}


\begin{deluxetable}{cccccccccc}
\tablecaption{Individual Fields.\label{tab3}}
\footnotesize
\tablewidth{0pt}
\tablehead{\colhead{Field} & \colhead{SOD} & \colhead{DIA} & 
\colhead{New}& \colhead{$l$ $(\arcdeg)$} &
\colhead{$b$ $(\arcdeg)$} & \colhead{$r$} & \colhead{$p$ ($\%$)} & \colhead{$\tau$ ($10^{6}$)}}
\startdata
$101$ & 6  & 11  & 6  & 3.728 & $-$3.021 & 1.12 & 11.8 & 1.72$^{_{+0.60}}_{^{-0.48}}$ \nl
$104$ & 10 & 16  & 10 & 3.109 & $-$3.008 & 1.17 & 10.6 & 4.18$^{_{+1.62}}_{^{-1.35}}$ \nl
$108$ & 12 & 16  & 8  & 2.304 & $-$2.649 & 1.43 & 11.4 & 2.39$^{_{+0.67}}_{^{-0.57}}$ \nl
$113$ & 9  & 17  & 9  & 1.629 & $-$2.781 & 1.57 & 8.9 & 1.96$^{_{+0.54}}_{^{-0.45}}$ \nl
$118$ & 8  & 13  & 8  & 0.833 & $-$3.074 & 1.41 & 10.0 & 2.64$^{_{+0.89}}_{^{-0.78}}$ \nl
$119$ & 6  & 12  & 7  & 1.065 & $-$3.831 & 1.00 & 9.3 & 2.43$^{_{+0.86}}_{^{-0.72}}$ \nl
$128$ & 4  & 10  & 7  & 2.433 & $-$4.029 & 0.91 & 8.4 & 1.62$^{_{+0.62}}_{^{-0.47}}$ \nl
$159$ & 3  & 4   & 2  & 6.353 & $-$4.402 & 0.48 & 9.3 & 1.06$^{_{+0.84}}_{^{-0.66}}$ \nl
\enddata
\tablecomments{
Col. (1), field number Id.
Col. (2), number of events which pass cuts performed on SodoPhot photometry as of 
August 1999.
Col. (3), number of events from the Difference Image Analysis.
Col. (4), number of new events from this analysis.
Cols. (5) \& (6), field location.
Col. (7), Number of clump stars relative to field 119 (BW).
Col. (8), Percentage of disk stars in each field to $V=23$.
Col. (9), microlensing optical depth.}
\end{deluxetable}



\makeatletter
\def\jnl@aj{AJ}
\ifx\revtex@jnl\jnl@aj\let\tablebreak=\nl\fi
\makeatother

\begin{deluxetable}{lccccc}
\footnotesize
\tablecaption{Candidate Microlensing Events.\label{tab4}}
\tablewidth{0pt}
\tablehead{\colhead{Event Id} & \colhead{DIA Id} & \colhead{MACHO Id} &
\colhead{RA (J2000)} & \colhead{Dec (J2000)} & \colhead{S ($\arcsec$)}}
\startdata
95-BLG-d6\tablenotemark{v} & 101.14.1893 & 101.21688.5320 & 18 07 02.02 & -27 32 40.2 & 1.47\nl
95-BLG-30\tablenotemark{f} & 101.15.3933 &  101.21821.128 & 18 07 04.27 & -27 22 06.3 & 0.16\nl
95-BLG-d7 & 101.15.3935 & 101.21950.1897 & 18 07 25.07 & -27 24 41.1 & 1.28\nl
95-BLG-s8 & 101.15.3936 &  101.21691.836 & 18 06 52.77 & -27 23 19.8 & 0.55\nl
96-BLG-d2 & 101.19.3670 & 101.21171.4799 & 18 05 38.12 & -27 23 07.8 & 0.51\nl
97-BLG-24 & 101.21.3714 & 101.20650.1216 & 18 04 20.26 & -27 24 45.8 & 0.51\nl
97-BLG-42 & 101.22.3422 & 101.20914.3873 & 18 04 56.56 & -27 10 43.2 & 3.75\nl
95-BLG-5 & 101.23.3319 & 101.20658.2639 & 18 04 22.40 & -26 53 15.8 & 0.56\nl
97-BLG-s4 & 101.24.2939 & 101.21174.3417 & 18 05 38.80 & -27 08 29.5 & 1.06\nl
96-BLG-d3 & 101.24.2940 & 101.21174.2131 & 18 05 47.03 & -27 08 54.8 & 0.43\nl
95-BLG-15 & 101.26.2507 & 101.21564.4657 & 18 06 28.79 & -27 09 35.9 & 0.85\nl
95-BLG-s9 & 104.14.5859 & 104.21161.1997 & 18 05 34.46 & -28 02 51.7 & 0.24\nl
96-BLG-d4 & 104.15.7362 & 104.21162.3642 & 18 05 47.75 & -27 56 32.9 & 0.01\nl
96-BLG-d5 & 104.15.7365 &  104.21423.530 & 18 06 09.00 & -27 53 39.1 & 1.42\nl
96-BLG-d6 & 104.15.7366 & 104.21033.4316 & 18 05 29.53 & -27 54 00.6 & 2.38\nl
96-BLG-14 & 104.16.4493 & 104.21032.4118 & 18 05 15.39 & -27 58 24.4 & 0.71\nl
97-BLG-d5 & 104.16.4494 & 104.20901.1319 & 18 05 02.68 & -28 00 47.7 & 0.56\nl
96-BLG-12\tablenotemark{p} & 104.19.5184 &  104.20382.803 & 18 03 53.20 & -27 57 35.7 & 0.14\nl
96-BLG-1 & 104.19.5185 & 104.20645.3129 & 18 04 26.19 & -27 47 35.0 & 0.25\nl
97-BLG-38 & 104.19.5186 & 104.20514.1500 & 18 04 06.10 & -27 48 26.9 & 0.72\nl
95-BLG-d10 & 104.19.5187 &  104.20643.299 & 18 04 25.12 & -27 54 31.6 & 0.93\nl
97-BLG-18 & 104.20.5880 & 104.20121.1692 & 18 03 15.26 & -28 00 13.9 & 0.13\nl
97-BLG-58 & 104.24.4584 &  104.20515.498 & 18 04 09.68 & -27 44 35.1 & 0.27\nl
95-BLG-d11 & 104.24.4585 &  104.20517.707 & 18 04 06.15 & -27 39 21.4 & 0.21\nl
96-BLG-26 & 104.25.4571 & 104.20388.2766 & 18 03 53.97 & -27 33 30.5 & 0.08\nl
97-BLG-2 & 104.26.4393 & 104.20775.2644 & 18 04 50.73 & -27 45 57.3 & 0.09\nl
96-BLG-d7\tablenotemark{b} & 104.27.4089 & 101.20908.1433 & 18 04 57.73 & -27 33 18.3 & 0.71\nl
97-BLG-d6\tablenotemark{v} & 113.14.6365 &  113.19454.768 & 18 01 35.64 & -29 08 39.3 & 0.87\nl
95-BLG-d12 & 113.14.6367 & 113.19322.2128 & 18 01 15.47 & -29 18 06.6 & 1.33\nl
95-BLG-s13 & 113.16.6650 & 113.18934.4131 & 18 00 28.87 & -29 09 34.9 & 0.63\nl
96-BLG-d8 & 113.16.6651 &  113.19192.228 & 18 01 04.67 & -29 17 31.1 & 0.52\nl
96-BLG-d9 & 113.16.6652 & 113.18932.3227 & 18 00 26.34 & -29 17 36.1 & 3.01\nl
97-BLG-1\tablenotemark{b} & 113.18.6227 &  113.18674.756 & 17 59 53.38 & -29 09 07.8 & 0.47\nl
95-BLG-4 & 113.18.6228 & 113.18804.1061 & 18 00 03.41 & -29 11 04.3 & 0.73\nl
96-BLG-21 & 113.21.5667 & 113.18156.1823 & 17 58 43.15 & -29 00 30.0 & 1.83\nl
97-BLG-d7\tablenotemark{b} & 113.21.5669 &  113.18286.536 & 17 59 02.71 & -29 03 02.5 & 0.24\nl
96-BLG-s10 & 113.22.6004 & 113.18420.5494 & 17 59 25.01 & -28 46 32.9 & 0.61\nl
95-BLG-1 & 113.23.5372 & 113.18292.2374 & 17 59 00.57 & -28 36 57.3 & 0.53\nl
96-BLG-20 & 113.24.6037 & 113.18550.1664 & 17 59 40.59 & -28 47 24.9 & 0.53\nl
96-BLG-10 & 113.25.5974 & 113.18680.3511 & 18 00 02.01 & -28 45 17.6 & 0.68\nl
95-BLG-23 & 113.25.5975 & 113.18812.4511 & 18 00 03.61 & -28 39 14.8 & 1.15\nl
97-BLG-d8 & 113.26.5353 & 113.18938.3003 & 18 00 32.48 & -28 53 22.7 & 0.83\nl
95-BLG-d14 & 113.26.5354 & 113.18940.3399 & 18 00 34.56 & -28 47 06.5 & 2.35\nl
95-BLG-d15 & 113.26.5357 & 113.19070.2853 & 18 00 46.72 & -28 46 45.0 & 0.79\nl
96-BLG-8 & 118.15.7509 & 118.19184.3770 & 18 00 58.17 & -29 49 50.5 & 0.50\nl
96-BLG-d11 & 118.17.3390 & 118.18663.1884 & 18 00 01.63 & -29 52 19.7 & 0.76\nl
97-BLG-16 & 118.17.6294 & 118.18662.2180 & 17 59 56.37 & -29 56 37.5 & 0.70\nl
97-BLG-4 & 118.18.5693 & 118.18270.3615 & 17 59 04.71 & -30 07 06.5 & 0.90\nl
96-BLG-d12 & 118.18.6885 & 118.18531.1816 & 17 59 37.67 & -30 00 53.1 & 0.53\nl
97-BLG-8\tablenotemark{p} & 118.18.6886 &  118.18529.538 & 17 59 35.35 & -30 08 48.1 & 0.36\nl
95-BLG-d16 & 118.19.7905 &  118.18404.992 & 17 59 18.58 & -29 47 49.4 & 0.52\nl
95-BLG-d17\tablenotemark{v} & 118.20.1711 & 118.19050.2888 & 18 00 42.46 & -30 06 37.2 & 2.16\nl
96-BLG-d12 & 118.21.3144 & 118.18143.4794 & 17 58 34.55 & -29 53 13.4 & 2.98\nl
97-BLG-d9 & 118.23.7346 & 118.18019.3386 & 17 58 28.69 & -29 29 11.5 & 1.44\nl
95-BLG-10 & 118.23.7347 & 118.18018.2379 & 17 58 16.01 & -29 32 10.9 & 0.06\nl
96-BLG-d13 & 118.25.5470 & 118.18539.3614 & 17 59 34.95 & -29 30 04.2 & 0.86\nl
97-BLG-26\tablenotemark{p} & 118.26.5695 & 118.18797.1397 & 18 00 06.94 & -29 38 06.0 & 0.01\nl
97-BLG-d10 & 119.14.4347 &  119.20479.459 & 18 04 13.52 & -30 09 25.9 & 0.16\nl
95-BLG-11 & 119.14.4348 & 119.20738.3418 & 18 04 37.26 & -30 12 11.6 & 0.41\nl
97-BLG-14 & 119.15.5936 & 119.20480.2914 & 18 04 16.36 & -30 07 23.2 & 0.18\nl
97-BLG-37 & 119.17.5482 & 119.20352.2589 & 18 03 58.66 & -29 58 48.8 & 0.08\nl
95-OGLE-16 & 119.21.1551 & 119.19571.1616 & 18 02 07.62 & -30 01 12.7 & 0.46\nl
96-BLG-3\tablenotemark{b} & 119.22.1686 & 119.19444.2055 & 18 01 45.54 & -29 49 47.1 & 0.28\nl
95-BLG-d18 & 119.22.4857 & 119.19704.1819 & 18 02 22.70 & -29 50 35.2 & 0.74\nl
95-BLG-39 & 119.23.4960 & 119.19576.2024 & 18 02 04.76 & -29 43 15.6 & 0.27\nl
95-BLG-d19 &  119.23.541 &  119.19447.724 & 18 01 41.23 & -29 37 23.2 & 0.47\nl
95-BLG-d20 & 119.25.1896 & 119.20096.2073 & 18 03 07.84 & -29 40 09.7 & 1.04\nl
95-BLG-3 & 119.25.5509 & 119.19837.1072 & 18 02 37.52 & -29 39 35.9 & 0.12\nl
97-BLG-d11 & 119.26.5056 & 119.20223.2492 & 18 03 33.85 & -29 53 30.0 & 2.73\nl
96-BLG-d14 & 128.15.5153 & 110.22318.4078 & 18 08 24.14 & -28 54 59.9 & 2.21\nl
96-BLG-d15 & 128.16.4751 & 128.21923.1479 & 18 07 32.70 & -29 13 06.8 & 2.81\nl
96-BLG-s16 & 128.17.5429 & 128.22057.2384 & 18 07 38.96 & -28 57 11.8 & 0.05\nl
96-BLG-d17 & 128.20.3810 & 128.21145.1300 & 18 05 40.51 & -29 05 59.6 & 1.12\nl
96-BLG-d18 & 128.22.4279 & 128.21410.1924 & 18 06 07.41 & -28 47 25.5 & 0.25\nl
95-BLG-d21 & 128.23.4933 &  128.21153.867 & 18 05 36.73 & -28 32 41.5 & 0.08\nl
96-BLG-31 & 128.24.4809 & 128.21541.1133 & 18 06 42.39 & -28 41 15.9 & 0.12\nl
97-BLG-d12 & 128.24.4810 & 128.21671.4941 & 18 06 52.95 & -28 42 19.2 & 2.87\nl
95-BLG-d22 & 128.26.4587 &  128.21800.522 & 18 07 11.15 & -28 46 59.8 & 0.89\nl
95-BLG-18\tablenotemark{p} & 128.27.4562 & 128.21932.1362 & 18 07 20.56 & -28 36 51.1 & 0.57\nl
97-BLG-d13 & 159.14.3680 &  159.26652.168 & 18 18 21.08 & -25 56 37.1 & 0.43\nl
95-BLG-22 & 159.16.3693 & 159.26132.3182 & 18 17 14.80 & -25 55 58.2 & 0.20\nl
97-BLG-s14 & 159.21.3328 & 159.25486.1627 & 18 15 42.76 & -25 41 01.9 & 1.34\nl
97-BLG-d15\tablenotemark{v} & 159.25.4026 &  177.26012.459 & 18 16 51.10 & -25 19 03.8 & 1.10\nl
\enddata
\tablecomments{Col. (1) gives the event ID. Col. (2) gives the ID for DIA light curve. Col. (3)
gives the identification number of the nearest monitored SoDoPhot object. Cols. (4) \& (5) are 
the centroid location coordinates of the event. Col. (6), the seperation $S$ of the 
event centroid from the SoDoPhot object centroid.
We have not included events from field 108 as these have been presented in 
\shortciteN{ALC99c}. Events where the centroid offset $S$ is $> 1$ pixel from any SoDoPhot object are likely
Pixel lensing events.
}
\tablenotetext{f}{ Finite source event}
\tablenotetext{b}{ Binary lens event}
\tablenotetext{p}{ Parallax event}
\tablenotetext{v}{ Possible variable}
\end{deluxetable}



\makeatletter
\def\jnl@aj{AJ}
\ifx\revtex@jnl\jnl@aj\let\tablebreak=\nl\fi
\makeatother

\begin{deluxetable}{lccccccccc}
\footnotesize
\tablecaption{Parameters of Classical Microlensing Events.\label{tab5}}
\tablewidth{0pt}
\tablehead{\colhead{Event Id} & \colhead{$\hat t_{ns}$ (days)} & \colhead{$\hat t_{s}$ (days)} & \colhead{$t_{max}$} & \colhead{$\rm A_{ns}$} & \colhead{$\rm A_{s}$} & \colhead{$\rm V_{ns}$} & \colhead{$\rm V_{s}$} & \colhead{$(\rm V-R)_{ns}$} & \colhead{$(\rm V-R)_{s}$}}
\startdata
95-BLG-30 & $ 60.62\pm{0.03 }$ & $ 72.2\pm{0.3 }$ & $1321.38\pm{0.01 }$ & $ 14.06$ & $ 24.77$& $ 16.27$ & $ 16.59^{_{+0.01 }}_{^{-0.01 }}$ & $ 1.37$ & $ 1.39$\nl
95-BLG-s8 & $ 28.6\pm{0.2 }$ & $ 41.9\pm{1.8 }$ & $ 836.14\pm{0.06 }$ & $ 1.76$ & $ 2.94$& $ 17.17$ & $ 18.37^{_{+0.11 }}_{^{-0.10 }}$ & $ 0.55$ & $ 0.72$ \nl
97-BLG-24 & $ 9.7\pm{0.2 }$ & $ 25.4\pm{4.5 }$ & $1594.29\pm{0.11 }$ & $ 2.99$ & $ 42.80$& $ 17.98$ & $ 20.00^{_{+0.34 }}_{^{-0.26 }}$ & $ 0.59$ & $ 0.64$ \nl
95-BLG-5 & $ 17.7\pm{0.4 }$ & $ 17.6\pm{2.7 }$ & $ 827.90\pm{0.06 }$ & $ 6.80$ & $ 6.73$& $ 20.71$ & $ 20.62^{_{+0.29 }}_{^{-0.23 }}$ & $ 0.91$ & $ 0.84$ \nl
95-BLG-s9 & $ 30.4\pm{0.2 }$ & $ 41.0\pm{1.4 }$ & $ 925.21\pm{0.02 }$ & $ 5.99$ & $ 8.77$& $ 18.63$ & $ 19.10^{_{+0.05 }}_{^{-0.05 }}$ & $ 0.74$ & $ 0.74$ \nl
97-BLG-d5 & $ 59.1\pm{2.3 }$ & $111.3\pm{21.8}$ & $1527.99\pm{0.89 }$ & $ 1.36$ & $ 3.09$& $ 18.20 $ & $ 19.96^{_{+0.54 }}_{^{-0.36 }}$ & $ 0.76 $ & $ 0.72$ \nl
96-BLG-12 & $236.7\pm{0.9 }$ & $297.9\pm{11.0}$ & $1399.31\pm{0.32 }$ & $ 16.46$ & $ 70.65$& $ 17.77$ & $ 18.19^{_{+0.09 }}_{^{-0.08 }}$ & $ 0.89$ & $ 0.86$ \nl
96-BLG-1 & $147.4\pm{1.3 }$ & $158.2\pm{10.0}$ & $1180.88\pm{0.42 }$ & $ 1.69$ & $ 1.83$& $ 17.70$ & $ 17.89^{_{+0.20 }}_{^{-0.17 }}$ & $ 1.04$ & $ 1.03$ \nl
97-BLG-18 & $140.7\pm{0.9 }$ & $161.1\pm{11.4}$ & $1609.24\pm{0.20 }$ & $ 2.14$ & $ 2.56$& $ 18.57$ & $ 18.96^{_{+0.17 }}_{^{-0.15 }}$ & $ 0.82$ & $ 0.88$ \nl
97-BLG-58 & $ 52.2\pm{5.5 }$ & $ 61.9\pm{278.5}$ & $1683.67\pm{7.40 }$ & $ 1.64$ & $ 2.19$& $ 17.63$ & $ 18.42^{_{\; \dots}}_{^{-3.39 }}$\tablenotemark{a} & $ 0.93$ & $ 0.90$ \nl
95-BLG-d11 & $111.7\pm{21.4}$ & $124.8\pm{632.2}$ & $ 715.01\pm{41.88}$ & $ 1.51$ & $ 1.54$& $ 17.34$ & $ 18.35^{_{\; \dots}}_{^{-3.45 }}$ & $ 0.85$ & $ 0.97$ \nl
96-BLG-26 & $ 55.4\pm{0.7 }$ & $105.1\pm{8.3 }$ & $1315.03\pm{0.08 }$ & $ 5.03$ & $ 11.15$& $ 19.83$ & $ 20.54^{_{+0.12 }}_{^{-0.11 }}$ & $ 0.90$ & $ 0.56$ \nl
97-BLG-2 & $ 38.8\pm{0.7 }$ & $ 55.4\pm{6.9 }$ & $1523.27\pm{0.07 }$ & $ 4.61$ & $ 7.14$& $ 20.03$ & $ 20.29^{_{+0.21 }}_{^{-0.17 }}$ & $ 0.86$ & $ 0.64$ \nl
97-BLG-1 & $ 76.0\pm{0.2 }$ & $ 44.6\pm{0.9 }$ & $1511.65\pm{0.03 }$ & $ 4.76$ & $ 2.18$& $ 17.39$ & $ 16.05^{_{+0.06 }}_{^{-0.06 }}$ & $ 0.90$ & $ 0.84$ \nl
97-BLG-d7 & $ 14.7\pm{0.2 }$ & $ 22.1\pm{3.5 }$ & $1618.44\pm{0.14 }$ & $ 1.14$ & $ 1.45$& $ 17.04$ & $ 18.21^{_{+0.72 }}_{^{-0.43 }}$ & $ 0.83$ & $ 0.74$ \nl
95-BLG-1 & $ 40.6\pm{0.2 }$ & $ 57.5\pm{2.8 }$ & $ 816.80\pm{0.03 }$ & $ 8.56$ & $ 13.13$& $ 19.16$ & $ 19.61^{_{+0.07 }}_{^{-0.07 }}$ & $ 1.00$ & $ 0.94$ \nl
96-BLG-20 & $ 40.6\pm{0.6 }$ & $ 61.0\pm{6.0 }$ & $1273.57\pm{0.26 }$ & $ 1.83$ & $ 3.21$& $ 18.43$ & $ 19.33^{_{+0.25 }}_{^{-0.20 }}$ & $ 0.89$ & $ 0.69$ \nl
97-BLG-8 & $ 98.2\pm{0.2 }$ & $159.9\pm{2.1 }$ & $1575.50\pm{0.01 }$ & $ 17.24$ & $ 31.48$& $ 19.53$ & $ 20.18^{_{+0.02 }}_{^{-0.02 }}$ & $ 1.02$ & $ 1.02$ \nl
95-BLG-d16 & $ 8.1\pm{0.6 }$ & $ 18.1\pm{12.6}$ & $ 896.33\pm{0.40 }$ & $ 1.26$ & $ 4.16$& $ 18.23$ & $ 20.94^{_{\; \dots}}_{^{-0.89 }}$ & $ 0.85$ & $ 0.72$ \nl
95-BLG-10 & $ 79.3\pm{0.7 }$ & $ 89.3\pm{5.8 }$ & $ 869.15\pm{0.17 }$ & $ 2.31$ & $ 2.71$& $ 19.04$ & $ 19.29^{_{+0.16 }}_{^{-0.14 }}$ & $ 0.77$ & $ 0.70$ \nl
97-BLG-26 & $111.4\pm{0.2 }$ & $127.7\pm{1.9 }$ & $1636.63\pm{0.02 }$ & $ 7.05$ & $ 8.38$& $ 19.38$ & $ 19.47^{_{+0.02 }}_{^{-0.02 }}$ & $ 1.26$ & $ 1.16$ \nl
97-BLG-d10 & $ 70.0\pm{0.6 }$ & $ 87.2\pm{7.8 }$ & $1636.49\pm{0.24 }$ & $ 1.22$ & $ 1.45$& $ 16.91$ & $ 17.73^{_{+0.34 }}_{^{-0.26 }}$ & $ 0.73$ & $ 0.75$ \nl
95-BLG-11 & $ 22.5\pm{0.2 }$ & $ 31.2\pm{1.9 }$ & $ 853.36\pm{0.02 }$ & $ 22.89$ & $ 33.02$& $ 20.04$ & $ 20.58^{_{+0.09 }}_{^{-0.08 }}$ & $ 0.90$ & $ 0.98$ \nl
97-BLG-14 & $ 47.4\pm{1.4 }$ & $ 60.6\pm{12.1}$ & $1561.32\pm{0.36 }$ & $ 1.83$ & $ 2.40$& $ 19.72$ & $ 20.37^{_{+0.66 }}_{^{-0.41 }}$ & $ 0.79$ & $ 0.91$ \nl
97-BLG-37 & $ 21.6\pm{0.3 }$ & $103.9\pm{9.7 }$ & $1610.86\pm{0.08 }$ & $ 1.74$ & $ 13.88$& $ 17.48$ & $ 20.62^{_{+0.14 }}_{^{-0.12 }}$ & $ 0.54$ & $ 0.69$ \nl
95-OGLE-16 & \nodata\tablenotemark{b} & $ 41.8\pm{26.3}$ & $ 884.38\pm{0.68 }$ & \nodata & $ 1.35$ & \nodata & $ 19.45^{_{\; \dots}}_{^{-1.20 }}$ & \nodata & $ 0.89$ \nl
96-BLG-3 & $ 46.4\pm{0.2 }$ & $ 71.7\pm{3.8 }$ & $1167.33\pm{0.02 }$ & $ 14.38$ & $ 24.48$& $ 19.15$ & $ 19.85^{_{+0.07 }}_{^{-0.07 }}$ & $ 0.70$ & $ 0.81$ \nl
95-BLG-39 & $ 39.3\pm{0.7 }$ & $ 53.5\pm{5.2 }$ & $ 993.97\pm{0.19 }$ & $ 2.21$ & $ 3.66$& $ 18.77$ & $ 19.51^{_{+0.22 }}_{^{-0.19 }}$ & $ 0.65$ & $ 0.53$ \nl
95-BLG-d19 & $ 55.7\pm{1.6 }$ & $104.2\pm{31.7}$ & $ 936.83\pm{0.82 }$ & $ 1.15$ & $ 1.98$& $ 17.92$ & $ 19.84^{_{+1.51 }}_{^{-0.61 }}$ & $ 0.75$ & $ 0.53$ \nl
95-BLG-3 & $ 2.28\pm{0.04 }$ & $ 2.9\pm{0.6 }$ & $ 809.29\pm{0.02 }$ & $ 4.80$ & $ 7.05$& $ 18.93$ & $ 19.47^{_{+0.52 }}_{^{-0.35 }}$ & $ 1.01$ & $ 1.05$ \nl
96-BLG-s16 & $ 74.8\pm{2.2 }$ & $ 73.3\pm{18.0}$ & $1339.17\pm{0.77 }$ & $ 1.55$ & $ 1.51$& $ 19.12$ & $ 19.09^{_{+1.43 }}_{^{-0.60 }}$ & $ 0.51$ & $ 0.56$ \nl
96-BLG-d18 & $ 5.4\pm{0.2 }$ & $ 6.5\pm{2.7 }$ & $1255.38\pm{0.07 }$ & $ 1.69$ & $ 2.11$& $ 19.11$ & $ 19.62^{_{\; \dots}}_{^{-0.77 }}$ & $ 0.82$ & $ 0.63$ \nl
95-BLG-d21 & $ 31.8\pm{0.6 }$ & $ 35.6\pm{9.8 }$ & $ 872.37\pm{0.29 }$ & $ 1.32$ & $ 1.46$& $ 18.02$ & $ 18.35^{_{+1.87 }}_{^{-0.65 }}$ & $ 0.85$ & $ 0.79$ \nl
96-BLG-31 & $ 49.6\pm{0.4 }$ & $ 54.4\pm{1.8 }$ & $1373.14\pm{0.05 }$ & $ 4.80$ & $ 6.09$& $ 18.80$ & $ 18.88^{_{+0.08 }}_{^{-0.07 }}$ & $ 0.92$ & $ 0.80$ \nl
95-BLG-18 & $ 83.1\pm{0.9 }$ & $ 98.7\pm{10.1}$ & $ 904.18\pm{0.29 }$ & $ 1.70$ & $ 2.18$& $ 18.93$ & $ 19.55^{_{+0.31 }}_{^{-0.24 }}$ & $ 0.79$ & $ 0.93$ \nl
97-BLG-d13 & $ 5.82\pm{0.09 }$ & $ 25.7\pm{3.2 }$ & $1607.01\pm{0.05 }$ & $ 1.13$ & $ 6.39$& $ 16.75$ & $ 20.65^{_{+0.23 }}_{^{-0.19 }}$ & $ 0.74$ & $ 0.70$ \nl
95-BLG-22 & $ 19.9\pm{0.7 }$ & $ 20.1\pm{4.3 }$ & $ 899.82\pm{0.13 }$ & $ 4.97$ & $ 4.85$& $ 20.75$ & $ 20.80^{_{+0.54 }}_{^{-0.36 }}$ & $ 0.70$ & $ 0.79$ \nl
\enddata
\tablecomments{The results of two types of fits. One where the source flux is assumed to be that 
of the nearest SoDoPhot source. The other where the source flux is also used as a fit parameter.  The parameters from
the fits are denoted by the subscipts $ns$ (no source fit) ans $s$ (source fit) respectively.
We have not include events from field 108 as these have been presented in \shortciteN{ALC99c}.
Col. (1), the DIA light curve identification number. 
Cols. (2) \& (3), the fit timescale of the event. 
Col. (4), the time of maximum amplifiction (JD $- 2449000$).
Cols (5) \& (6), the amplificiations of these two fits. 
Col. (7), the baseline magnitude of the SoDoPhot object. 
Col. (8), the fitted source magnitude. 
Col. (9), the colour of the nearest SoDoPhot object.
Col. (10), the fit colour of the object.
In all cases the presented uncertainties are the formal $1 \sigma$ errors from the fitting process.
}
\tablenotetext{a}{ Events where a lower limit to the source flux was not determined.}
\tablenotetext{b}{ Nearest SoDoPhot source has $R_{m}$ band data only.}
\end{deluxetable}



\makeatletter
\def\jnl@aj{AJ}
\ifx\revtex@jnl\jnl@aj\let\tablebreak=\nl\fi
\makeatother
\begin{deluxetable}{lccccc}
\footnotesize
\tablecaption{Parameters of Pixel Microlensing Events.\label{tab6}}
\tablewidth{0pt}
\tablehead{\colhead{Event Id} & \colhead{$\hat t$ (days)} & \colhead{$t_{max}$} &
\colhead{A} & \colhead{V$_{fit}$} & \colhead{(V-R)$_{fit}$}}
\startdata
95-BLG-d6 & $ 9.5\pm{13.8}$ & $ 873.46\pm{0.60 }$ & $ 1.73$ & $ 20.56^{_{\; \dots}}_{^{-1.76 }}$ & $ 0.56$ \nl
95-BLG-d7 & $ 34.4\pm{16.8}$ & $ 872.24\pm{0.06 }$ & $ 17.36$ & $ 22.59^{_{+0.93 }}_{^{-0.49 }}$ & $ 0.91$ \nl
96-BLG-d2 & $ 30.0\pm{11.9}$ & $1222.84\pm{0.23 }$ & $ 3.99$ & $ 21.51^{_{+1.25 }}_{^{-0.57 }}$ & $ 1.02$ \nl
97-BLG-42 & $ 37.5\pm{12.5}$ & $1635.06\pm{0.34 }$ & $ 2.21$ & $ 20.41^{_{+1.67 }}_{^{-0.63 }}$ & $ 0.67$ \nl
97-BLG-s4 & $ 25.8\pm{5.2 }$ & $1533.26\pm{0.15 }$ & $ 7.68$ & $ 20.21^{_{+0.42 }}_{^{-0.30 }}$ & $ 0.73$ \nl
96-BLG-d3 & $107.7\pm{14.6}$ & $1172.75\pm{0.09 }$ & $ 19.03$ & $ 21.20^{_{+0.21 }}_{^{-0.18 }}$ & $ 0.57$ \nl
95-BLG-15 & $ 37.6\pm{4.0 }$ & $ 853.84\pm{0.04 }$ & $ 22.80$ & $ 21.16^{_{+0.16 }}_{^{-0.14 }}$ & $ 0.67$ \nl
96-BLG-d4 & $202.5\pm{72.2}$ & $1319.32\pm{0.78 }$ & $ 9.66$ & $ 22.10^{_{+0.73 }}_{^{-0.43 }}$ & $ 0.82$ \nl
96-BLG-d5 & $ 62.6\pm{15.8}$ & $1258.99\pm{0.39 }$ & $ 4.04$ & $ 21.15^{_{+0.64 }}_{^{-0.40 }}$ & $ 0.85$ \nl
96-BLG-d6 & $ 10.0\pm{2.2 }$ & $1247.00\pm{0.03 }$ & $ 49.41$ & $ 21.43^{_{+0.36 }}_{^{-0.27 }}$ & $ 0.83$ \nl
96-BLG-14 & $ 37.9\pm{6.7 }$ & $1244.09\pm{0.17 }$ & $ 3.26$ & $ 20.13^{_{+0.46 }}_{^{-0.32 }}$ & $ 0.68$ \nl
97-BLG-38 & $ 12.2\pm{3.3 }$ & $1616.52\pm{0.01 }$ & $ 33.85$ & $ 21.13^{_{+0.51 }}_{^{-0.35 }}$ & $ 0.86$ \nl
95-BLG-d10 & $ 84.7\pm{7.1 }$ & $ 866.39\pm{0.08 }$ & $ 10.93$ & $ 20.45^{_{+0.13 }}_{^{-0.12 }}$ & $ 0.73$ \nl
96-BLG-d7 & $ 18.9\pm{10.2}$ & $1221.76\pm{0.25 }$ & $ 2.12$ & $ 20.22^{_{\; \dots}}_{^{-0.88 }}$ & $ 0.86$ \nl
97-BLG-d6 & $ 12.4\pm{13.0}$ & $1632.81\pm{0.39 }$ & $ 1.58$ & $ 19.96^{_{\; \dots}}_{^{-1.51 }}$ & $ 0.58$ \nl
95-BLG-d12 & $ 91.4\pm{23.5}$ & $ 804.21\pm{0.90 }$ & $ 2.58$ & $ 20.15^{_{+0.92 }}_{^{-0.49 }}$ & $ 0.96$ \nl
95-BLG-s13 & $ 25.0\pm{15.2}$ & $806.19\pm{0.14 }$ & $ 87.45$ & $ 22.25^{_{+2.11 }}_{^{-0.67 }}$ & $ 0.96$ \nl
96-BLG-d8 & $ 31.4\pm{3.6 }$ & $1225.88\pm{0.02 }$ & $ 50.90$ & $ 22.51^{_{+0.15 }}_{^{-0.13 }}$ & $ 1.32$ \nl
96-BLG-d9 & $ 10.3\pm{3.9 }$ & $ 895.32\pm{0.02 }$ & $ 18.94$ & $ 22.07^{_{+0.62 }}_{^{-0.39 }}$ & $ 0.94$ \nl
95-BLG-4 & $ 13.7\pm{7.3 }$ & $ 790.09\pm{0.10 }$ & $ 4.44$ & $ 20.31^{_{+1.97 }}_{^{-0.66 }}$ & $ 0.91$ \nl
96-BLG-21 & $ 41.9\pm{7.2 }$ & $1272.15\pm{0.14 }$ & $ 6.56$ & $ 20.96^{_{+0.33 }}_{^{-0.25 }}$ & $ 0.78$ \nl
96-BLG-s10 & $ 35.5\pm{4.3 }$ & $1169.70\pm{0.03 }$ & $ 9.80$ & $ 20.40^{_{+0.19 }}_{^{-0.16 }}$ & $ 0.59$ \nl
96-BLG-10 & $ 62.0\pm{7.7 }$ & $1236.97\pm{0.07 }$ & $ 10.92$ & $ 21.05^{_{+0.21 }}_{^{-0.18 }}$ & $ 0.81$ \nl
95-BLG-23 & $ 21.1\pm{7.8 }$ & $ 900.93\pm{0.25 }$ & $ 4.03$ & $ 21.15^{_{+1.24 }}_{^{-0.56 }}$ & $ 0.83$ \nl
97-BLG-d8 & $ 18.0\pm{8.7 }$ & $1525.98\pm{0.21 }$ & $ 3.51$ & $ 20.87^{_{+2.61 }}_{^{-0.70 }}$ & $ 0.86$ \nl
95-BLG-d14 & $ 24.5\pm{3.4 }$ & $ 816.36\pm{0.04 }$ & $ 11.70$ & $ 20.10^{_{+0.24 }}_{^{-0.20 }}$ & $ 0.71$ \nl
95-BLG-d15 & $ 47.6\pm{10.0}$ & $ 886.35\pm{0.35 }$ & $ 1.90$ & $ 19.44^{_{+0.81 }}_{^{-0.46 }}$ & $ 0.77$ \nl
96-BLG-8 & $ 39.7\pm{9.8 }$ & $1224.63\pm{0.05 }$ & $ 20.82$ & $ 22.24^{_{+0.39 }}_{^{-0.29 }}$ & $ 1.00$ \nl
96-BLG-d11 & $ 49.3\pm{28.1}$ & $1249.74\pm{0.38 }$ & $ 5.98$ & $ 22.50^{_{+2.15 }}_{^{-0.67 }}$ & $ 0.79$ \nl
97-BLG-16 & $ 38.3\pm{8.6 }$ & $1567.20\pm{0.30 }$ & $ 3.13$ & $ 21.05^{_{+0.67 }}_{^{-0.41 }}$ & $ 1.12$ \nl
97-BLG-4 & $ 25.4\pm{9.0 }$ & $1518.15\pm{0.04 }$ & $ 51.38$ & $ 22.62^{_{+0.56 }}_{^{-0.37 }}$ & $ 0.97$ \nl
96-BLG-d12 & $171.0\pm{65.4}$ & $1183.51\pm{1.29 }$ & $ 4.42$ & $ 21.92^{_{+1.17 }}_{^{-0.55 }}$ & $ 0.64$ \nl
95-BLG-d17 & $ 12.0\pm{13.3}$ & $ 853.74\pm{0.50 }$ & $ 1.40$ & $ 18.67^{_{\; \dots}}_{^{-1.66 }}$ & $ 0.72$ \nl
96-BLG-d12 & $ 27.8\pm{13.1}$ & $1250.07\pm{0.19 }$ & $ 5.69$ & $ 21.78^{_{+1.38 }}_{^{-0.59 }}$ & $ 0.62$ \nl
97-BLG-d9 & $ 90.1\pm{32.3}$ & $1546.12\pm{0.30 }$ & $ 10.96$ & $ 22.84^{_{+0.75 }}_{^{-0.44 }}$ & $ 0.86$ \nl
96-BLG-d13 & $ 33.5\pm{10.1}$ & $1317.54\pm{0.15 }$ & $ 7.94$ & $ 21.71^{_{+0.62 }}_{^{-0.39 }}$ & $ 0.87$ \nl
\tablebreak
95-BLG-d18 & $ 20.2\pm{13.1}$ & $ 954.71\pm{0.24 }$ & $ 4.12$ & $ 21.75^{_{\; \dots}}_{^{-0.83 }}$ & $ 0.65$ \nl
95-BLG-d20 & $ 26.7\pm{4.7 }$ & $ 869.10\pm{0.06 }$ & $ 12.31$ & $ 20.76^{_{+0.30 }}_{^{-0.23 }}$ & $ 0.75$ \nl
97-BLG-d11 & $ 49.3\pm{10.1}$ & $1607.01\pm{0.31 }$ & $ 2.90$ & $ 20.13^{_{+0.60 }}_{^{-0.38 }}$ & $ 0.63$ \nl
96-BLG-d14 & $ 93.5\pm{15.7}$ & $1140.66\pm{12.00}$ & $ 14.31$ & $ 19.43^{_{+1.70 }}_{^{-0.63 }}$ & $ 0.82$ \nl
96-BLG-d15 & $ 26.3\pm{6.1 }$ & $1332.95\pm{0.06 }$ & $ 12.74$ & $ 21.57^{_{+0.46 }}_{^{-0.32 }}$ & $ 0.81$ \nl
96-BLG-d17 & $ 42.5\pm{7.3 }$ & $1257.79\pm{0.23 }$ & $ 2.70$ & $ 19.73^{_{+0.49 }}_{^{-0.34 }}$ & $ 0.93$ \nl
97-BLG-d12 & $ 11.5\pm{6.3 }$ & $1630.10\pm{0.11 }$ & $ 4.47$ & $ 21.72^{_{+2.75 }}_{^{-0.71 }}$ & $ 0.74$ \nl
95-BLG-d22 & $ 24.5\pm{5.3 }$ & $ 938.76\pm{0.04 }$ & $ 19.08$ & $ 21.12^{_{+0.40 }}_{^{-0.29 }}$ & $ 0.62$ \nl
97-BLG-s14 & $134.8\pm{16.9}$ & $1633.64\pm{0.19 }$ & $ 6.71$ & $ 21.48^{_{+0.22 }}_{^{-0.18 }}$ & $ 0.73$ \nl
97-BLG-d15 & $ 27.7\pm{19.4}$ & $1569.43\pm{0.65 }$ & $ 2.84$ & $ 21.41^{_{\; \dots}}_{^{-1.00 }}$ & $ 0.68$ \nl
\enddata
\tablecomments{Events where offset $> 1$ pixel from any SoDoPhot object or the fitted baseline magnitude 
is below the detection threshold for SoDoPhot. Parameters come from fits to the DIA photometry
light curves. In all cases the presented uncertainties are the 
formal $1 \sigma$ errors from the fitting process. Values of $t_{max}$ are (JD $- 2449000$).
Events for field 108 have been presented in \shortciteN{ALC99c}.
}
\tablenotetext{a}{ Events where a lower limit to the source flux was not determined.}
\end{deluxetable}

\begin{deluxetable}{clc}
\footnotesize
\notetoeditor{Table \ref{pars} should be rotated and occupy an entire page
if too big.}
\tablewidth{0pt}
\tablecaption{Measurements of Bar Orientation.\label{tab7}}
\tablehead{\colhead{Ref} & \colhead{Method} & \colhead{Inclination Angle ($\theta$)}
}
\scriptsize
\startdata
1 & Gas Dynamics ($\rm H_{I}$) & 30$-$45 \nl
2 & Gas Dynamics ($\rm H_{I}$,CO,CS) & 16 $\pm$ 2 \nl
3 & Main Sequence Star CMD & 18 $\pm$ 3 \nl
4 & Dirbe non-parametric deprojection & 10$-$40\nl
5 & 2-D Gas Simulations ($\rm H_{I}$, CO) & $>25$\nl
6 & Dirbe L,M-band deprojection &  15$-$35\nl
7 & COBE K-band constrained N-Body Sim& 28 $\pm$ 8\nl
8 & Star Counts & 19$\pm$1\nl
8 & Star Counts & 24$\pm$2\nl
9 & Red-clump giant numbers & 10$-$45\nl
10 & OGLE+MACHO Microlensing & $<$ 30\nl 
11 & Fux N-body \& OH/IR stars & 44 $\pm \sim$ 5\nl 
12 & 2-D simulations \& $\rm H_{I}$ data & 35$\pm$5 \nl 
\enddata
\tablerefs{
(1) \citeNP{DEV64}; (2) \citeNP{BGSBU91};
(3) \citeNP{BERT95}; (4) \citeNP{DAH95};
(5) \citeNP{WES96}; (6) \citeNP{BGS97};
(7) \citeNP{Fux97}; (8)  \citeNP{NL99};
(9) \citeNP{SUSK97};  (10) \citeNP{Gyuk99};
(11) \citeNP{SSV99}; (12) \citeNP{WES99}}.
\tablecomments{Col. (1), reference. 
Col. (2), the observation or method used in determination. 
Col. (3), the bar inclination angle relative to Sun-GC line-of-sight with 
uncertainties if given.}
\end{deluxetable}

\begin{deluxetable}{clcc}
\footnotesize
\tablecaption{Optical Depths from Models\label{tab8}}
\tablewidth{0pt}
\tablehead{\colhead{Ref} & \colhead{Model} 
& \colhead{$\rm M_{b} (10^{10} M_{\odot})$} & \colhead{ $\tau\; (10^{-6})$}
}
\startdata
1  & double exp disk & \nodata & 0.4$-$0.8\nl 
2  & double exp disk + halo & \nodata &  0.5$-$1.1\nl
3 &  symetric bulge + massive disk & 1.9 & 1.9\nl
4 &  bar + truncated disk & 2.0& 2.2 $\pm$ 0.45\nl 
5 &  bar + double exp disk & 1.8 &  1.9\nl
6 &  N-body model (m08t3200)  & 3.0\tablenotemark{a} & 1.8\nl
6 &  N-body model (m04t3000)  & 5.0\tablenotemark{a} & 2.0\nl
7 &  non-bisymmetric disk  & 1.65 & 1.1$-$1.8\nl
7 &  bisymmetric disk  & 1.65 & 1.1$-$1.6\nl
8 &  bar + nucleus + dble exp disk & 2.2 & 1.54\nl 
8 &  bar + nucleus + dble exp disk & 3.3 & 2.14\nl 
9 &  Fux N-body \& OH/IR stars & 2.0 & 2.2\nl 
10 &  maximum likelihood  & \nodata & $1.93\pm0.39$\nl
11 &  thick \& thin disk & 1.8 & 1.9\nl
12 & Schwarzchild orbits & \nodata & 1.4\nl
\enddata
\tablerefs{
(1) \citeNP{PAC91}; (2) \citeNP{GAAB91};
(3) \citeNP{Evans94};
(4) \citeNP{ZSR95}; (5) \citeNP{ALC97e};
(6) \citeNP{Fux97} (7) \citeNP{NL99};
(8) \citeNP{Peale98}; (9) \citeNP{SSV99};
(10) \citeNP{Gyuk99}; (11) \citeNP{GJSDP99};
(12) \citeNP{HAF99}.}
\tablecomments{Col. (1), reference.
Col. (2), characteristic feature of model. Col. (3), Bulge/Bar
mass used in model (if known). Col. (4), optical depth obtain from model.
}
\tablenotetext{a}{ Spheroid plus nucleus mass}
\end{deluxetable}

\end{document}